\documentclass[useAMS,usenatbib]{mn2e}

\usepackage{graphicx}
\usepackage{txfonts}
\usepackage{amssymb,alltt}
\usepackage{color}
\usepackage{hyperref}
\usepackage{supertabular}
\hypersetup{colorlinks, linkcolor=blue}

\usepackage{caption}
\usepackage{subcaption}
\usepackage{multicol}
\usepackage{float}

\def\Lya{\mbox{Ly\,$\alpha$ }}
\def\HII{\mbox{H\,{\sc ii}}}
\def\SiIIa{\mbox{Si\,{\sc ii}}~$\lambda 1260$ }
\def\SiIIb{\mbox{Si\,{\sc ii}}~$\lambda 1304$ }
\def\CII{\mbox{C\,{\sc ii}}~$\lambda 1334$ }
\def\SiIIc{\mbox{Si\,{\sc ii}}~$\lambda 1526$ }
\def\OI{\mbox{Si\,{\sc ii}}~$\lambda 1302$ }

\def\HeII{\mbox{He\,{\sc ii}}~$\lambda 1640$ }
\def\OIIIUV{\mbox{O\,{\sc iii}}]~$\lambda \lambda 1661,66$ }
\def\OIII{\mbox{O\,{\sc iii}}]~$\lambda 5009$ }
\def\CIII{\mbox{C\,{\sc iii}}]~$\lambda \lambda 1907,09$}
\def\NIV{\mbox{N\,{\sc iv}}]~$\lambda \lambda 1483,86$ }
\def\NIVratio{\mbox{N\,{\sc iv}}]~$\lambda 1483$/$\lambda 1486$ }
\def\CIIIratio{\mbox{C\,{\sc iii}}]~$\lambda 1907$/$\lambda 1909$ }
\def\SiIIIratio{\mbox{Si\,{\sc iii}}]~$\lambda 1883$/$\lambda 1892$ }
\def\OIIratio{\mbox{O\,{\sc ii}}]~$\lambda 3729$/$\lambda 3726$ }
\def\SiIV{\mbox{Si\,{\sc iv}}~$\lambda 1394,1403$ }
\def\CIV{\mbox{C\,{\sc iv}}~$\lambda 1548,51$ }
\def\SiIII{\mbox{Si\,{\sc II}}~$\lambda 1882,92$}

\title[]{A young star-forming galaxy at $z=3.5$ with an extended Lyman $\alpha$ halo seen with MUSE}
\author[Patr\'icio et al. 2015]
{Vera Patr\'icio$^{1}$\thanks{E-mail:
vera.patricio@univ-lyon1.fr}, Johan Richard$^{1}$, Anne Verhamme $^{1,2}$, Lutz Wisotzki$^{3}$, Jarle Brinchmann$^{4,5}$,
\newauthor
Monica L. Turner$^{4}$, Lise Christensen$^{6}$, Peter M. Weilbacher$^{3}$, J\'er\'emy Blaizot$^{1}$, Roland Bacon$^{1}$
\newauthor
Thierry Contini$^{7,8}$, David Lagattuta$^{1}$,  Sebastiano Cantalupo$^{9}$, Benjamin Cl\'ement$^{1}$, 
 \newauthor
 Genevi\`eve Soucail$^{7,8}$\\
 \\
$^{1}$CRAL, Observatoire de Lyon, Universit\'e Lyon 1, 9 Avenue Ch. Andr\'e, 69561 Saint Genis Laval Cedex, France\\
$^{2}$ Observatoire de Gen\`eve, Universit\'e de Gen\`eve, 51 Ch. des Maillettes, 1290 Versoix, Switzerland\\
$^{3}$ AIP, Leibniz-Institut f\"ur Astrophysik Potsdam (AIP) An der Sternwarte 16, D-14482 Potsdam, Germany\\
$^{4}$ Leiden Observatory, Leiden University, P. O. Box 9513, NL-2300 RA Leiden, Netherlands\\
$^{5}$ Instituto de AstrofÕ\'isica e Ci\^encias do Espa\c{c}o, Universidade do Porto, CAUP, Rua das Estrelas, PT4150-762 Porto, Portugal\\
$^{6}$ Dark Cosmology Centre, Niels Bohr Institute, University of Copenhagen, Juliane Maries Vej 30, 2100 Copenhagen, Denmark\\
$^{7}$ IRAP, Institut de Recherche en Astrophysique et Plan\'etologie, CNRS, 14, avenue Edouard Belin, F-31400 Toulouse, France\\
$^{8}$ Universit\'e de Toulouse, UPS-OMP, Toulouse, France
$^{9}$ ETH Zurich, Institute of Astronomy, HIT J 12.3 , Wolfgang-Pauli-Str. 27, CH-8093 Zurich, Switzerland\\
}

\begin{document}

\date{Accepted later. Received earlier; in original form one of these days.}

\pagerange{\pageref{firstpage}--\pageref{lastpage}} \pubyear{2015}

\maketitle

\label{firstpage}

\begin{abstract}

Spatially resolved studies of high redshift galaxies, an essential insight into galaxy formation processes, have been mostly limited to stacking or unusually bright objects. We present here the study of a typical (L$^{*}$, M$_\star$ = 6 $\times 10^9$ $M_\odot$) young lensed galaxy at $z=3.5$, observed with MUSE, for which we obtain 2D resolved spatial information of \Lya and, for the first time, of {\mbox{C\,{\sc iii}]}} emission. The exceptional signal-to-noise of the data reveals UV emission and absorption lines rarely seen at these redshifts, allowing us to derive important physical properties (T$_e\sim$15600 K, n$_e\sim$300 cm$^{-3}$, covering fraction f$_c\sim0.4$) using multiple diagnostics. Inferred stellar and gas-phase metallicities point towards a low metallicity object (Z$_{\mathrm{stellar}}$  =  $\sim$ 0.07 Z$_\odot$ and Z$_{\mathrm{ISM}}$ $<$ 0.16 Z$_\odot$). The \Lya emission extends over $\sim$10 kpc across the galaxy and presents a very uniform spectral profile, showing only a small velocity shift which is unrelated to the intrinsic kinematics of the nebular emission. The \Lya extension is $\sim$4 times larger than the continuum emission, and makes this object comparable to low-mass LAEs at low redshift, and more compact than the Lyman-break galaxies and Ly$\alpha$ emitters usually studied at high redshift. We model the \Lya line and surface brightness profile using a radiative transfer code in an expanding gas shell, finding that this model provides a good description of both observables. 

\end{abstract}

\begin{keywords}
techniques: imaging spectroscopy; gravitational lensing: strong; galaxies: high-redshift; galaxies: abundances; galaxies: individual: SMACSJ2031.8-4036
\end{keywords}


\begin{figure*}
\begin{center}
	\begin{subfigure}[b]{0.65\textwidth}
		\includegraphics[height=\textwidth]{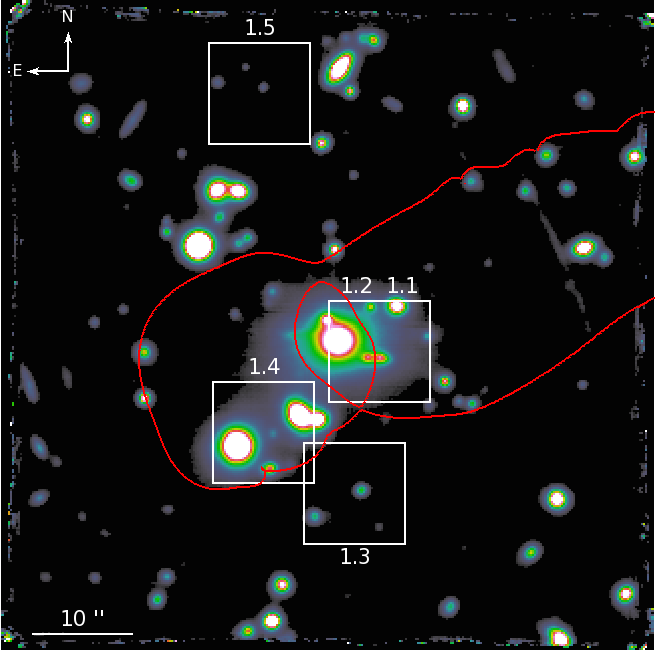}\\
	\end{subfigure}
		\quad
	\begin{subfigure}[b]{0.161\textwidth}
		\includegraphics[height=\textwidth]{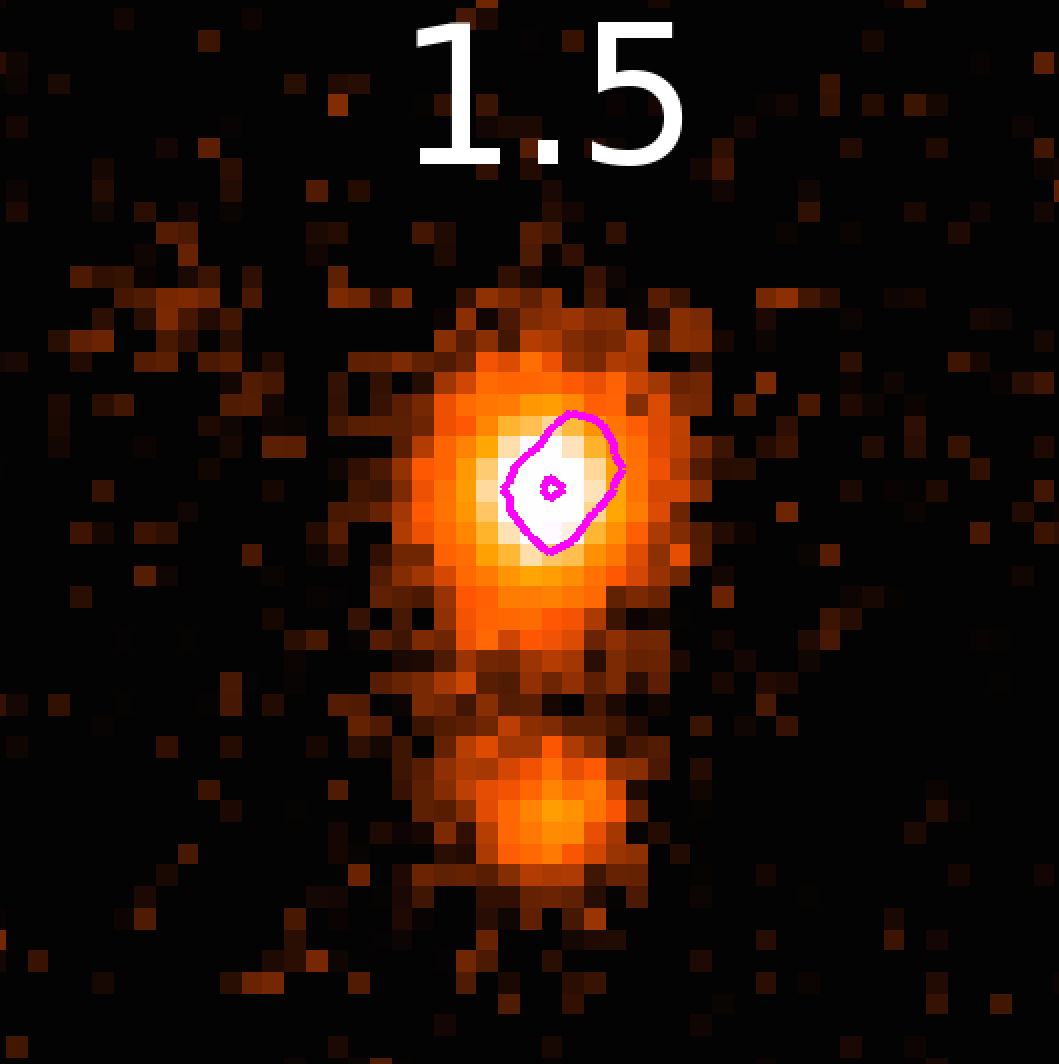}\\
		\includegraphics[height=\textwidth]{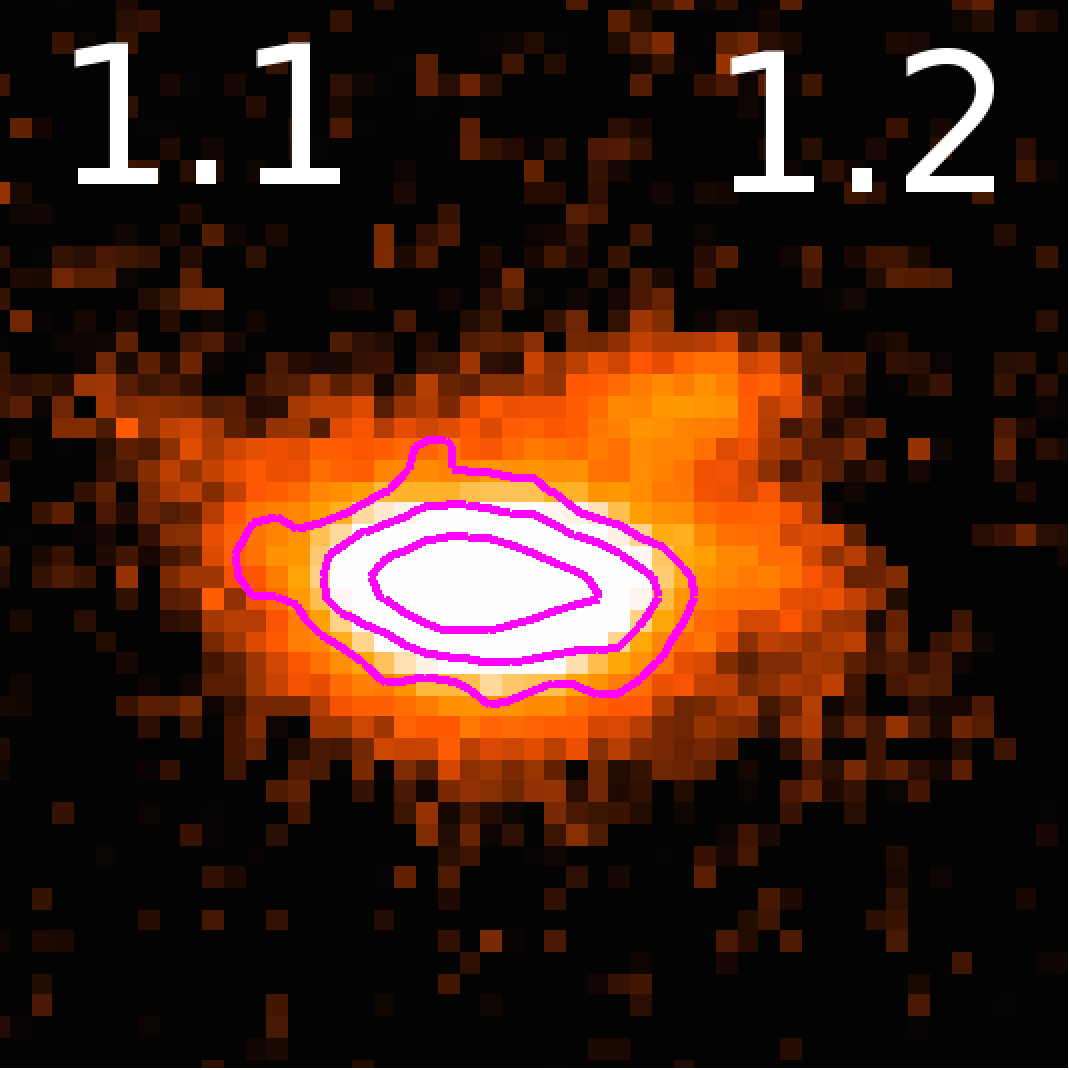}\\
		\includegraphics[height=\textwidth]{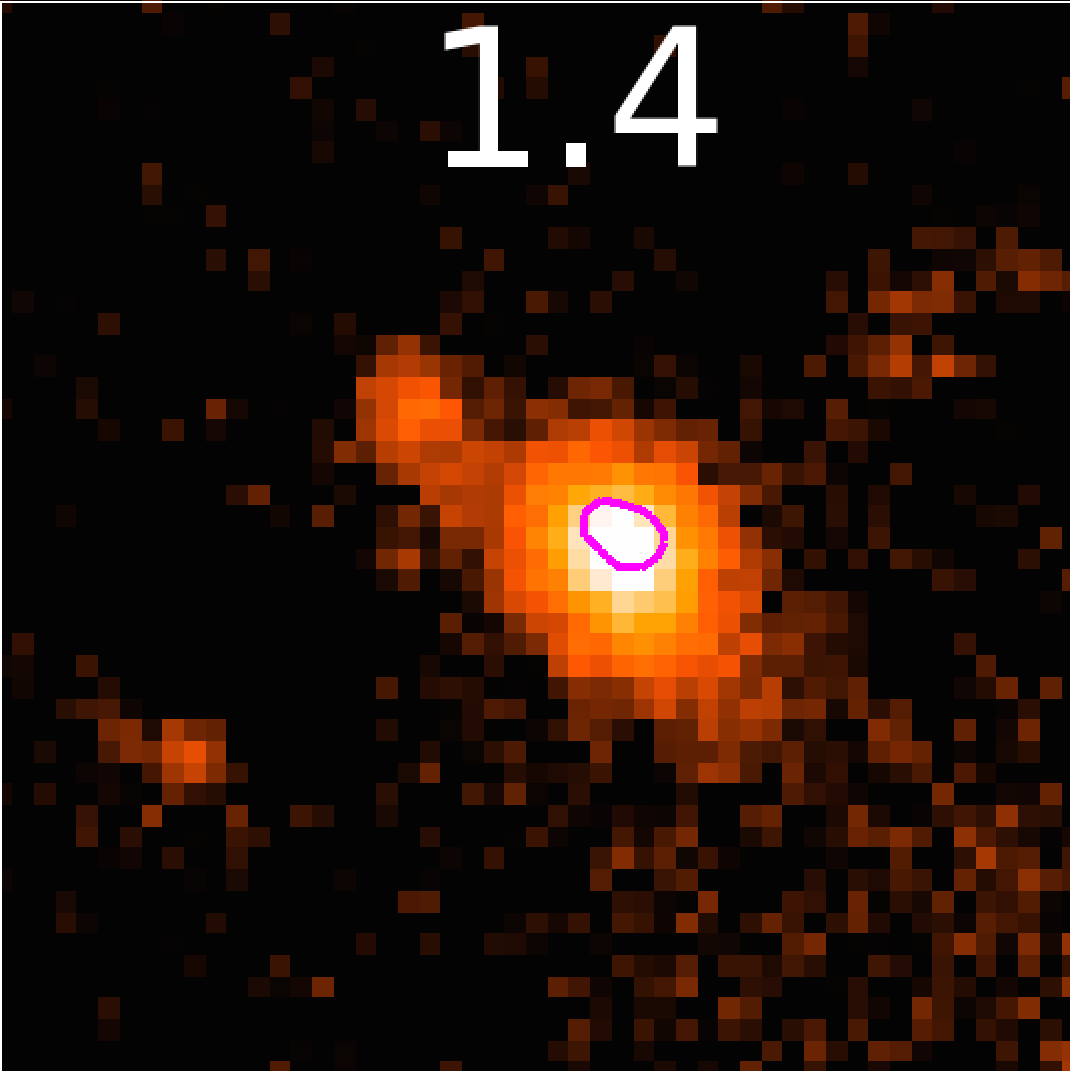}\\
		\includegraphics[height=\textwidth]{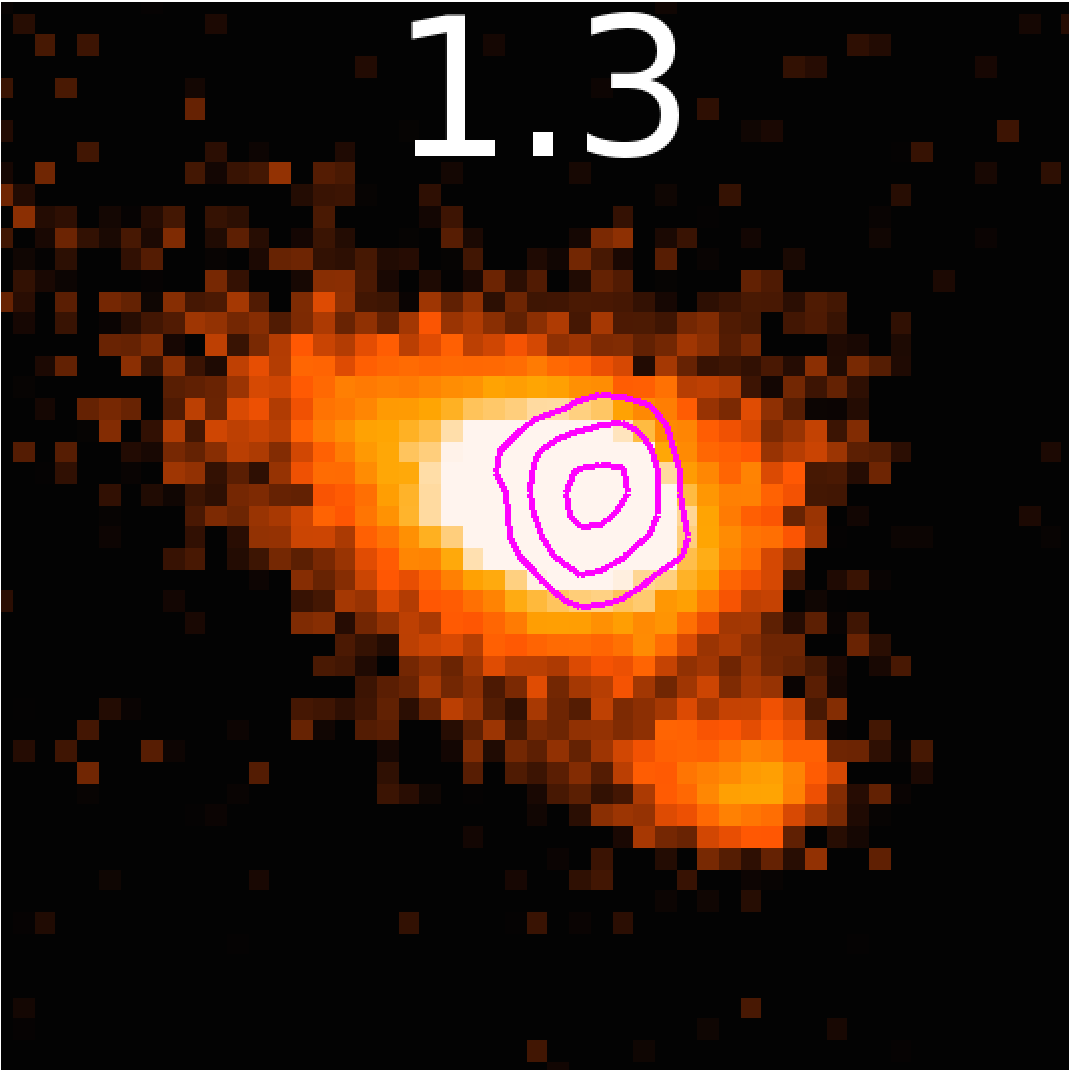}\\	
	\end{subfigure}
	\caption[FOV]{Left: MUSE white light (4750--9350 $\AA$) image of the SMACS2031 cluster core. \textit{Red line}: Critical line of the mass model (from \citealt{Richard+15}), \textit{white squares:} 10'' $\times$ 10'' regions around multiple sources that were zoomed in to produce the right side images. Right: \Lya  (5476--5494 $\AA$) line images of system 1 multiple images with $\mbox{C\,{\sc iii}}]$ (8578--8607 \AA) contours overplotted in magenta. All five multiple images are seen both in \Lya and $\mbox{C\,{\sc iii}}]$ line images. The \Lya emission extends up to 4'' and a small companion, previously unknown, can be seen at about $3''$ of the main body in the \Lya line image.}
\label{white}
\end{center}
\end{figure*}

\section{Introduction}

During the past few decades, our understanding of galaxy formation and evolution has made significant progress thanks to the hundreds of high redshift ($z>3$) galaxies which 
have been detected in dedicated observing campaigns \citep[e.g.][]{Shapley+03,Vanzella+09,Stark+13}. The main spectral feature used to confirm the distances of these galaxies is the \Lya emission, since it is the brightest emission line we can observe in distant sources. Unfortunately, many of these objects are too faint to show any other emission line at rest-frame UV wavelengths or, even less likely, continuum and absorption lines, offering a limited picture of the characteristics of high redshift galaxies. The complexity of the \Lya resonant process, which depends not only on the gas dynamics but also on gas density and dust content, also requires the observation of non resonant lines in order to robustly probe the physical properties of such galaxies.

Stacking techniques, which combine spectra or images of dozens or even hundreds of objects in order to increase signal-to-noise (e.g. \citealt{Shapley+03,Hayes+14,Momose+14,Erb+14}), allow the study of statistically significant properties of high redshift galaxies. In particular their rest-frame UV and optical nebular lines as well as their spatial extension both in continuum and \Lya can be probed \citep{Steidel+11}, providing an essential clue to understand when most galaxies gather their mass. However, these techniques erase the structure, kinematics and other resolved properties of individual sources, and with it the possibility of learning something new about the detailed physical processes that shape galaxy evolution. Studies of the full 2D extent of \Lya emission in individual sources have been implemented with narrow-band imagers or integral field unit spectrographs (IFU), but have generally focused on extreme objects such as giant \Lya blobs or powerful radio galaxies. The intrinsic brightness of these objects makes it possible to probe deeply into the physics of the gas, even to resolve the spatial variation of the \Lya line at kpc scales (e.g. \citealt{Weijmans+10,Swinbank+15,Prescott+15}), but  ideally, one would like to pursue similar studies on typical high redshift galaxies, which requires a very high signal-to-noise only achievable with the next generation of ground-based optical and near-infrared telescopes.

One way forward is to focus on high-redshift lensed galaxies, since gravitational lensing not only boosts the total observed flux of the sources but also enlarges them, making them ideal targets for resolved properties studies. There is already a small but growing collection of such lensed galaxies studied in rest-frame UV from $z=1.4$ to 4.9 \citep{Pettini+02,Fosbury+03,Villar-Martin+04,Cabanac+08,Quider+09,Dessauges-Zavadsky+10,Christensen+12a,Christensen+12b,Bayliss+14}, that probe L$^{\star}$ type galaxies at a resolution impossible to achieve without the lensing effect. To date, many of these studies have been performed with long-slit spectroscopy, due to technological constraints, which allows properties such as metallicity, star formation rate and age to be derived but with limited spatially-resolved information. IFU observations are therefore desirable in order to obtain a full 2D picture of such resolved properties in individual galaxies. However, even with magnification produced by lensing, this is also challenging from a technological point of view, and currently studies have focused mainly on $z<3$ galaxies in the near-infrared \citep{Stark+08,Yuan+11,Jones+13a}. So far there have only been two examples of such studies at $z>3$: the $z=5$ lensed galaxies in MS1358.4+6245 and RCS0224-0002 \citep{Swinbank+07,Swinbank+09}.

Combining gravitational lensing magnification with the unique efficiency of the MUSE (Multi Unit Spectroscopic Explorer) integral-field spectrograph it is already possible to obtain increase the number of resolved studies in high redshift sources. Here we present the 2D morphology and kinematics of a $M_\star =10^9 M_\odot$ galaxy as well as the analysis of emission and absorption features seldom accessible at these redshifts. This galaxy is a strongly-lensed system of 5 images at $z=3.5$ (Fig.~\ref{white}), first reported by \citet{Christensen+12a}. It was detected in the HST image of the massive cluster 
SMACSJ2031.8-4036 obtained as part of SMACS, the Southern extension of the MAssive Cluster Survey (MACS, \citealt{Ebeling+01}). New spectroscopic data were obtained 
during MUSE commissioning time, giving rise to an improved lens model of the cluster core \citep{Richard+15}.

The paper is organised as follows: in Sect. \ref{dataset} we detail the observations and data reduction; in Sect. \ref{dataanalysis} we describe the data analysis, including the extraction of spectra and line images from the MUSE data cube. The results and discussion of spectral and morphological features are presented in Sect. \ref{results}, while detailed spectral modelling of the \Lya profiles and its interpretation is shown in Sect. \ref{Lyafit}. In Sect. \ref{summary} we summarise and conclude. Throughout this paper, we adopt a $\Lambda$-CDM cosmology with $\Omega$=0.7, $\Omega_m$=0.3 and H = 70 km\,s$^{-1}$\,Mpc$^{-1}$ and use proper transverse distances for sizes and impact parameters.

\section{MUSE observations and data reduction}
\label{dataset}

MUSE is an integral field optical spectrograph (4750 to 9350 $\AA$ in nominal mode) with a field of view of $1 \times$1 arcmin$^2$ \citep{Bacon+10}. The field of view is divided into 24 sub-fields (\textit{channels}) and each of them is sliced into 48 which are dispersed by a grating and projected onto the detector (one per channel). The data has a spatial sampling of $0.2''\times0.2''$ and a spectral sampling of 1.25 $\AA$ for each voxel (volumetric pixel) in the final data cube. MUSE large field of view and medium resolution spectroscopy ($\sim$1500--3500), makes it particularly efficient in the study of gravitational lenses, since it simultaneously provides both the spatial location and redshift of multiply lensed images, which are essential for constraining the lensing models.

The data used in this paper were obtained over several nights of the second MUSE commissioning run, between April 30 and May 7 2014. The observing strategy combined a small dithering pattern with rotations of the field of view. This was done in order to avoid systematics by minimising the slice pattern that arises during the image reconstruction process from the small gaps between slices at the CCD level. The pointing coordinates were randomly selected from a box of $1.2''\times1.2''$ centred on the cluster and the rotation angle was varied by 90 degrees between two consecutive pointings. The instrumental wobble introduces an additional random offset of 0.1-0.3$''$. In total, 33 exposures of 1200 s each were acquired with a measured maximum offset of $0.82''$ in RA and $0.98''$ in DEC from the central point.\\

The data reduction was performed with version 1.0 of the MUSE pipeline (Weilbacher al., in prep). Basic calibration --- which includes bias subtraction, flat fielding correction, wavelength and geometrical calibrations --- is applied to individual exposures using the \textit{muse\_bias}, \textit{muse\_flat}, \textit{muse\_wavecal} and \textit{muse\_scibasic} pipeline recipes. Calibration data were chosen to match, closely as possible, the instrument temperature of each exposure, since channel alignment is sensitive to this temperature. After these first processing step, exposures are saved as 24 individual pixel tables, one per channel, where listing the calibrated fluxes at each CCD pixel. 

The next step is applying flux and astrometric corrections and combines the several channels of each exposure with the \textit{muse\_scipost} recipe. The pipeline performs sky subtraction by fitting a model of the sky spectrum to the data and subtracting it. Several tests were performed with different fractions of the field-of-view used to estimate the sky spectrum and different spectral samplings of the continuum and lines, but since there were no significant quality differences between the several trials, we adopted the default parameters: 30\% of sky fraction, continuum and line sampling of 1.25 \AA. This process results in 33 individual data cubes with $312\times 322 \times 3681$ voxels (spatial $\times$ spatial $\times$ spectral). Telluric correction, derived from the standard star, is also applied at this point of the reduction. Sky subtraction is furthermore improved using {\sc zap} (Zurich Atmosphere Purge, Soto et al., in prep.), a principal component analysis method, aimed at removing systematics remaining after sky subtraction. {\sc zap} estimates sky residuals in spatial regions free from bright continuum objects and subtracts these residuals in the entire field of view. Finally, several standard stars were reduced and their respective response curves produced and compared. We used them to calibrate the flux in science frames, rejecting response curves taken under non photometric conditions.

Before combining all 33 individual cubes into one single data cube, the sky transmittance during each exposure was estimated by measuring the integrated flux of the brightest stars in the field. The highest transmittance was taken as the photometric reference and in the final combination each exposure's flux was rescaled to match the reference. The positions of the stars were measured and used to correct the small offsets between exposures (due to the aforementioned instrumental wobbling) before averaging all cubes into a single data cube. A 3 $\sigma$-clipping rejection for cosmic rays was used during the combination and the combined cube was also corrected with {\sc zap}. Since the background spectral shape was found not to be flat we estimated the median value of the background on each wavelength plane, by masking bright objects and averaging the 100 nearest planes (50 to the red and 50 to the blue), then subtracted this value from the respective wavelength plane.

We measured the seeing by fitting a 2D circular moffat profile to the brightest star, obtaining FWHM values in the combined cube of 0.72'' and 0.63" at 6000 and 9000 $\AA$ respectively (see Table \ref{table_seeing} for details of individual exposures), with a wavelength dependence similar to the deep MUSE data from \cite{Bacon+15}. Photometry was checked to be in accordance with the HST data up to 8$\%$ by comparing the flux of the brightest star measured on HST data and on an equivalent image produced by integrating the MUSE cube using the HST transmittance filter. The variance is propagated by the pipeline trough the reduction process for each individual CCD pixel, resulting in a final cube with a variance value associated to each voxel.\\

\begin{table}
\begin{center}
    \begin{tabular}{|c|c|c|cl}
                \hline  
        DATE &  Exposure time (s) & Seeing ('')  & Conditions\\   
                \hline \hline
 		2014-04-30	&	6	$\times$ 	1200		& 	0.96		&	Photometric\\
		2014-05-01	&	4	$\times$	 1200	&	1.09		&	Clear \\
		2014-05-02	&	4	$\times$	 1200	&	1.23		&	Clear\\
		2014-05-04	&	3	$\times$  	1200		&	0.85		&	Clear\\
		2014-05-05	&	4	$\times$	1200		&	0.86		&	Photometric\\
		2014-05-06	&	7	$\times$	 1200	&	0.67		&	Photometric\\
		2014-05-07	&	5	$\times$ 	1200		&	0.58		&	Photometric\\
		Combined Cube &	33    $\times$  	1200		&	0.72          &			\\     
	         \hline  
     \end{tabular}
	 \caption[MUSE observations]{Exposure list with mean seeing of each night exposure measured around 6000 \AA.}      
     \label{table_seeing}          
\end{center}
\end{table}


\section{Data Analysis}
\label{dataanalysis}

In this section we describe the steps taken to extract emission line images and spectra needed to measure the spectrophotometric properties of the system from the MUSE data cube. Since this is a lensed system, we also detail the corrections applied to recover the undistorted morphology of the galaxy as well as its intrinsic flux.

\subsection{Line image analysis}

In order to study the \Lya emission morphology we extract an image around the \Lya line (5480--5490 \AA, observed) from the data cube, choosing the spectral width that maximise its signal-to-noise ratio. We correct this image by subtracting a continuum line image, produced by averaging two other images, one blueward and the other redward of \Lya. An equivalent procedure is used to create a line image of the $\mbox{C\,{\sc iii}}]$ emission (8580--8595 \AA), carefully choosing the red and blue continuum images to avoid emission lines and sky residuals. 

\begin{figure}
	\includegraphics[width=0.5\textwidth]{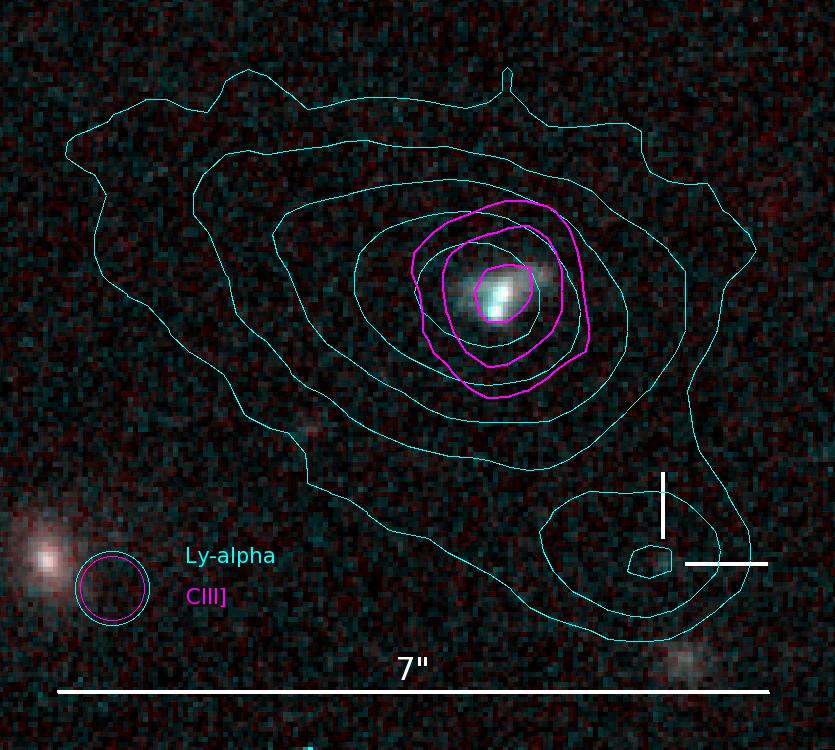}\\		
\caption{Image 1.3 in the HST V+I band, with \Lya (cyan) and $\mbox{C\,{\sc iii}}]$ (magenta) MUSE contours in geometric scale from 3 to 50 $\sigma$. The white cross marks the location of the faint companion in HST (F814W $\sim$28 AB), clearly detected with MUSE. Although $\mbox{C\,{\sc iii}}]$ peaks at the same location has the HST continuum, \Lya appears offseted by $\sim$0.4". The bottom left circles show the FWHM of the MUSE PSF.}
\label{im3}
\end{figure}

All five multiple images of system 1 detected with HST are seen both in \Lya and $\mbox{C\,{\sc iii}}]$ line images, with smaller spatial extension in $\mbox{C\,{\sc iii}}]$ and an offset of up to $\sim$ 0.4" between the \Lya and $\mbox{C\,{\sc iii}}]$ peaks (see Fig.~\ref{im3}). We adopt the same nomenclature as \cite{Richard+15} to refer to the multiple images, from the most magnified image --- 1.1$\&$1.2 --- to the least. All multiple images show consistent spatially extended emission in \Lya, up to 4'' in image 1.3. A small companion (identified as system 2 in \citealt{Richard+15}) is also clearly visible near each image, about 2.3--3.2'' away from the main component, and is associated with a very faint (F814W $\sim$ 28 AB) HST point source (Fig.~\ref{im3}). 

\begin{figure*}
	\begin{subfigure}[b]{0.24\textwidth}
		\includegraphics[width=\textwidth]{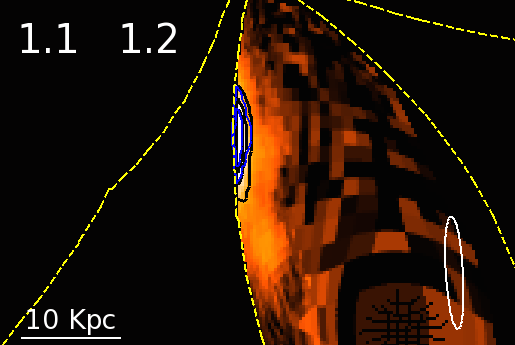}\\
	\end{subfigure}	
	\begin{subfigure}[b]{0.24\textwidth}
			\includegraphics[width=\textwidth]{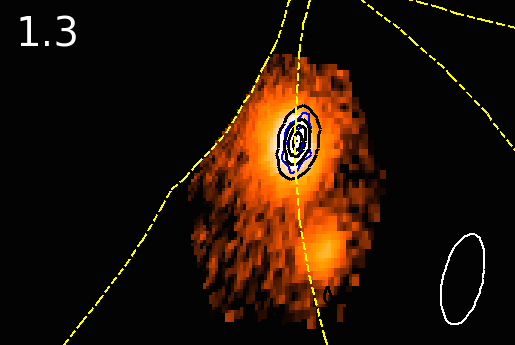}\\		
	\end{subfigure}
	\begin{subfigure}[b]{0.24\textwidth}
		\includegraphics[width=\textwidth]{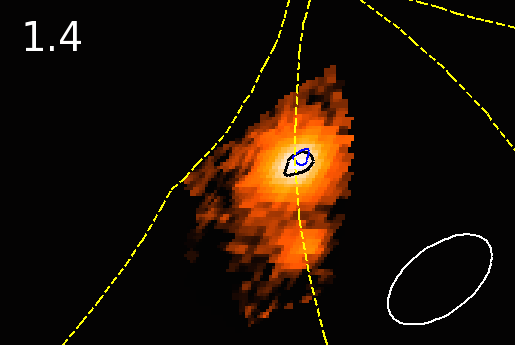}\\
	\end{subfigure}
	\begin{subfigure}[b]{0.24\textwidth}
		\includegraphics[width=\textwidth]{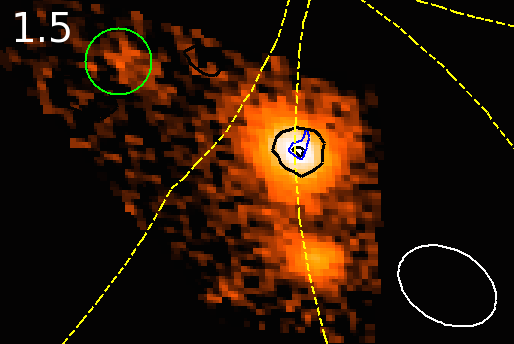}\\
	\end{subfigure}.
	\caption{Source plane reconstruction of the \Lya line image. \textit{Dashed Yellow:} Caustic lines. \textit{Blue contours}: $\mbox{C\,{\sc iii}}]$ line image reconstruction. \textit{Black contours}: continuum near \Lya image reconstruction. The ellipses on the lower right corners correspond to the source plane PSF, obtained by 	reconstructing the 2D moffat profile of the measured seeing (at the wavelength of \Lya ) into the source plane. The small differences between the source plane images is well explained by the different PSF shapes. Note that images 1.1 and 1.2 are radial images and only cover the western half of the source.}
\label{source_plane}
\end{figure*}

\subsection{Source plane reconstruction}

We use the well-constrained mass model derived from the 12 multiple image systems identified in the HST image and detected in the MUSE data cube \citep{Richard+15} to demagnify the images and recover the intrinsic source morphology. This is done by inverting the lens equation and putting the observed pixels on a regular source plane grid, while conserving surface brightness. All 5 multiple images are comparable to each other in the source plane, though some small differences in configuration arise due to the position of the caustic lines along the extension of the galaxy and the source plane PSF shape (see Fig.~\ref{source_plane}). The small offset between \Lya and continuum peak emission is confirmed in all images ($\sim$0.6 kpc). The relative position of the companion in all 5 multiple images is well reproduced by the model, which confirms that this smaller source is physically close to the main body ($\sim$11.2 kpc). The companion is $\sim$ 7 times intrinsically fainter in \Lya than the main body and it is not detected in the $\mbox{C\,{\sc iii}}]$ line image. This is confirmed in the 1D spectrum, where no other emission line besides \Lya is detected (see Sect. \ref{specfeatures}).

Although images 1.1$\&$1.2 have a very high magnification factor, since they are very close to the critical line, neither is a complete image of the original source. Conversely, image 1.3, the second brightest image, is very nearly a complete image of the source, thus we will focus the spatial analysis on this observation. Image 1.5 is the only truly complete image of the galaxy, but is also the least magnified. In this multiple image, an additional, faint, \Lya structure can be seen towards the north-east (green circle in the right panel of Fig.~\ref{source_plane}), which is not expected to be seen in images 1.1, 1.2 and 1.3 since it lies outside of the caustic line. 

\subsection{Spectrum extraction and magnification factors}

To maximise the signal-to-noise ratio in the extracted spectrum, we use the $\mbox{C\,{\sc iii}}]$ line image to define a 3$\sigma$ surface brightness threshold at $\sim 2.8 \times 10^{-19}\ \mathrm{erg/s/cm}^2/ \mathrm{arcsec}^2$. This level encloses three compact regions at the peak of images 1.1, 1.2 and 1.3, all probing the same physical region in the source plane (within 500 pc) despite having different magnification factors. Light contamination by cluster members is a concern at the location of images 1.1$\&$1.2 and we correct it by subtracting a scaled cluster member spectrum, using 
the spectral slope of image 1.3 (further from cluster members) as a reference. We find that the continuum slope of this decontaminated spectrum is in good agreement with the fit from \cite{Christensen+12a}, who obtained a slit spectrum of the central part of image 1.2.

After extracting this high signal-to-noise spectrum (which throughout the analysis is refered to as the \textit{combined spectrum}), we investigate differences between spectra extracted from different locations, focusing on image 1.3. Inspecting the spectral profile of \Lya on a pixel-by-pixel basis, we find that --- remarkably --- all profiles have a similar appearence, regardless of their location in the galaxy. This motivated us to extract spectra from regions where physical differences would be expected. In particular, we choose to explore two different regions: the central part of the galaxy, with strong $\mbox{C\,{\sc iii}}]$ and continuum emission, and the outer part (halo), containing 
\Lya emission.

\begin{figure}
\begin{center}
	\includegraphics[width=0.50\textwidth]{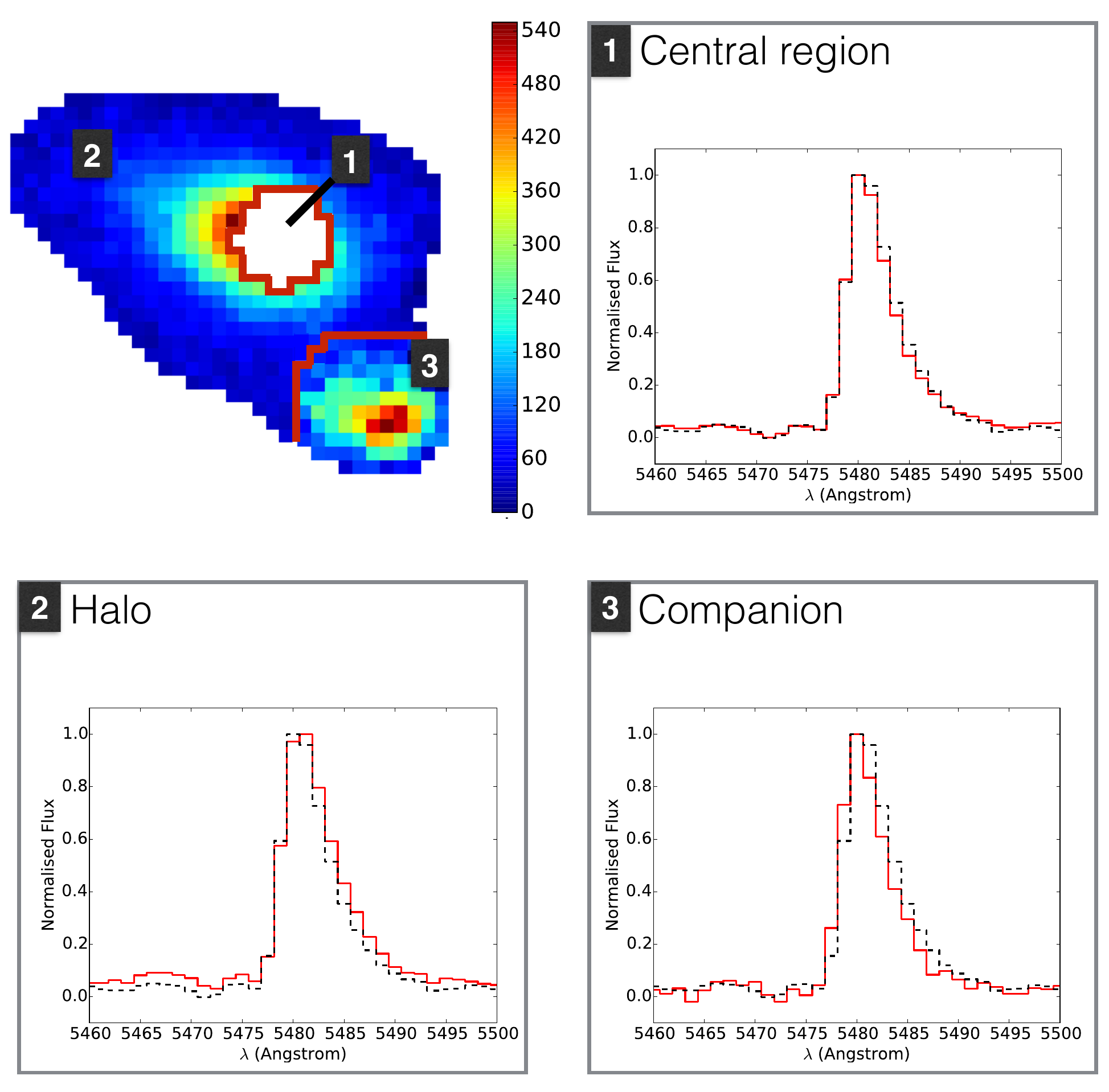}
\caption{ The three extraction regions shown for image 1.3: central (1) in white, halo (2) in colour and companion (3) also in colour (colour bar encodes the observed \Lya surface brightness). The extraction region of the total spectrum corresponds to the sum of regions 1 and 2. The panels show the \Lya line of regions 1 to 3 and the one extracted from the total region (in dashed black). }
\label{regions}
\end{center}
\end{figure}

We start with the previously defined $\mbox{C\,{\sc iii}}]$ regions and extract a \textit{central} spectrum in image 1.3. We then define in a similar way a more extended region containing the full extent of detected \Lya emission, excluding the companion. A 2$\sigma$ threshold in the \Lya line image ($\sim 3.75 \times 10^{-19}\ \mathrm{erg/s/cm}^2 /\mathrm{arcsec}^2$), defines this \textit{total} spectrum (see upper left panel of Fig.~\ref{regions}). The \textit{halo} spectrum is defined by excluding the central region from this total spectrum. Due to the seeing, we estimate 38\% of the central continuum emission to be present in the halo spectrum, that we correct by rescaling the central spectrum flux to include this missing fraction, and correct the halo spectrum to remove the contribution from the central region. Finally, the spectrum of the companion is selected by defining a circular region of 1.2" radius centred on the brightest pixel of image 1.3, and the respective spectra are extracted. 

In total, the extraction results in the 5 following spectra that we use in the next sections to derive the physical properties of the system:

\begin{itemize}
\item \textbf{Combined:} with maximum signal-to-noise, achieved by combining the central region of images 1.1\&1.2 and 1.3;
\item \textbf{Total}: extracted from an extended region of image 1.3 based on the \Lya line image;
\item \textbf{Central:} extracted from the central region of image 1.3 based on the $\mbox{C\,{\sc iii}}]$ line image;
\item \textbf{Halo:} extracted from the total region of image 1.3 but excluding the central region;
\item \textbf{Companion:} nearby compact source detected in \Lya line image (see Fig.~\ref{im3}).
\end{itemize}

The magnification factors of the several regions needed to recover the intrinsic fluxes of these spectra were calculated using the \cite{Richard+15} mass model. Images 1.1\&1.2 are a special case, since they do not image the entire galaxy: the central region corresponds only to 43\% of the total image and the extended region (defined in \Lya ) to 37\% . Taking these factors into account, we estimate a total magnification factor of 26 for the combined spectra. After magnification correction, the central region contains $40\pm2$\% of the total flux and the halo $60\pm2$\%.

\begin{table}
\begin{center} 
\begin{tabular}{l r | r | r | r |r|r| }
                \hline  
         &  Flux  10$^{-18}$& EW & FWHM & $\Delta$v&$\mu$ \\   
        		    &   [$\mathrm{erg/s/cm}^2] $   &    [$\AA$]          &  [km\,s$^{-1}$] & [km\,s$^{-1}$] & \\
              \hline \hline               
Total  		&	$409\pm11$  		&	$32\pm4$ 			&	248	&	$176\pm9$	&	8.5\\
Centre		&	$205\pm8$ 		& 	$20\pm2$				&	225	&	$172\pm9$	&	7.4\\
Halo			&	$206\pm4$		&	$49\pm12$ 			&	260	&	$178\pm9$	&	8.6\\
Companion	&	$41\pm6$		&	$^{(*)}360\pm70$			&	247	&	$156	\pm10$&	5.1\\
\hline
Combined 		&	$526\pm6$	 	&	$18\pm24$			&	216	&	$186\pm9$	&	26\\
                 \hline  
     \end{tabular}
	 \caption[Lyman alpha tables]{\Lya line measurements on image 1.3 (total, centre and halo, see Fig. \ref{regions}) and on the combined (high signal-to-noise) spectrum from all multiple images. Observed fluxes are \textit{uncorrected} for magnification and EW are given in rest frame. $\Delta$v is relative to the systemic redshift. $\mu$ is the magnification factor of the corresponding region. (*) Estimated from the F814W photometry}      
     \label{table_lya}          
\end{center}
\end{table}

In the following sections the detailed physical properties of this object will be discussed in detail. With a magnification corrected rest-frame UV magnitude of -21.14 AB, this galaxy has a typical L$^*$ luminosity at this redshift \citep{Steidel+11}. It has a global equivalent width of 32 $\AA$, classifying it as a \Lya emitter, and a \Lya luminosity of 5.6$\times$10$^{42}$ erg s$^{-1}$, similar to the typical luminosity of narrow band surveys for redshifts z$\sim$3 \citep{Garel+15}.

\section{Results and discussion}
\label{results}

In this section we present and discuss the integrated properties that can be derived from the spectral analysis of the combined spectrum, such as temperature, density and metallicity. We then focus on the resolved properties of image 1.3, comparing \Lya with continuum and $\mbox{C\,{\sc iii}}]$ emission.

\subsection{Spectral features}
\label{specfeatures}

\begin{table*}
\begin{center}
    \begin{tabular}{l l | r | r | r | r | }
                \hline  
        Line & $ \lambda_{rest}  $ & $ \lambda_{obs}  $ & Flux $\times 10^{-20}$& $\sigma$   & EW   \\   
        		&[$\AA$]  &[$\AA$] 			&[$\mathrm{erg/s/cm}^2$]	& 	[km\,s$^{-1}$] &[$\AA$] \\
                \hline \hline
\Lya 						& 	1215.67	&	$5479.02\pm0.25$	&	$51100\pm165$	&	$110.5\pm0.8$	&	$17.22\pm2.44$	\\
\mbox{N\,{\sc v}}			& 	1242.80	&	$5604.83\pm0.41$	&	$2560\pm138$		&	$257.9\pm14.0$&	$0.67\pm0.04$	\\
\mbox{Si\,{\sc ii}}$^*$		&	1264.74	&	$5699.55\pm0.97$	&	$417	\pm103$		&	$24.8\pm20.2$	&	$0.11\pm	0.01$	\\
$[$\mbox{N\,{\sc iv}}]			&	1483.32 	&	$6683.47\pm0.25$	&	$613	\pm58$		&	$104.0\pm4.2$	&	$0.21\pm0.02$	\\
\mbox{N\,{\sc iv}}]			&	1486.50	&	$6697.77\pm0.25$	&	$2996\pm80$		&	$103.7\pm4.2$	&	$1.04\pm0.05$	\\
\mbox{He\,{\sc ii}}			&	1640.42 	&	$7390.39\pm0.28$	&	$2301\pm114$		&	$76.9\pm6.7$	&	$0.99\pm0.10$	\\
\mbox{O\,{\sc iii}}]	 		&	1660.81 	&	$7481.77\pm0.27$	&	$1479\pm92$		&	$58.6\pm6.2$	&	$0.65\pm0.05$	\\
\mbox{O\,{\sc iii}}]			&	1666.15	&	$7505.92\pm0.25$	&	$3002\pm56$		&	$54.1\pm1.4$	&	$1.33\pm0.09$	\\
$[$\mbox{Si\,{\sc iii}}] 		&	1881.96	&	$8481.49\pm0.26$	&	$1363\pm52$		&	$55.3\pm2.7$	&	$0.81\pm0.06$	\\
\mbox{Si\,{\sc iii}}]  			&	1892.03	&	$8523.66\pm0.28$	&	$1105\pm66$		&	$62.2\pm4.2$	&	$0.67\pm0.05$	\\
$[$$\mbox{C\,{\sc iii}}]$ 		&	1906.68	&	$8589.61\pm0.25$	&	$5090\pm72$	&	$60.0\pm1.0$	&	$3.14\pm0.19$	\\
$\mbox{C\,{\sc iii}}]$			&	1908.73 	&	$8598.84\pm0.25$	&	$3417\pm77$	&	$59.9\pm1.0$	&	$2.11\pm0.13$	\\                 \hline  
     \end{tabular}
	 \caption[Emission lines]{Emission line measurements from the central region spectrum combining signal from all multiple images. Flux corresponds to observed flux (without magnification correction). Observed wavelengths ($\lambda_{obs}$) are given in air and $\sigma$ was corrected for instrumental broadening. }      
     \label{table_emission}          
\end{center}
\end{table*}

To derive the integrated properties we make use of the combined spectrum, which corresponds to the central region of the galaxy, but has higher signal-to-noise that the central spectrum of image 1.3 alone (see Fig.~\ref{spectralines}). We measure a systemic redshift of $3.50618\pm0.00005$ by simultaneously fitting the strongest available emission lines --- \HeII, \OIIIUV and \CIII. \citet{Christensen+12a} report $z=3.5073$ from multiple emission lines but the small difference is due to an incorrect heliocentric-barycentric velocity correction in their analysis. Line properties are measured by fitting a Gaussian profile to emission lines and a constrained Gaussian fit to the $\mbox{C\,{\sc iii}}]$ doublet, fixing the wavelength ratio between lines to its theoretical value. As for the \Lya emission line, and some absorption lines where the profile was clearly asymmetric, we defined an \textit{asymmetric Gaussian profile} by allowing the FWHM of two half gaussians to freely vary, while forcing the peak value of both to coincide at the same wavelength. Instead of using the pipeline propagated variance to estimate the errors, as it does not include noise correlation, we generate five hundred realisations of the observed spectrum, by randomly picking the flux level at each wavelength from a Gaussian distribution. The mean of this gaussian distribution is the observed spectrum intensity and its sigma is the pipeline propagated error. The absolute value of the error was scaled so that the realisations would keep the same signal-to-noise ratio as the observed continuum spectrum, so that we empirically reproduce pixel correlation. The wavelength calibration error (0.03 \AA) was added in quadrature to error estimations of the line peaks. Results for the combined spectrum are listed in Table \ref{table_emission}. 

The most noticeable characteristic of the four \Lya profiles from the different regions defined in image 1.3 --- central, halo, total and companion (see Fig.~\ref{regions}) --- is how similar they all look: the line profile is clearly asymmetric with very close velocity shifts relative to the systemic redshift ($\sim178$ km\,s$^{-1}$), except the companion, which is slightly blue-shifted ($\sim20$km\,s$^{-1}$) relative to the other 3 spectra (see Table \ref{table_lya}). A similar result has been reported by \cite{Swinbank+07} in a $z = 4.88$ lensed galaxy and in the local galaxy Haro 2 \citep{Mas-Hesse+03}.

Unlike the companion and halo spectra, which do not show any other prominent line besides \Lya, the central spectrum displays a wealth of other emission lines, that can be used to derive many integrated properties. We identify the following lines: the \OIIIUV and \CIII\ doublets which are clearly resolved, the \NIV doublet and the \HeII line, previously undetected in \citealt{Christensen+12a} (see Fig.~\ref{spectralines}) and \SiIII. Also worth noticing is the {\mbox{Si\,{\sc ii}}}$^*$ emission line, a fine structure transition line also reported by \cite{Erb+10} in a low metallicity $z=2.3$ lensed galaxy and in local galaxies (e.g. LARS --- \citealt{Rivera-Thorsen+15}) and Green Peas galaxies \cite{Henry+15}.

The average intrinsic velocity dispersion, after correcting for the line spread function, is $61\pm8$ km\,s$^{-1}$, excluding the $\mbox{N\,{\sc v}}]$ line and the $\mbox{N\,{\sc iv}}]$ doublet that show a higher dispersion. Our data also confirm that there is no AGN present in this system, since the measure a  $\mbox{C\,{\sc iii}}]$ / \Lya ratio of 0.07, much lower than the 0.125 value measured in a sample of narrow-line AGN LBG \citep{Shapley+03}. Finally, we derive a rest-frame UV absolute magnitude of $-21.1$ AB (magnification corrected and assuming no extinction), equivalent to an L$^*$ galaxy at this redshift. Using the \citet{Kennicutt98} conversion (SFR = L$_{\nu} \times 1.4\, 10^{-28} / \mu $), assuming a Salpeter IMF, we obtain a current, magnification corrected, SFR of $\sim17.5$ $M_{\odot}$/yr.

\begin{figure}
	\includegraphics[width=0.5\textwidth]{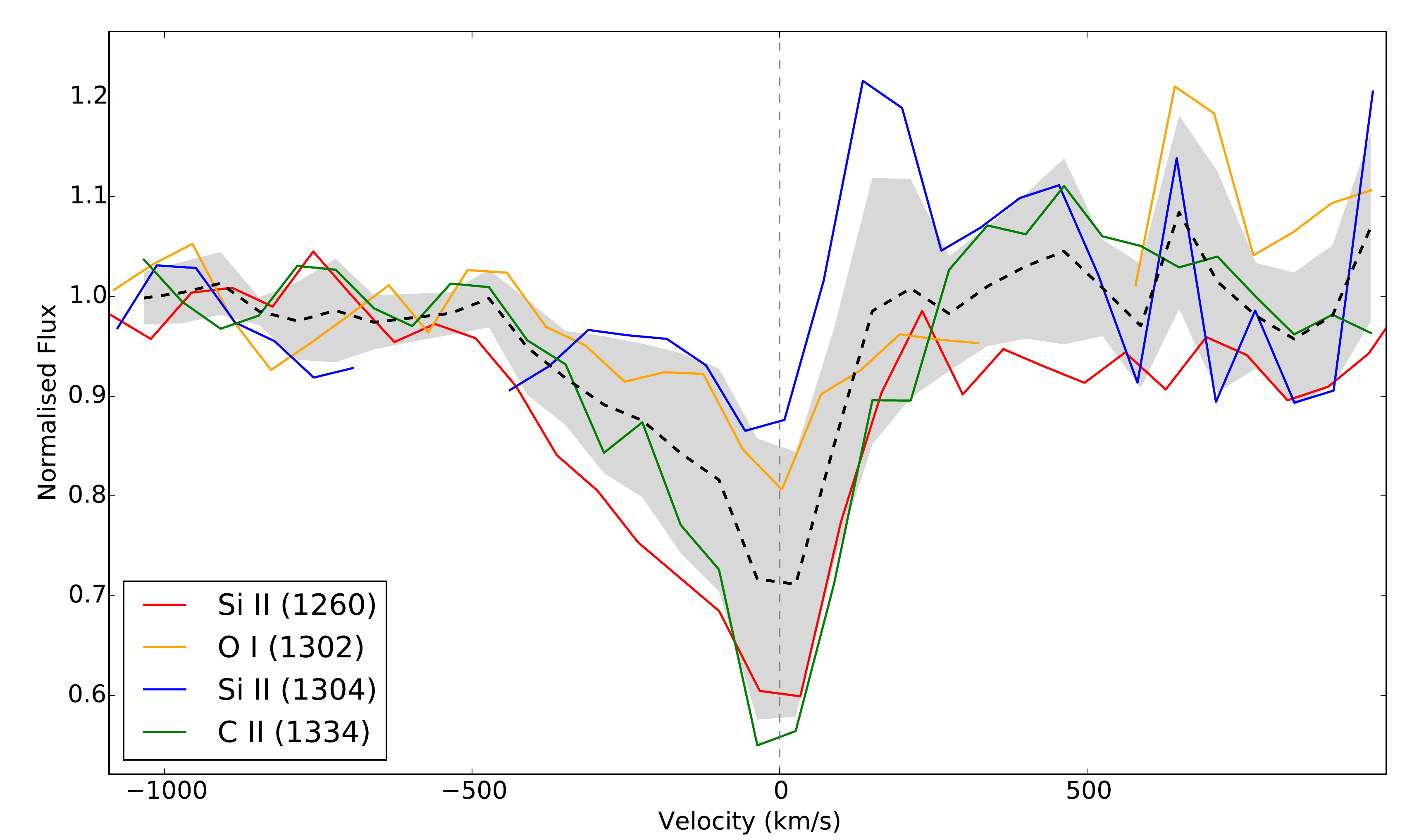}\\	
	\includegraphics[width=0.5\textwidth]{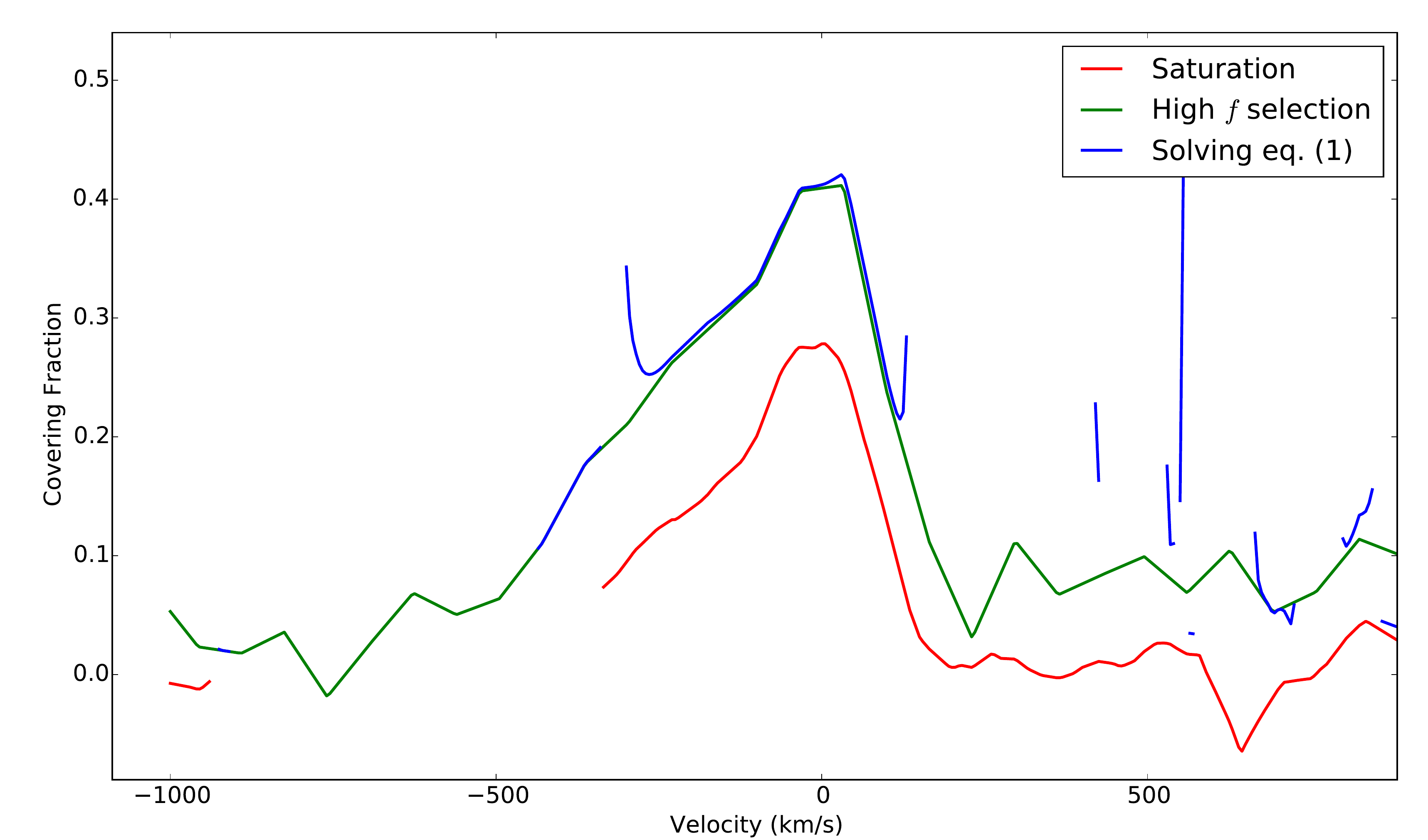}\\		
\caption{Upper panel: Low ionisation absorption lines on rest velocity frame. \textit{Black}: Mean profile with standard deviation in grey shadow. The maximum absorption at zero velocity suggests that most of the neutral gas is at rest, although the extended blue wind indicates that some outflowing gas is also present. Lower panel: covering fraction derived by 3 different methods. \textit{Blue}: directly fitting equation \ref{eq} to the \SiIIa, \SiIIb, \SiIIc lines. \textit{Red}: using all low ionisation lines from the first panel and calculating covering fraction assuming saturation. \textit{Green}: Selecting only the lines with the highest oscillator strengths (\SiIIa ($f\sim1.22$), \CII ($f\sim0.13$) and \SiIIc ($f\sim 0.13$), and assuming saturation. Comparing the solution found by directly solving equation (1) with all available lines under the saturation assumption, it is clear that not all elements are saturated. By selecting the elements with the highest oscillator strength (hence more easily saturated) a much better agreement is achieved.}
\label{absorption}
\end{figure}

\begin{figure*}
\begin{center}
	\begin{subfigure}[b]{\textwidth}
			\includegraphics[width=\textwidth]{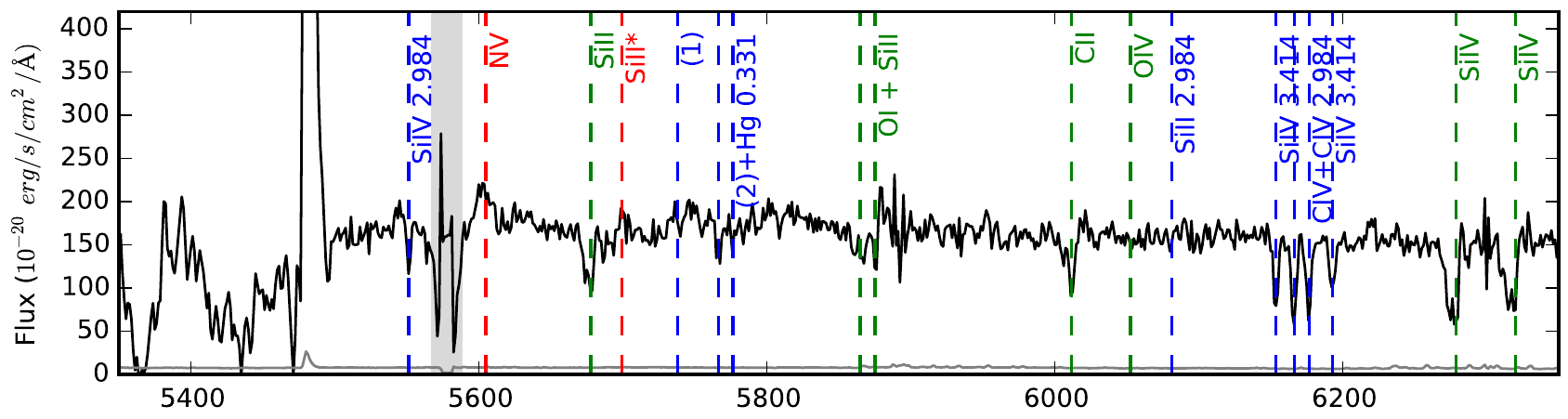}\\
	\end{subfigure}	
	\begin{subfigure}[b]{\textwidth}	
		\includegraphics[width=\textwidth]{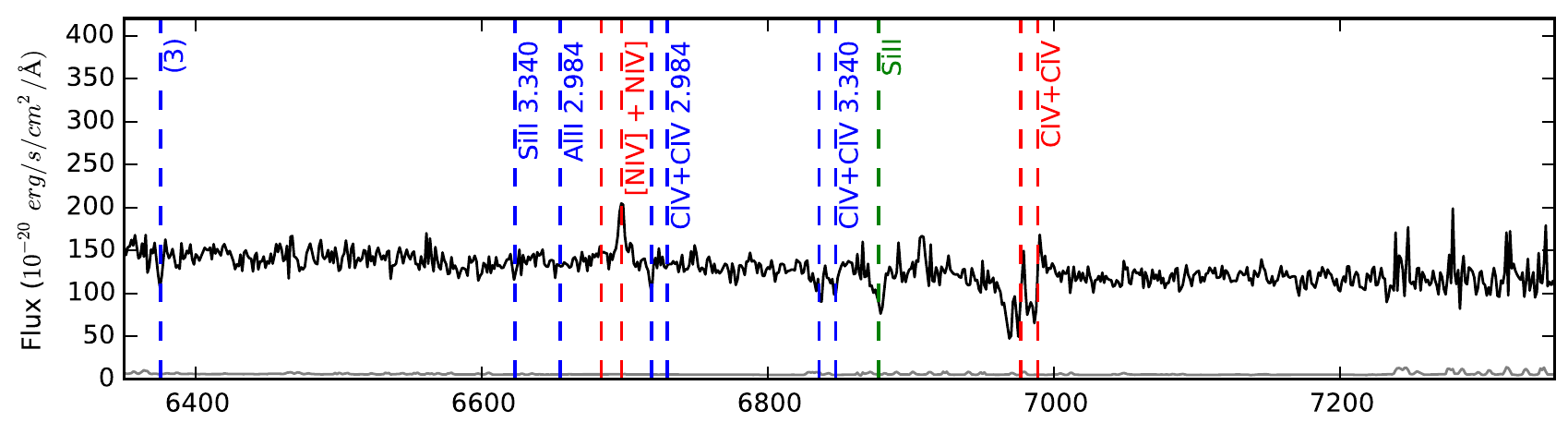}\\
	\end{subfigure}
	\begin{subfigure}[b]{\textwidth}	
		\includegraphics[width=\textwidth]{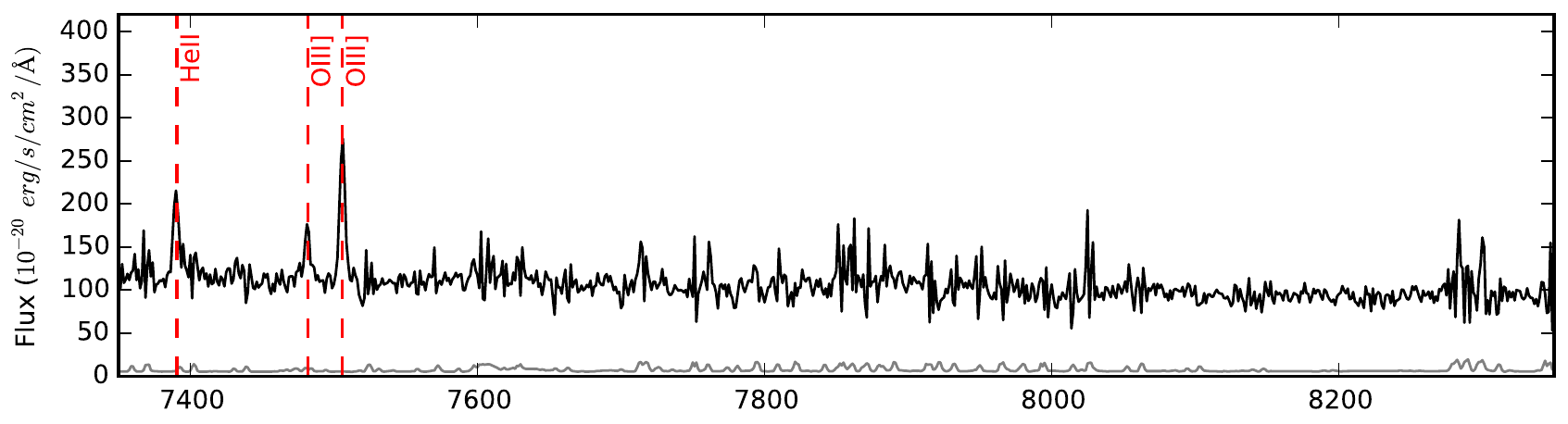}\\
	\end{subfigure}
	\begin{subfigure}[b]{\textwidth}	
		\includegraphics[width=\textwidth]{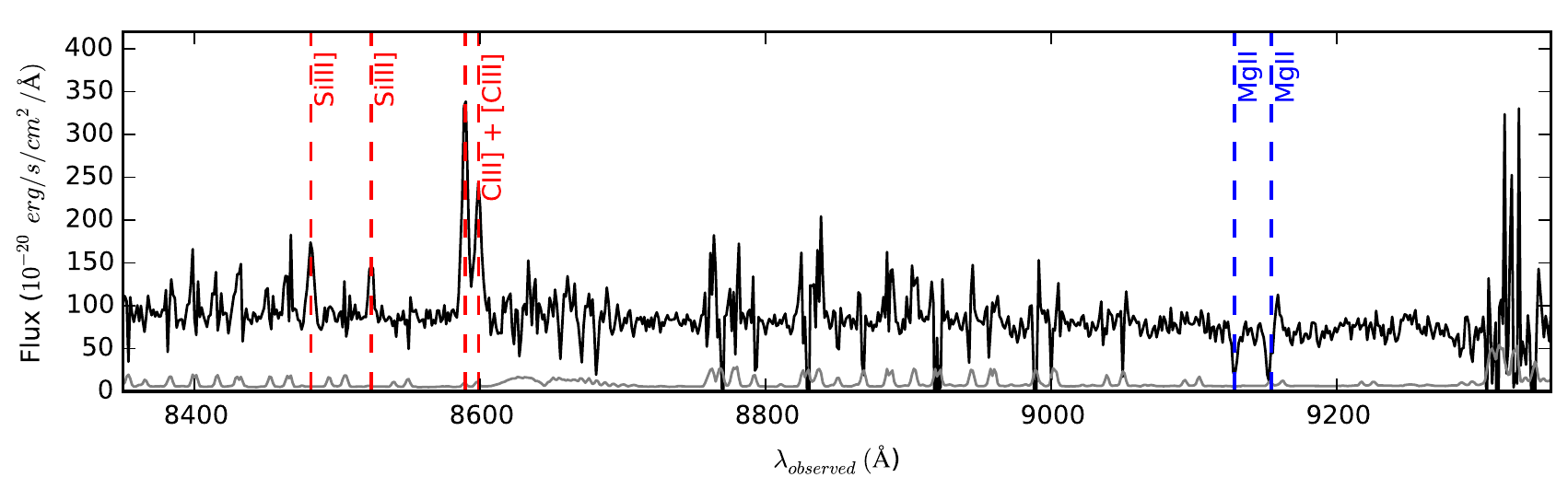}\\
	\end{subfigure}
	\caption{Spectral lines identified in the central region spectrum. \textit{Black:} Observed spectrum central region. \textit{Gray Line:} Pipeline propagated variance. \textit{Shaded gray}: Strong sky residuals. \textit{Green dashed lines:} absorption lines associated with the $z=3.5$ galaxy. \textit{Red dashed lines:} emission lines associated with the same system. \textit{Blue dashed lines:} other absorption lines, some of which we associate with intervening systems over the line of sight (see Appendix A). Besides the \Lya, $\mbox{C\,{\sc iii}}]$ and \mbox{O\,{\sc iii}}] emission lines, several other emission lines, more rarely seen at higher redshifts, were identified, such as the $\mbox{Si\,{\sc iii}}]$, $\mbox{Si\,{\sc ii}}^*$ and the $\mbox{N\,{\sc iv}}]$ doublet. The magnification provided by lensing as well as the combination of spectra from several multiple images, provides a high signal-to-noise also in the continuum, allowing the confident identification of several Si, C and O absorption lines.}
\label{spectralines}
\end{center}
\end{figure*}

The extraordinary quality of the data also reveals absorption lines at high signal-to-noise, many of them sufficiently resolved to clearly display an asymmetric profile, with an elongated blue tail, generally interpreted as tracing absorption by outflowing material (e.g. \cite{Shapley+03}). In our case, the maximum of absorption seems to arise from static gas: we did not find any velocity shift of the peak of the low ionisation absorption lines (\SiIIa, \OI, \SiIIb, \CII and \SiIIc) with respect to the systemic redshift (see Fig.~ \ref{absorption}). We follow an equivalent procedure to fit the absorption lines as the one described above for the emission lines, except that we used an asymmetric profile, constructed as two half gaussians with independent FWHM but the same peak intensity, to fit the strongest absorption lines. From these measurements we derive a larger velocity dispersion for the absorptions ($150\pm18$ km\,s$^{-1}$) compared to the emission lines. 

A number of strong absorption lines appear in the spectrum at wavelengths which do not match with any feature of the 3.5062 redshift system. They are most likely due to absorptions produced by systems within the line of sight. Indeed, we confidently identified three pairs of absorption lines as \SiIV and \CIV doublets, due to their characteristic line ratios, and the derived redshifts are a good match to the redshift of single image s2 ($z=2.982$), multiple-image systems 4 ($z=3.340$) and 12 ($z=3.414$) from \cite{Richard+15}. We present a brief analysis of these intervening absorbers in Appendix A.

\begin{table}
    \begin{tabular}{| l | r | r| r | r | }
                \hline  
        Line & $ \lambda_{rest}  $ & $ \lambda_{obs}  $ & $\sigma$   & EW   \\   
        		&[$\AA$] 	&[$\AA$]		& 	[km\,s$^{-1}$] &[$\AA$] \\
                \hline \hline															
\mbox{Si\,{\sc ii}}	& 	1260.42		&	$5677.84\pm0.49$	&	$166.1\pm23.9$	&	$0.70\pm0.036$	\\
\mbox{O\,{\sc i}}	&	1302.17		&	$5864.89\pm0.27$	&	$218.6\pm12.6$	&	$0.31\pm0.031$	\\
\mbox{Si\,{\sc ii}}	&	1304.37		&	$5875.20\pm0.43$	& 	$14.9\pm15.2$		&	$0.09\pm0.003$	\\
\mbox{C\,{\sc ii}}	&	1334.53		&	$6011.75\pm0.35$	&	$119.0\pm14.9$	&	$0.61\pm0.033$	\\
\mbox{O\,{\sc iv}}	&	1342.99		&	$6052.54\pm1.80$	&	$148.0\pm60.7$	&	$0.22\pm0.021$	\\
\mbox{Si\,{\sc iv}}	&	1393.76		&	$6278.77\pm0.28$	&	$153.2\pm7.5$		&	$1.21\pm0.052$	\\
\mbox{Si\,{\sc iv}}	&	1402.77		&	$6319.90\pm0.51$	&	$130.7\pm13.7$	&	$0.94\pm0.099$	\\
\mbox{Si\,{\sc ii}}	&	1526.71		&	$6877.29\pm0.30$	&	$120.1\pm4.9$		&	$0.47\pm0.215$	\\
                 \hline             
     \end{tabular}
	 \caption[Emission lines]{Absorption line measurements from the central region spectrum combining signal from all multiple images. Flux corresponds to observed flux (without magnification correction). Observed wavelengths ($\lambda_{obs}$) are given in air and $\sigma$ is corrected for instrumental broadening.}    
     \label{table_absorption}          
\end{table}

\subsection{Physical parameters from integrated spectrum}

The next subsections are dedicated to the analysis of the properties of the emission and absorption lines, from which we derive the gas density, temperature and covering fraction, as well as the analysis of the continuum, from which we derive the stellar population content of the galaxy and estimate metallicity. Within this Section we use the combined spectrum, described in Sect. \ref{dataanalysis}.

\subsubsection{Temperature and density from line ratio diagnostics}
\label{temden}

From the MUSE data, we have access to four line ratio diagnostics, which we complement with previous X-Shooter results \citep{Christensen+12b}  to obtain a more comprehensive view. Three electron density diagnostics come from MUSE data exclusivelly ---  \NIVratio, \SiIIIratio and \CIIIratio --- for which we measure ratios of $0.204\pm0.025$, $1.228\pm0.120$ and $1.490\pm0.055$, respectively. We update the previous electron temperature measurement from the \OIIIUV / \OIII diagnostic by combining both data sets. To do this, we rescale the X-Shooter line fluxes so that the strong $\mbox{C\,{\sc iii}}]$ doublet has the same intensity as the MUSE spectrum, and we measure an \OIIIUV / \OIII ratio of $0.061\pm0.003$ (reddening corrected $E(B-V)$=0.03 obtained by \cite{Christensen+12b}). Using the {\sc pyneb} package \citep{pyneb} we study the possible temperature and density combinations for these line ratios also including the \OIIratio diagnostic reported in \cite{Christensen+12b} (see Fig.~\ref{TN}). We find that the \OIIratio, \OIIIUV / \OIII and \CIIIratio diagnostics are compatible and indicate a gas temperature of 16500 $\pm 200 $ K and a density of $\sim300$ $\pm$ 700 $\mathrm{cm}^{-3}$ where the errors were estimated from the region limits where the several diagnostics (except for $\mbox{N\,{\sc iv}}]$) agree. This temperature is well within the typical values found for other high-redshift lensed galaxies --- 13000 to 25000 K \citep{James+14,Villar-Martin+04,Erb+10}. On the other hand, the density is lower than what was measured in other lensed galaxies, that nevertheless span a large range of values: from 600 to 5000 cm$^{-3}$ for the Clone, the Cosmic Horseshoe and J0900+2234 lensed galaxies \citep{Wuyts+12b,Bian+10,Hainline+09}. The $\sim300$ $\mathrm{cm}^{-3}$ we obtain are more typical of giant {H\,{\sc ii} regions in the nearby universe and are also consistent with the typical density (100 cm$^{-3}$) found in low metallicity local galaxies with compact {H\,\sc ii} regions \citep{Morales-Luis+14}. 

Contrastingly, the \NIV electron density diagnostic points to a much higher density ($> 10^5 \mathrm{cm}^{-3}$) at this temperature. Discrepancies between line diagnostics have been previously reported in high redshift studies (e.g. \citealt{Villar-Martin+04}) with one possible explanation for this being the local variations in density and temperature within the galaxy, that are not disentangled in an integrated spectrum, resulting in incompatible diagnostic results. To test if any spatial variation of local density and temperature could be seen within the MUSE data, we measure the \CIIIratio ratio at several spatial positions in the central part of the galaxy, but found no significant change, with a ratio varying between 1.533 to 1.536. Another possible explanation is that N$^{3+}$ is a higher ionisation line, thus probing higher ionisation zones, that can have a considerably higher density, which could explain why this diagnostic points to a higher density compared to the ones from lower ionisation lines. 

\begin{figure}
\begin{center}
\includegraphics[width=0.50\textwidth]{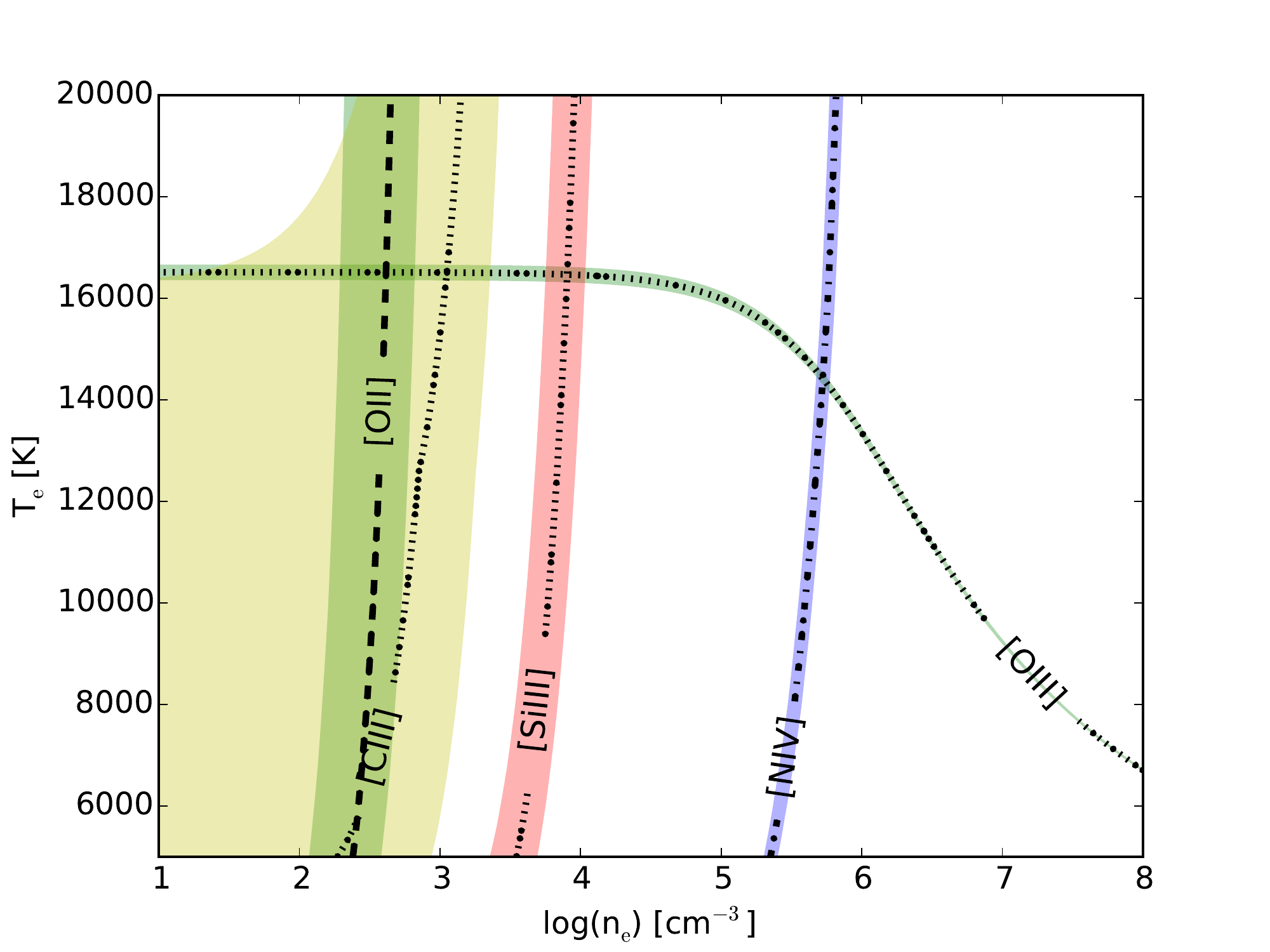}\\
\caption{Range of electron density and temperature allowed by our emission line ratio diagnostics. Most diagnostics agree on a temperature of $\sim$16500 K and an electron density of $\sim$~300 cm$^{-3}$. The discrepancy of the \NIV diagnostic can be due to its higher ionisation energy, that would limit it to probe higher density regions (closer to the ionising source).}
\label{TN}
\end{center}
\end{figure}

\subsubsection{Neutral gas covering fraction}
\label{covfrac}

In this subsection we derive the neutral gas covering fraction from absorption lines depth. For a particular velocity bin, the intensity of the absorption lines can be related to the covering fraction by the following equation:

\begin{equation}
I = 1 - f_c  \left( 1 - \exp\,\left( \frac{-f\  \lambda\  N}{3.768 \times 10 ^ {14}} \right) \right)
\label{eq}
\end{equation}

\noindent where $I$ is the observed line intensity normalised to the continuum level, $f_c$ the covering fraction, $f$ the oscillator strength, $ \lambda$ the transition wavelength and N the column density. \cite{Jones+13b} discuss two possible approaches to derive the covering fraction with this equation: using at least two low-ionisation lines and solving the system for the density and covering fraction or assuming that the lines are saturated ($\lambda N \gg 3.768 \times 10^{14}$) which simplifies expression (\ref{eq}) to $I = 1 - f_c$. We follow both procedures, using \SiIIa, \SiIIb and \SiIIc lines to derive a covering fraction solving equation (\ref{eq}) (see Fig.~\ref{absorption}). For the saturation assumption, we made a first test averaging all low ionisation lines: the three aforementioned {\mbox{Si\,{\sc ii}}}  lines as well as \OI and \CII and directly calculating the covering fraction. From this average we obtain a lower covering fraction for every velocity bin, which indicates that the saturation assumption is not valid, at least, for all lines. We repeat the same method, this time only using the lines with higher oscillator strengths (hence more easily saturated): \SiIIa, \SiIIb and \SiIIc lines. The covering fraction obtained with the average profile of these lines (lower panel of Fig.~\ref{absorption})  is in good agreement with the result obtained without the approximation, suggesting that these three {\mbox{Si\,{\sc ii}}}  lines must be close to saturation. 

We obtain a maximum covering fraction of 0.4 at v= 0 km\,s$^{-1}$ which suggests most of the neutral gas in this galaxy is at rest. However, it should be noticed that a covering fraction smaller than 1 in each velocity bin does not allow to conclude that the global covering fraction of the gas traced is lower than 1 (see figure 17 in \citealt{Rivera-Thorsen+15}, for an illustration). Given the shape of the \Lya profile emergent from this galaxy, the covering fraction of the scattering medium is most certainly 1, since the profile does not peak at the systemic redshift, but is redshifted by $\sim$170 km/s, which means no intrinsic \Lya emission is seen \citep{Behrens+14, Verhamme+15}.

We also compare this system with the sample of $z\sim4$ lensed galaxies analysed by \cite{Jones+13b}, where the evolution of the maximum outflow velocity and maximum absorption depth  with redshift is analysed (figure 5 of the cited paper). The values that we obtain for this particular galaxy, $-500$ km\,s$^{-1}$ for the maximum velocity where outflowing gas is still present and maximum covering fraction of 0.4, are in good agreement with the $z\sim4$ sample presented. In the same work, a correlation of the maximum absorption depth with \Lya  equivalent width is presented and, once more, our results agree with the derived correlation.

\subsubsection{Stellar populations}
\label{stellarpop}

The continuum signal from this magnified galaxy is also high enough to allow us to probe stellar populations. We chose to do so with the full spectral fitting technique, using the {\sc starlight} code \citep{Cid+05}, that produces a model spectrum from a linear combination of simple stellar population spectra, convolved with a kinematic kernel and corrected for extinction, to find the stellar population model that best matches an observed spectrum. Starburst99 \citep{Leitherer+99} stellar populations were used as templates to fit the spectrum, with absolute metallicities of 0.001, 0.002, 0.008, 0.014 and 0.040 and ages between 2 Myr and 3 Gyr. We prefer instantaneous burst models, since continuous star formation models are not simple to interpret in this linear combination method, and use the Geneva tracks with zero rotation and the \citet{Kroupa+01} IMF. The Calzetti \citep{Calzetti+00} extinction law is adopted in all fits. 

We build six different libraries of stellar populations, which are used to independently fit the observations: five with a single metallicity spanning ages from 2 Myr to 3 Gyr, and one with an equal number of models of each metallicity, but with ages only from 2 Myr to 600 Myr and a coarser time grid than the single metallicities libraries. To estimate errors, we generate 100 realisations of the observed spectra (using the same method as described in Sect. \ref{specfeatures}) and perform the fit with the same parameters, taking the errors as the standard deviation of the 100 results. All emission lines are masked before the fit, as well as all absorption lines confidently identified as features of systems in the line of sight. We also masked stellar winds and ISM lines (not easily reproduced by UV spectral models) but this did not affect the results of the fit.

All five model spectra from the single metallicity libraries provide similar fits to the data (upper panel of Fig.~\ref{S99}), which confirms that precise metallicity estimations are hard to obtain from full spectral fitting in the UV. The models fail to correctly reproduce the intensity of the observed {\mbox{N\,{\sc v}}} emission, where models predict a fainter emission, although as broad as seen in the observed spectrum. They also do not reproduce the complex {\mbox{C\,{\sc iv}}}] profile, which is not surprising given that this line is a mixture of stellar and ISM emission and highly dependent on stellar winds, which are still poorly modelled. On the other hand, other absorption features such as \SiIIa, \CII are well reproduced by all models. The different libraries also give comparable results: most of the flux can be reproduced by very young populations with ages up to 10 Myr (middle panel of Fig.~\ref{S99}), whereas most of the mass is created at $>$100Myr lookback time (lower panel of the same figure). The model with all available metallicities has a very similar spectral shape to the others, and also shows a strong burst at 10 Myr and most mass ($\sim98\%$) being created between 300 and 350 Myr, with exclusively the lowest metallicity populations (Z=0.001). This preference for lower metallicity stellar populations in the mixed library and the fact that the lowest $\chi^2$ of the fits is found using populations with Z=0.001 (0.07 Z$_\odot$) suggests that this is a low stellar metallicity object. All spectral models obtained have low extinction (see Table \ref{starlight_results}) which is in good agreement with the previous results in \cite{Christensen+12a}. 

We obtain current stellar masses --- the mass of the initial gas cloud assumed in burst models of stellar populations accounting for mass losses by winds and supernovae --- of M$_\star = 6.7\pm0.2$, $6.5\pm0.3$, $6.7\pm0.4$, $2.1\pm0.3$, $1.9\pm0.2$ $\times 10^9$ M$_{\sun}$ for the lowest to the highest single-metallicity libraries and M$_\star = 4.1\pm0.2 \times 10^9$ M$_{\sun}$ for the mixed library. To derive this, we use the combined spectrum to ensure the highest signal-to-noise possible, but rescale its continuum level to match the magnification corrected flux of the central spectrum, which contains the total continuum flux. The masses are marginally higher than the ones derived by \cite{Christensen+12a}. The difference probably arises from the improved cluster mass model (and consequently the magnification factors) and a better estimate of the total flux with IFU data. Throughout the rest of the analysis, we assume that the mass of the galaxy is $6.7\pm0.2\times 10^9$, the one obtained with the lowest metallicity library, the fit with the lowest $\chi2$.

Another possible approach to the study of stellar populations is to analyse the \HeII equivalent width that \cite{Brinchmann+08} suggest as a metallicity and age indicator. As a very high ionisation energy is needed to ionise \mbox{He\,{\sc ii}} , this line is an indicator of the presence of Wolf-Rayet stars, and can therefore be used as an age indicator. Because the equivalent width of \mbox{He\,{\sc ii}} will depend on the ratio of Wolf-Rayet and O stars which in turn depends on metallicity, it can also be used as a metallicity indicator. If we compare the measured equivalent width ($\sim1\, \AA$) with the expected EW for continuous star formation models (see figure 1 of \citealt{Brinchmann+08}) this value corresponds to stellar formation from about 3 Myr earlier. Although these are smoothed, simplified models of real star formation histories, this seems consistent with the younger stellar populations we found with the stellar population models fit (2--8 Myr). Nevertheless, an equivalent width of $1 \AA$ implies a much higher stellar metallicity for these continuous star formation models: at least 0.4 Z$_{\odot}$ compared with the preferred 0.07 Z$_{\odot}$ from the stellar populations modelling. \cite{Brinchmann+08} also explore the effects of bursts, concluding that the equivalent width of \mbox{He\,{\sc ii}} could be strongly boosted for a very short period of time right after a recent episode of star formation which, in this case, would mean a boost of about 50\% in the strength of the \mbox{He\,{\sc ii}} line in order to achieve the lowest metallicities considered in their models (0.2 Z$_\odot$). Although we should be careful about drawing conclusions from the \mbox{He\,{\sc ii}} equivalent width, since the study was made for continuous star formation and simplified star formation histories, it seems to also point to the strong and recent burst scenario found with the full spectrum fitting.

\begin{figure*}
	\begin{subfigure}[b]{\textwidth}
		\includegraphics[width=0.99\textwidth]{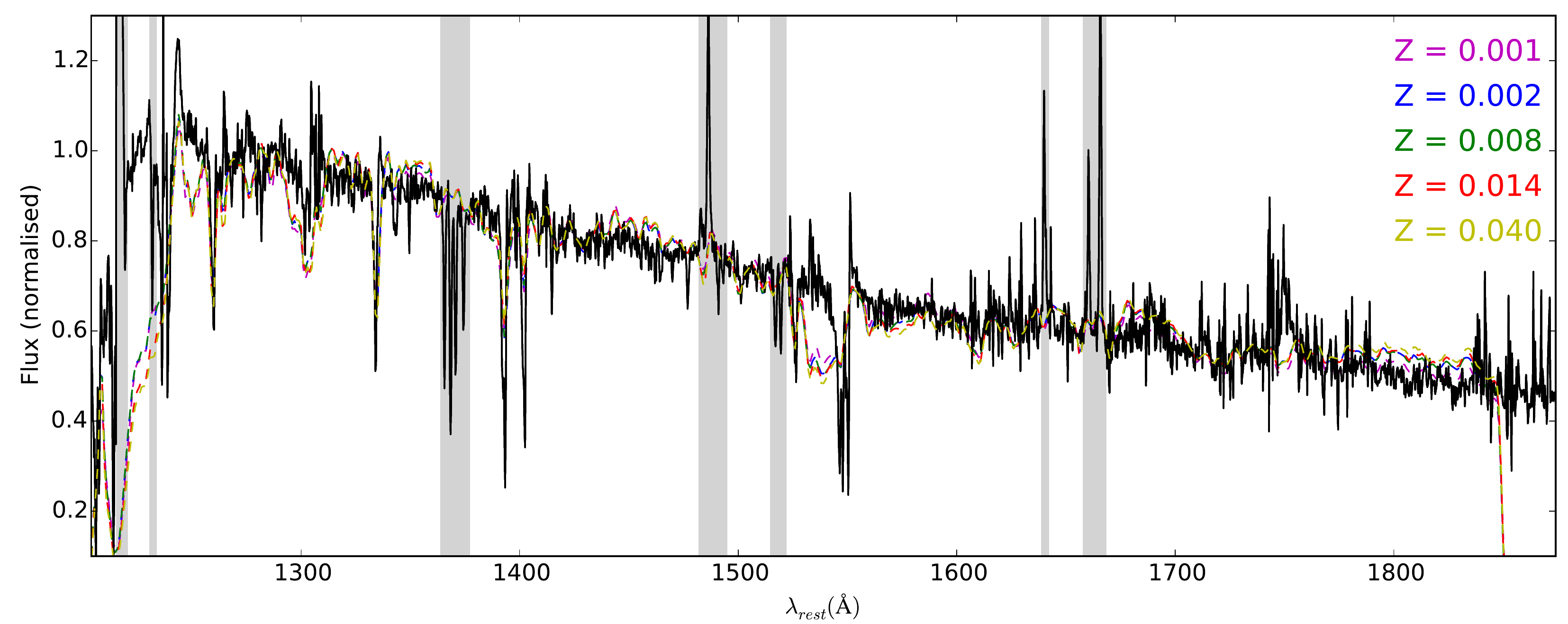}\\
		\includegraphics[width=\textwidth]{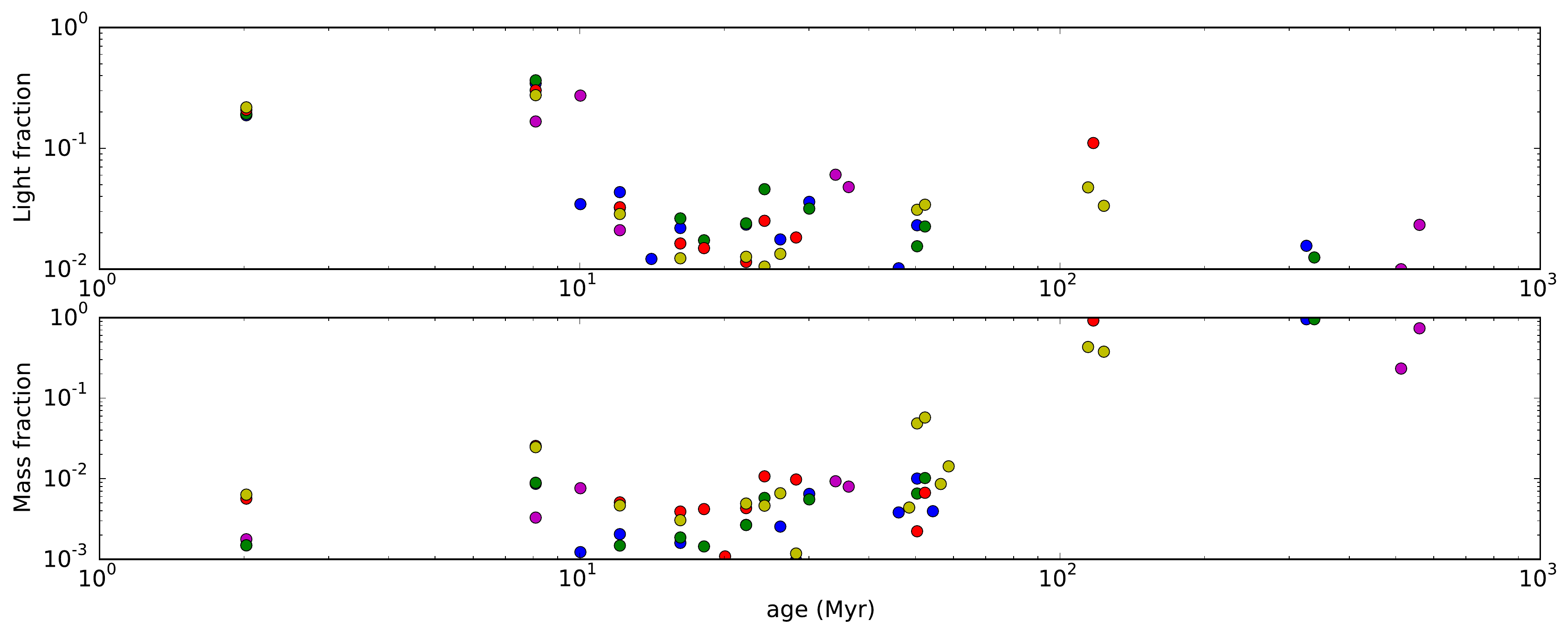}\\
	\end{subfigure}
\caption{\textit{Upper panel}: Five stellar population models obtained with the individual metallicity libraries. Model spectra in colour, observed spectrum in black and masked regions in shaded grey. Despite small differences, all models equally reproduce the continuum and fail to reproduce wind emission features such as \CIV \textit{Lower Panels}: Single stellar populations --- age in abscissa and metallicity colour coded --- used in each of the five stellar spectrum, with the same colour as the above panel. The middle panel presents the contribution to the model of each individual stellar population in flux and the lower panel the contribution in mass. Despite the different age distributions of each model they all display a strong star forming period at around 10 Myr and predict that most of the stellar mass ($>$80\%) was created 100 and 600 Myr of lookback time.}
\label{S99}
\end{figure*}

\begin{table}
\begin{center}
    \begin{tabular}{l c | c | c | c | c | c |}
                \hline  
       	Z 	& Z/Z$_\odot$	& 	$\chi^2$ 	& $10^9$ $M_{ini}/M_{\odot}$  & $10^9$ $M_{cur}/M_{\odot}$ & $E(B-V) (mag)$ \\
	\hline
	\hline
0.001	&	0.07		&	2.11	&	$8.2\pm0.2$	&	$6.7\pm0.2$	&	$0.016\pm0.001$\\
0.002	&	0.14		&	2.35	&	$7.6\pm0.3$	&	$6.5\pm0.3$	&	$0.000\pm0.002$\\
0.008	&	0.57 		&	2.34	&	$8.2\pm0.5$	&	$6.7\pm0.4$	&	$0.000\pm0.006$\\
0.014	&	1.00		&	2.53	&	$2.5\pm0.4$	&	$2.1\pm0.3$	&	$0.000\pm0.002$\\
0.040	&	2.86		&	2.50	&	$2.3\pm0.2$	&	$1.9\pm0.2$	&	$0.002\pm0.001$\\		
mixed 	&			&	2.15	&	$5.0\pm0.3$	&	$4.1\pm0.2$	&	$0.033\pm0.013$\\
        \hline  
     \end{tabular}
	 \caption[]{Stellar population fits results. From left to right: absolute and solar metallicities, reduced $\chi^2$, initial stellar mass and current stellar mass (accounting for mass losses due to SNe and winds), extinction. Errors are the standard deviation of fits performed on 100 realisations of the observed spectrum (see Sect. \ref{specfeatures})}      
     \label{starlight_results}          
\end{center}
\end{table}

\subsubsection{Stellar and ISM metallicity from absorption features}

From the previous stellar population analysis we find that it is not possible to constrain the stellar metallicity with confidence. This prompts us to explore another method, based on empirically calibrated equivalent widths of faint absorption UV features. Of the 5 possible metallicity indicators explored by \cite{Sommariva+12} - F1370, F1425, F1460, F1501, F1978 - we discard the first due to contamination by an intervening absorber and the last since sky residuals in this region are too high. To calculate the equivalent width of the remaining features, we normalise the spectrum following by the procedure described in \cite{Rix+04}: several continuum points, free of emission and absorption lines, are used to estimate the continuum level using a low-order spline. The equivalent width of the features within the spectral windows defined by \cite{Sommariva+12} are then estimated and the pseudo continuum calibrations used to obtain the metallicity (see Fig.~\ref{index}). The error is estimated by removing one of the continuum intervals at a time and recalculating the normalisation and the equivalent width measurement, and the standard deviation this sample is taken as the error. The metallicities from the F1425 and F1460 indices agree to within an order of magnitude, but for the F1501 index the obtained metallicity is unphysical (Table \ref{Sommariva}). The obtained metallicities ---  Z = $0.0424\pm0.0085$ Z${_\odot}$ and $0.0678\pm0.0078$ Z${_\odot}$ --- are lower but still comparable with our best measurement from the stellar population analysis (0.07 Z${_\odot}$), which is also the lowest possible metallicity in the Starburst99 models, and which once more suggests that this is a low stellar metallicity system. 

\begin{table}
\begin{center}
    \begin{tabular}{l r| r | r| r  | }
                \hline  
       index & $\lambda$ range & EW (\AA) & Z/Z$_{\odot}$  \\  
          \hline \hline 
       	F1425 &  1415-1435 	&	 $0.3940\pm0.0313$ 	& 	$0.0424\pm0.0085$ \\
	F1460 &  1450-1470 	& 	 $0.4870\pm0.0334$ 	& 	$0.0678\pm0.0078$ \\
	F1501 &  1496-1506 	& 	 $-0.0271\pm0.0316$ 	&	INDEF  \\
	        \hline  
     \end{tabular}
	 \caption[]{Direct metallicity estimations following the work of \cite{Sommariva+12}}      
     \label{Sommariva}          
\end{center}
\end{table}

\begin{figure}
	\begin{subfigure}[b]{0.48\textwidth}
		\includegraphics[width=\textwidth]{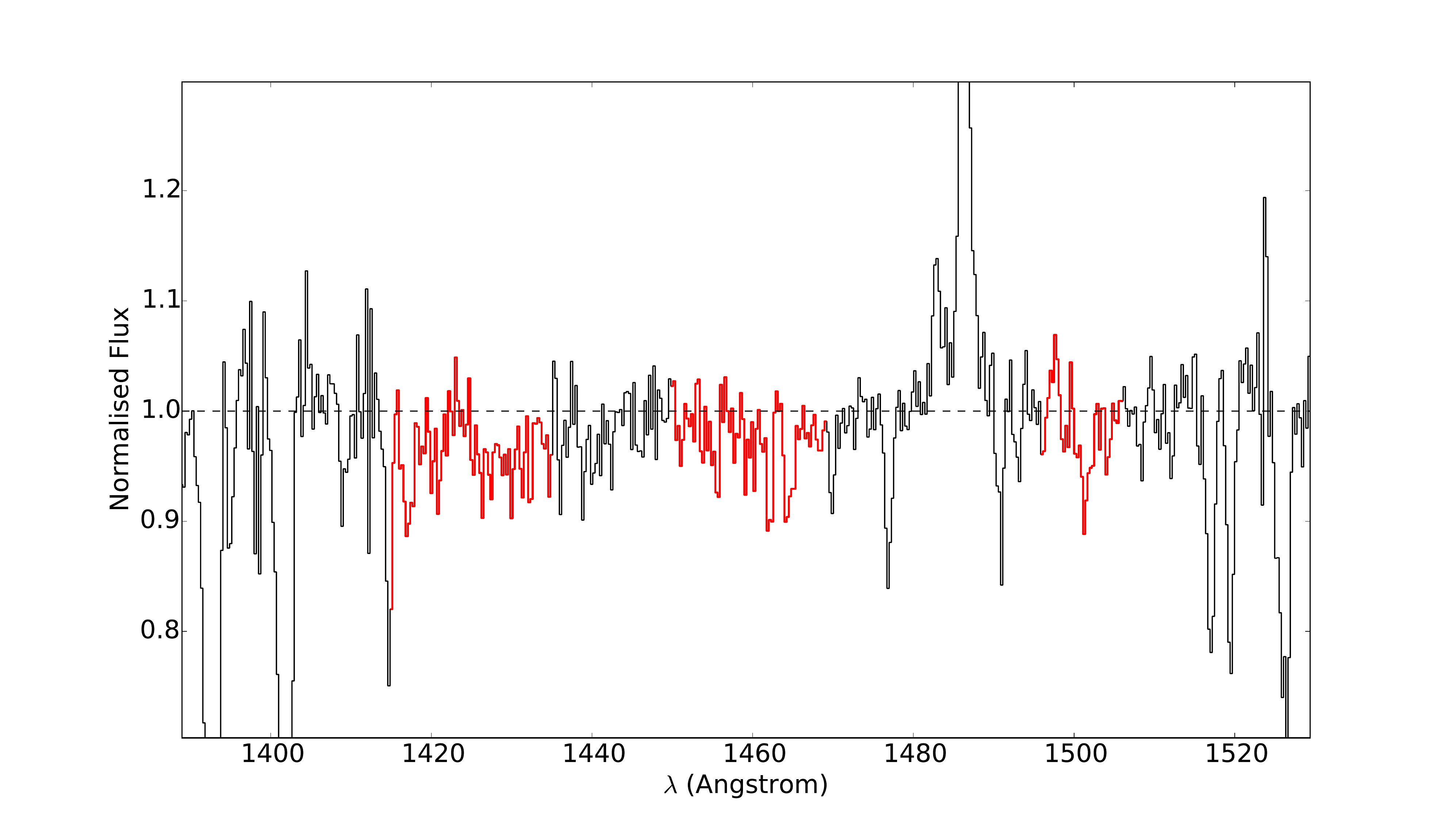}\\
	\end{subfigure}
\caption{Equivalent widths of faint absorption features were estimated in three spectral windows (in red) of the normalised spectrum (in black). These were used as direct stellar metallicity estimators, following the work from Sommariva et al. (2012). Applying their calibrations, we derive a low stellar metallicity for the first two windows (F1425 and F1460 indices) of 0.0424 and 0.0678 Z$_\odot$, respectively.}
\label{index}
\end{figure}

We also use a similar method to explore the ISM metallicity, following \cite{Leitherer+11} prescription to calculate the equivalent width of several ISM absorption lines. In their work, a sample of local galaxies with known metallicities is used to study how these features would correlate with metallicity, concluding that the \SiIIa, \mbox{O\,{\sc i}} and \mbox{Si\,{\sc ii}} $\lambda$1303 blend, \CII and \SiIIc\ absorptions are correlated to galaxy metallicity. In Fig.~\ref{Leitherer} we present the local galaxy measurements from figure 15 of \cite{Leitherer+11}, adding in our object for comparision. Although it is not possible to directly estimate metallicity following this method, our equivalent width measurements seem to follow the overall correlation of galaxies at low metallicities, with 12 + $\log$(O/H) $\sim$ 7.5 -- 7.9, which corresponds to a solar metallicity of 0.065 to 0.16 Z${_\odot}$, a value that is comparable or slightly higher than the stellar metallicity, a trend that is found in other high redshift galaxies (figure 6 of \citealt{Sommariva+12}). Other gas metallicity indicators were reported in \citet{Christensen+12b}: R23, Ne3O2 and a direct measurement, yielding metallicities of $7.74\pm0.03$, $7.56\pm0.11$ and $7.76\pm0.03$ in 12+$\log$(O/H), respectivelly, compatible to our estimations from the comparison with the \citet{Leitherer+11} local galaxy sample. 

We compare system 1 with the mass-metallicity relation derived at high redshift by using equation (2) of \cite{Mannucci+11}, we independently confirm the \citealt{Christensen+12b} results placing this galaxy $0.6$ dex (about 2 $\sigma$) below the mass-metallicity relation. Mass-metallicity outliers have been studied in the past in the local universe. In particular, \cite{Peeples+09} investigated a set of massive galaxies, noticing that many had high SFRs and disturbed morphologies, indicating a possible merger. They argue that low metallicity gas inflow induced by the merging could explain the offset in the metallicity-mass relation of those outliers. The close presence of the companion (see Sect. \ref{dataanalysis}) and high SFR might suggest that an earlier interaction could be a valid scenario for this high-redshift galaxy, but we cannot exclude other possible explanations.

\begin{figure}
	\begin{subfigure}[b]{0.24\textwidth}
		\includegraphics[width=\textwidth]{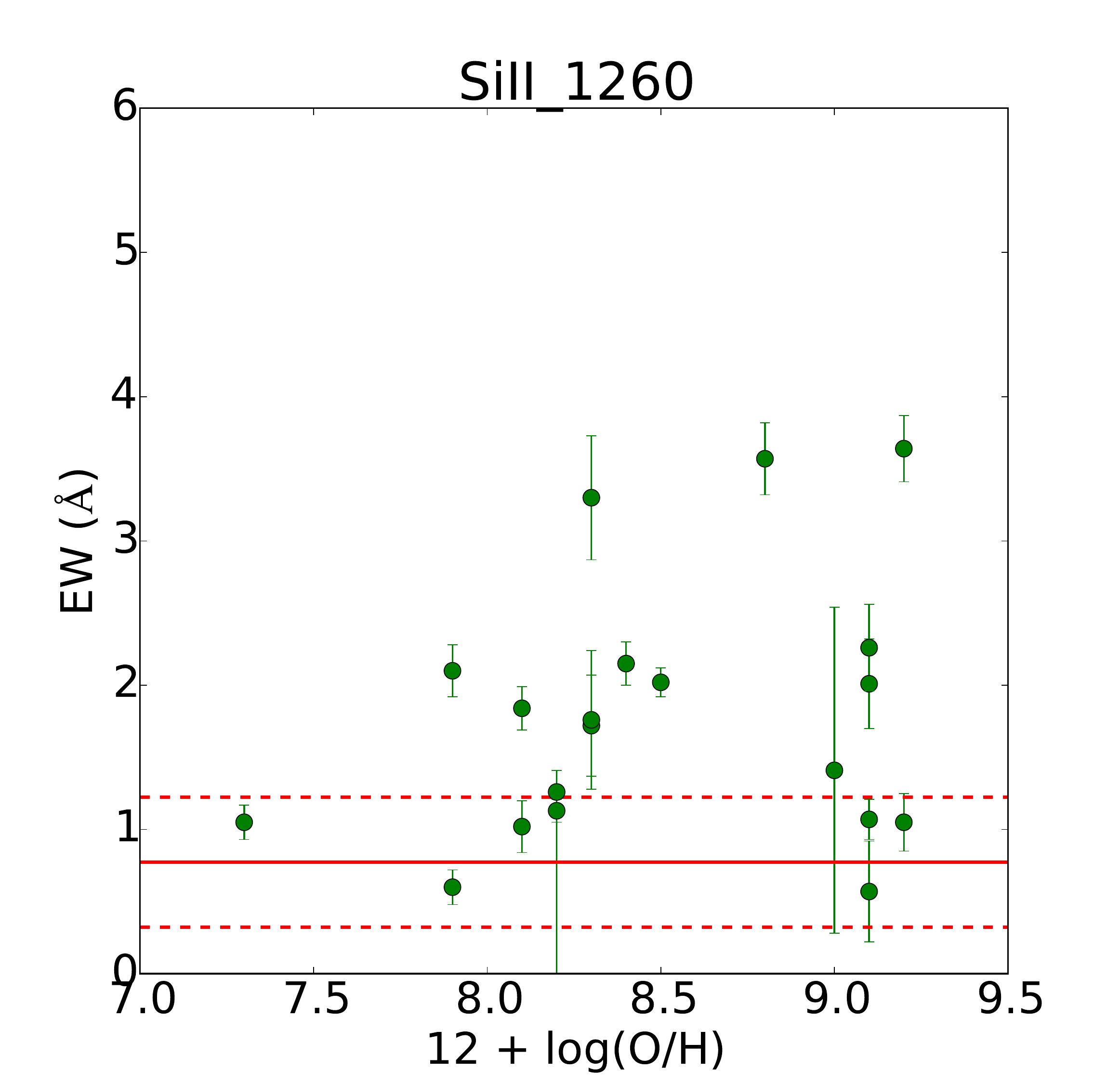}\\
		\includegraphics[width=\textwidth]{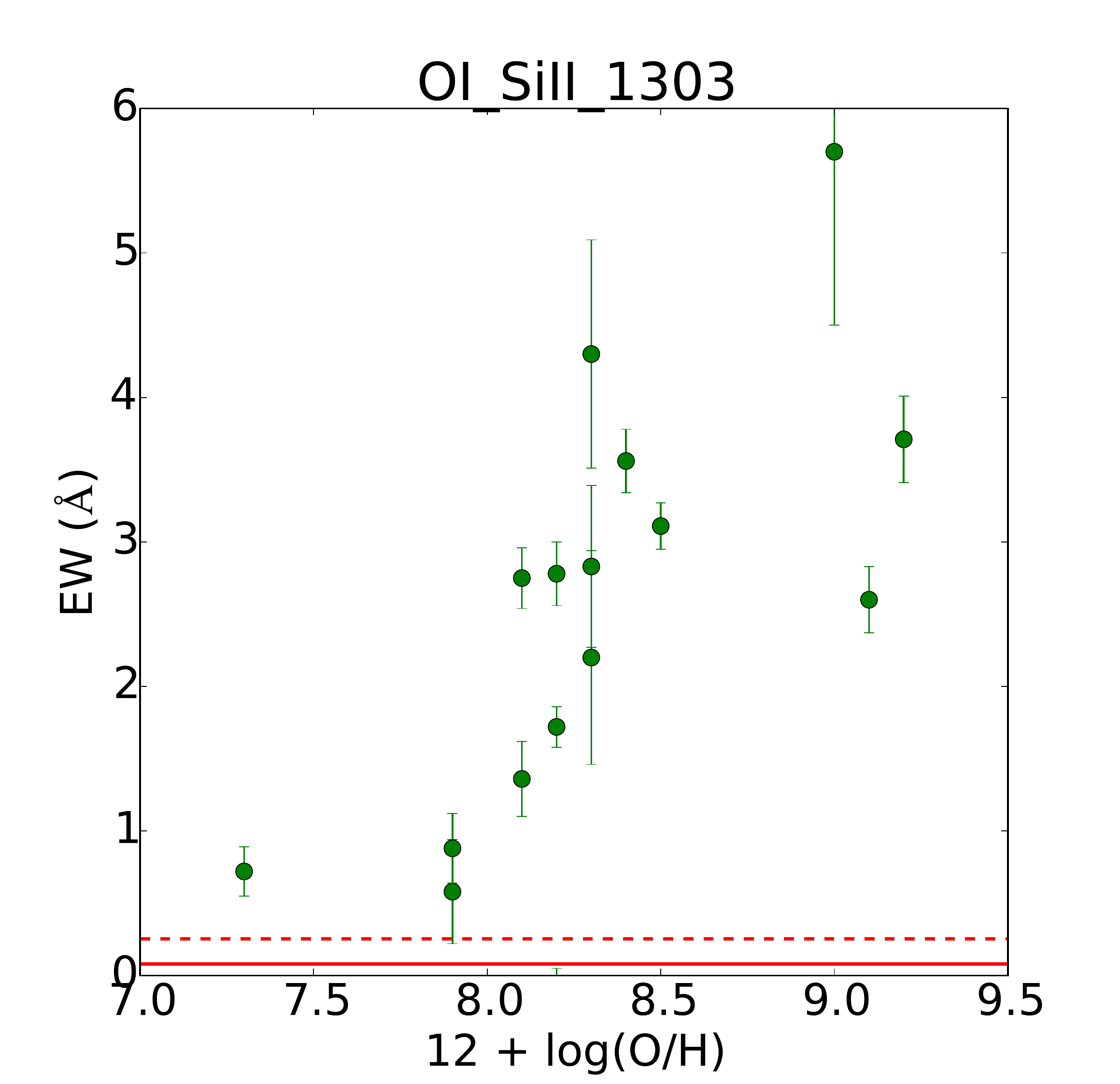}\\
	\end{subfigure}
	\quad
	\begin{subfigure}[b]{0.24\textwidth}
		\includegraphics[width=\textwidth]{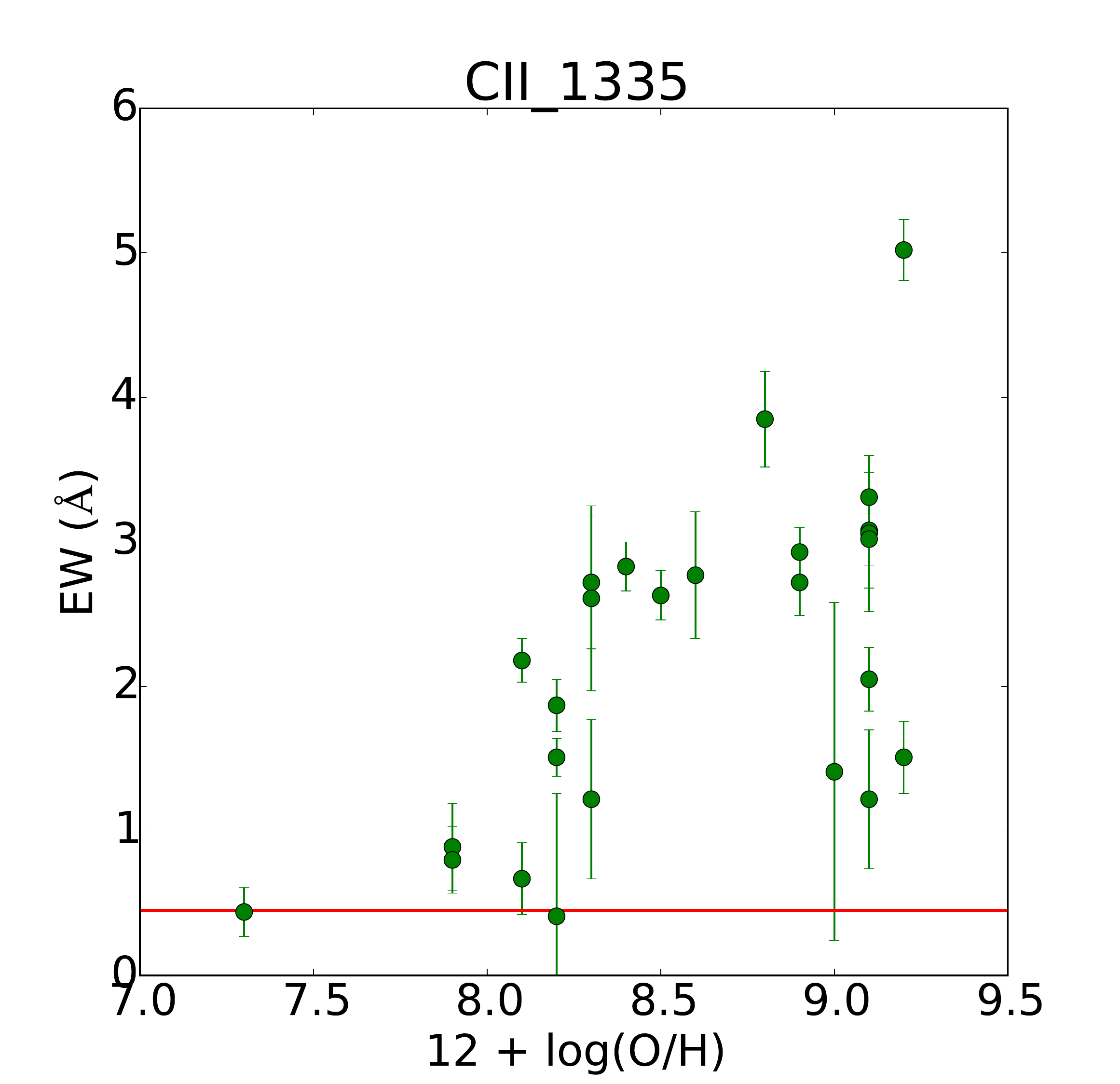}\\
		\includegraphics[width=\textwidth]{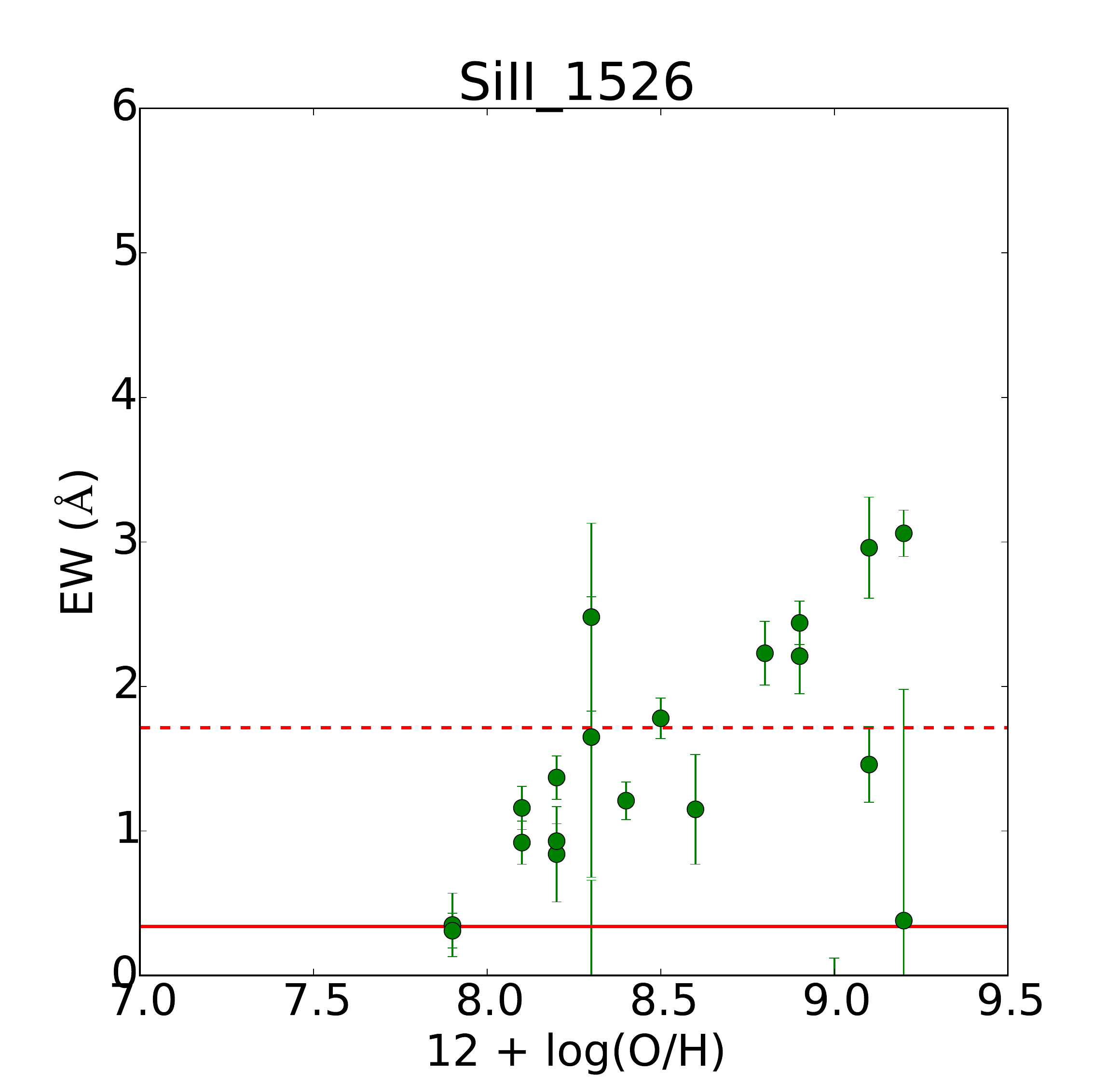}\\
	\end{subfigure}
\caption{ISM metallicity trends of various indices from \citealt{Leitherer+11}. \textit{Green dots}: data from local galaxies (figure 15 of \citealt{Leitherer+11} ) \textit{Red line}: Equivalent width obtained for system 1 (dashed lines correspond to the error). In order, the panels represent the \SiIIa, \mbox{O\,{\sc i}} and \mbox{Si\,{\sc ii}} $\lambda$1303 blend, \CII, \SiIIc features. All the above features point to a low metallicity object, when compared with the local sample.}
\label{Leitherer}
\end{figure}

\subsection{Resolved properties}
\label{resolved}

In this subsection, we make use of the full power of MUSE and explore the spatial extent of \Lya emission as well as the spectral variations seen in both \Lya and other non resonant lines across the galaxy. We focus this analysis on image 1.3, which covers the full extent of the source. 

\subsubsection{Emission line spatial variation} 

We define a small cube, centred on image 1.3, from which we extract the spectrum of each spaxel, fitting the \Lya line with an asymmetric gaussian profile (see Sect. \ref{specfeatures}), and other nebular lines (\CIII, \OIIIUV, \NIV) with a single gaussian. All nebular lines are fit simultaneously, using a single common redshift. We repeat this process following a quadtree method, spatially binning the small cube in $2\times2$ and $4\times4$ pixels, depending on the quality of the \Lya (or nebular lines) fit. This allows us to investigate fainter parts of the galaxy, although with a lower spatial resolution. The resulting maps can be seen in Fig.~\ref{maps}, with the peak shifts measured relative to the systemic redshift. 

Apart from the companion, which is consistently blue shifted relative to the central part of the main body by approximately 30 km\,s$^{-1}$, there is no clear trend in the \Lya map. The \CIII\, map (insert in Fig.~\ref{maps}) shows variations on small spatial scales, with hints of a velocity gradient along the SE/NW direction, with an amplitude of $\pm$25 km/s. In contrast, the central part of the \Lya map is very homogeneous with a mean redshift of 160 - 170 km/s. This is only in the external parts of the \Lya halo that the \Lya peak shows typical variations of at most $\pm$40 - 50 km/s. Assuming that the bulk of the \Lya emission in this galaxy is produced from recombination in \HII\, regions inside the ISM, we can interpret the velocity field shown by the \CIII\, map as the intrinsic velocity field of the \Lya emission, since \CIII\, is a nebular line produced in the same \HII\, regions as \Lya . In other words, if \Lya were non-resonant, it would display the same velocity map as \CIII. Undergoing multiple scatterings in the surrounding neutral gas, the \Lya radiation imprints the relative velocity of the scattering medium between its production site and its last scattering location: \Lya transfer in a static medium leads to a double-peaked profile; if the scattering medium is globally infalling onto the \Lya source, the emergent \Lya profile is blueshifted; and if the gas is outflowing, \Lya is redshifted (e.g. \citealt{Neufeld+90, Dijkstra+14}). The fact that the \Lya spectrum emerging from our halo is redshifted in every location is a probe that it scattered onto outflowing gas with respect to its production site. The asymmetric blue tail of the low ionisation absorption lines confirms that there is outflowing gas in the ISM of this galaxy, at least in front of the stellar continuum, although the strongest absorption is not blueshifted, tracing also static gas. On top of the bulk velocity of the scattering medium, the width and the shift of the maximum of the \Lya peak correlate with the column density of the scattering medium ( e.g. \cite{Verhamme+15}). From the velocity map, the shift is between $\sim$150 - 200 km/s everywhere, which is typical of high-redshift LAEs \citep{Hashimoto+15} tracing typical column densities of the order of log(NHI)$\sim$20.  

As previously stated, the companion \Lya peak shows a small but significant shift relative to the main body ($\sim$ 30 km/s). Assuming that the two bodies are embedded in the same gas envelope, with the same column density and dust distribution, this difference would imply that the companion is in motion relatively to the to the main body. However, as stated above, since we do not have non-resonant lines that directly trace the kinematics of the companion, it is also possible to image a scenario where medium around the companion is less optically thick than around the main body, which would equally explain the observed shift.

\begin{figure}
\begin{center}
	\begin{subfigure}[b]{0.48\textwidth}
		\includegraphics[width=\textwidth]{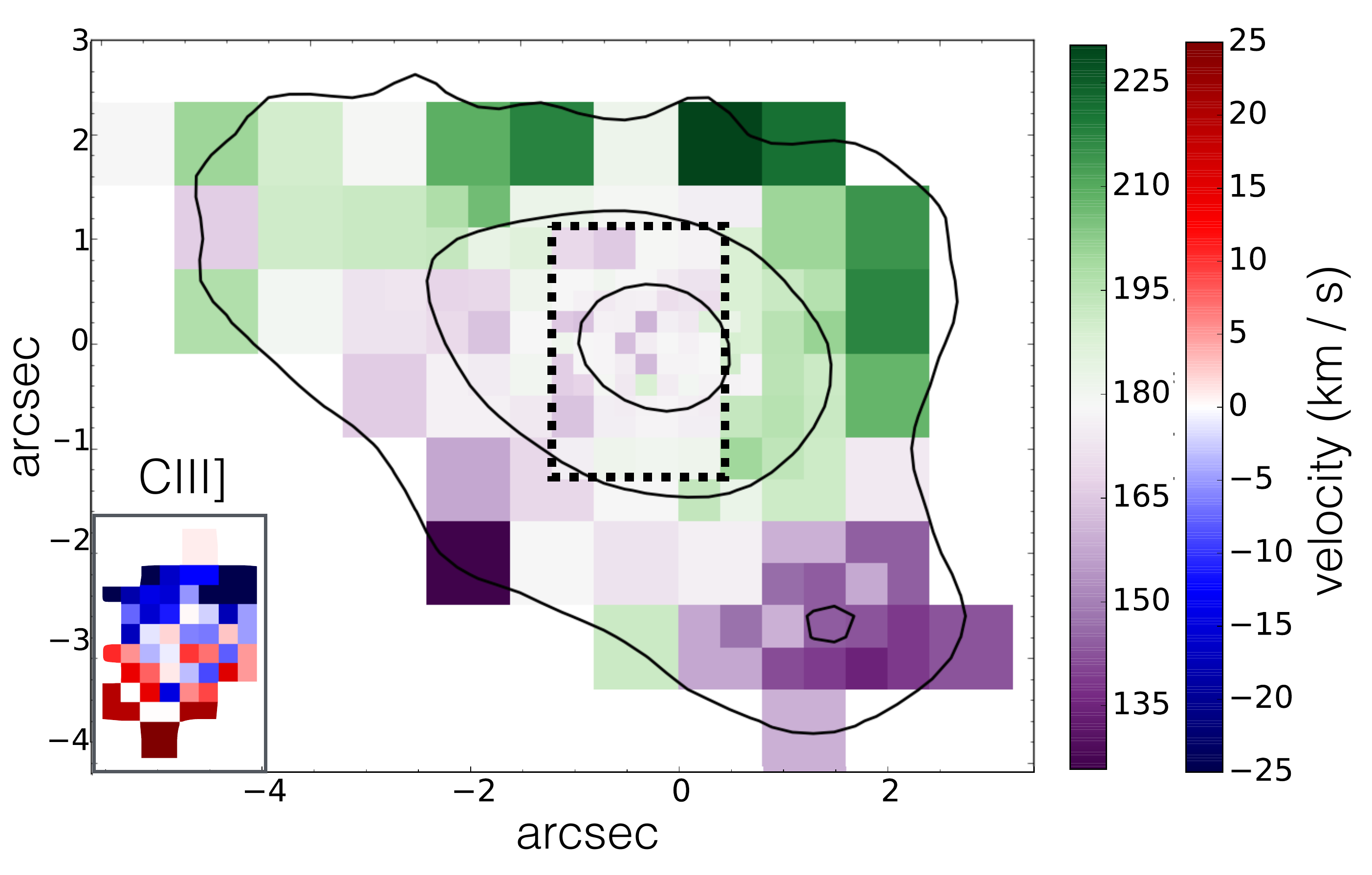}\\
	\end{subfigure}
	\caption[Maps]{\Lya velocity map relative to systemic velocity. Small insert on left bottom corner: $\mbox{C\,{\sc iii}}]$ velocity map, also relative to the systemic velocity, of the central part of the galaxy (marked with a black dotted rectangle in the \Lya image). It is worth noticing that the $\mbox{C\,{\sc iii}}]$ velocity (which traces the gas kinematics) is not reproduced in the \Lya map, since not only kinematics but also resonant processes influence the frequency of the emitted photons.}
\label{maps}
\end{center}
\end{figure}

\subsubsection{Surface Brightness Profiles}
\label{sb_obs}

To study the spatial extension of this object, we use the \Lya line image described in Sect. \ref{dataset} as well as a continuum image, defined after inspecting the combined spectrum and selecting contiguous wavelength ranges free of emission or absorption lines, with a total width of 140 $\AA$ redward of \Lya . To prevent contamination from the nearby companion, we mask out the corresponding region ( \# 3 in Fig.~\ref{regions}) in the images.

Both images are reconstructed in the source plane, where the galaxy shows an elliptical shape (mostly due to the PSF in the source plane). We measure the average surface brightness in elliptical annuli of 0.3 kpc width. The major/minor axis ratio of the ellipses is defined by the ratio of the FWHM of an elliptical 2D gaussian fit to the \Lya line source plane image. The radius of the ellipse is converted into a circularised effective radius ($r = \sqrt{(b / a)} \times a$; \citealt{Law+12}) in kpc and the mean surface brightness is measured at each elliptical aperture in the \Lya and continuum images (see Fig.~\ref{profile}). Since these measurements are done in source plane, neighbouring pixels are highly correlated, so the noise level is estimated in the image plane and corrected for magnification and pixel rescaling. We also look for differences between profiles extracted along different radial orientations, once more excluding the companion, but find no significant deviations from the mean profile, indicating that the \Lya emission has a nearly isotropic shape. 

In order to compare this result with other surface brightness studies it is important to disentangle the intrinsic size of the galaxy from what is observed due to the seeing. As a first approach, we assume that the intrinsic surface brightness profile can be well described with an exponential function: $s(r) = C \exp\,(- r / r_0)$, where $s$ is the surface brightness at radius $r$ and $r_0$ the scale length. To simulate the effects of the seeing, we build a sample of 2D exponential images with scale lengths varying from 0.2 to 2.0 kpc and convolve them with the source plane PSF. We then apply the same procedure used in the observations to obtain the surface brightness profile of these PSF convolved exponentials. Comparing the library of exponential radial profiles with the observations we find that the \Lya emission is best fit by a scale length of $0.98\pm0.03$ kpc and the continuum by $0.33\pm0.02$ kpc. The continuum surface brightness intensity does not deviate significantly from an exponential profile (see Fig.~\ref{profile}), whereas the \Lya emission is higher than this simple model at $r>4$ kpc. 

We compare this single exponential scale length with similar objects in the local Universe --- the LARS (Lyman Alpha Reference Sample, \citealt{Hayes+13}) --- a sample of local ($z<$0.18) star bursting galaxies. \cite{Hayes+13} measure the Petrosian radius ($R_{P20}$) --- the radius at which the local surface brightness is 20\% the average surface brightness inside of R --- of 14 local galaxies finding that these are generally substantially larger in \Lya than in the FUV. As in the case of the scale lengths, it is desirable to compare intrinsic (deconvolved from the PSF) $R_{P20}$ values, which we estimate by converting the intrinsic scale lengths with $R_{P20}=3.62\,r_0$. We calculate a $R^{Ly\alpha}_{P20}$ of 3.5 kpc, a value that is a lower limit due to the mismatch between model and observations on the outskirts of the galaxy, and 1.2 kpc for the continuum Petrosian radius. Comparing this with figure 3 from \citealt{Hayes+13}, we confirm that this galaxy is comparable to this sample of local galaxies regarding \Lya /continuum extension both in absolute and relative scales of \Lya and continuum.

We cannot directly compare these values with the ones obtained from stacking of LBGs and LAEs (e.g. \citealt{Steidel+11}), because in these studies only the more extended part of the profile are fitted, discarding the central region. Additionally, our previous approach does not fully reproduce the \Lya profile in the outer part of the galaxy, which prompts us to try to disentangle the contribution of the central part of the galaxy from the halo. To do this, we follow the approach introduced by \citealt{Wisotzki+15} and model the emission as the sum of two exponentials, one aiming to reproduce the central emission originating from the star forming regions and the other the more extended gaseous halo. We once more make a library of PSF convolved profiles, and fix the scale length of the inner part to the one obtained from the continuum emission fit (0.33 kpc), obtaining a \Lya 'halo' scale length of $1.51\pm0.18$ kpc (Fig.~\ref{double_exp}). This new model brings some improvements compared to the simple exponential: the $4<r<7$ kpc zone is now accurately reproduced. There is still a mismatch at larger radii, where the observations show an even more extended emission, which would require a scale length of $\sim5$ kpc, as estimated from the slope measured in the 8-12 kpc region. 

Comparing the scale lengths of 0.33 and 1.51 kpc (continuum and \Lya respectively) with typical values obtained in high redshift studies is not easy, since system 1 is less massive and smaller system than the high redshift sources usually studied. For example, the continuum scale length of this particular object is 10 times smaller than typical lengths of the \cite{Steidel+11} sample. In a similar study, \cite{Momose+14} also make a careful analysis of stacked images in redshift bins from 2 to 6.6, finding a \Lya scale length of 5 to 10 kpc, for a redshift close to 3.5, which is higher than the scale length we obtained from the double exponential fit. This is nevertheless comparable to what we directly measure on the 8-12 kpc region, although we can not confidently argue that this region in this particular object corresponds to the extended emission measured in \cite{Momose+14}. We choose to instead compare the r$_0$(\Lya )/r$_0$(Continuum) ratio, to test if the extent of \Lya relative to the continuum in our small object is comparable to that of bigger objects: i.e., does the \Lya extension depend on the nature/mass/environment of galaxies? Using the \Lya scale length from the double exponential model, we obtain a ratio of 4.6, whereas \cite{Steidel+11} obtain a ratio of 7.4 for the global sample of LBGs and 8.8 for a subsample of \Lya emitters, suggesting that this system \Lya emission is less extended than these bigger objects compared to the continuum emission. We note that some of the differences may arise from the surrounding Mpc-scale environment, particularly the density of the circum-galactic medium, as concluded by \cite{Matsuda+12}. \citealt{Wisotzki+15} recently used the newly obtained MUSE dataset on the Hubble Deep Field South \citep{Bacon+15} to study a relatively large (26) sample of \Lya emitters on an object to object basis. These are galaxies with comparable or slightly larger scale lengths both in \Lya and continuum emission. Directly comparing the scale lengths we find two other sources in the HDFS sample at $z=4$ with very similar values, although the \Lya / continuum ratio in our galaxy is slightly smaller than the typical ratio ($\sim5-10$) of the \citealt{Wisotzki+15} sample.

It is worth noticing that our error bars only include statistical errors. Systematics errors can affect the relative weights of the data points in the central part and the outskirts and be the source of the mismatch seen at $r>7$ kpc. To test this we increased the relative errors in the central region to a constant value of 5\%, matching the errors at 6 kpc where the observations start to deviate from the fit. Under this weighting scheme we would obtain a larger scale lenght ($r_o = 2.7$ kpc) for \Lya in the double-exponential model, but the fit still does not reproduce the observations at $r>7$ kpc. We alternatively tested the use of S\'ersic profiles with $1<n<10$ for the extended halo component but they did not provide a better fit in this region. 

\begin{figure}
	\begin{subfigure}[b]{0.49\textwidth}
		\includegraphics[width=\textwidth]{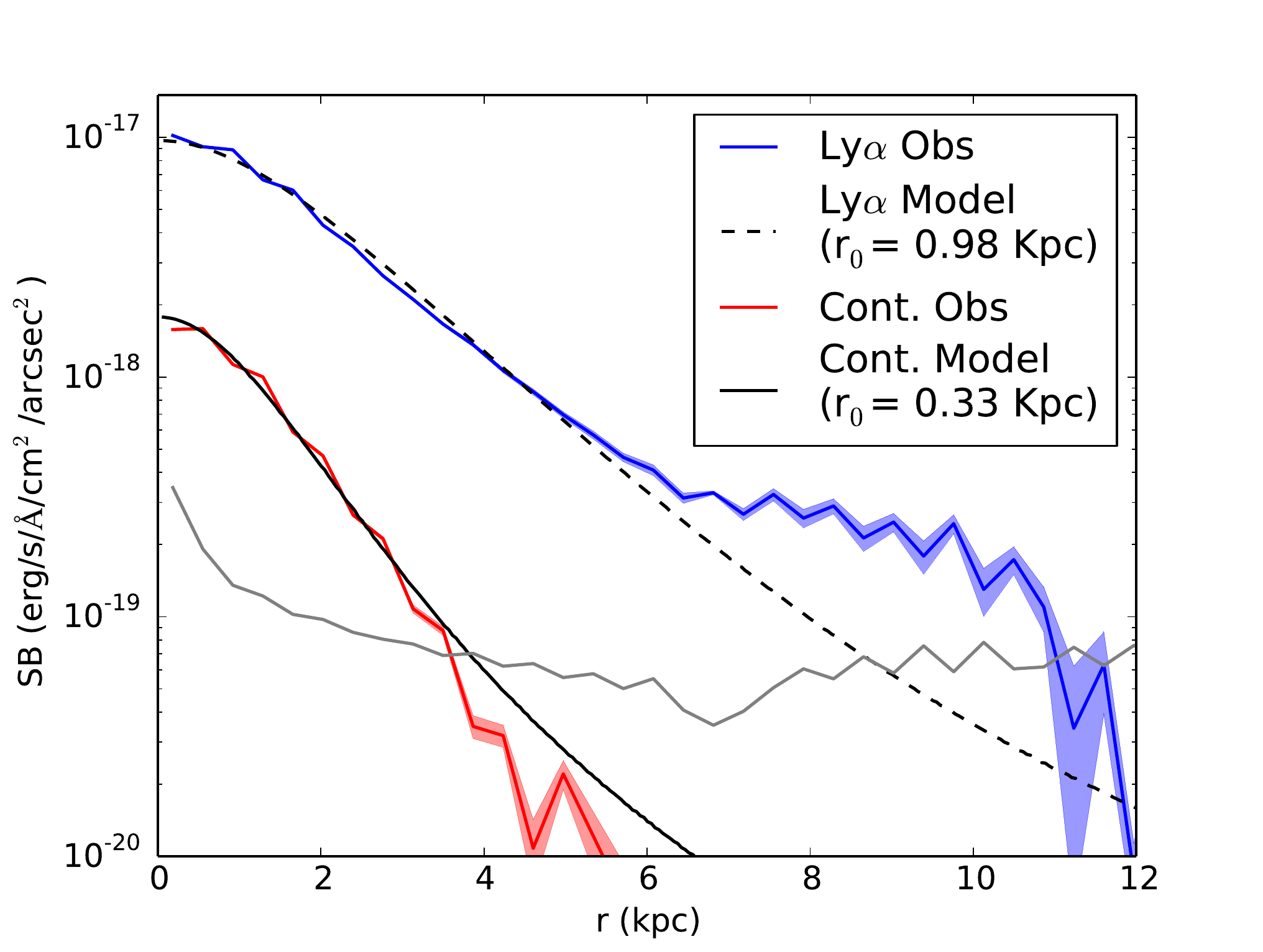}\\
	\end{subfigure}
\caption{\textit{Blue:} \Lya surface brightness profile measured in source plane, based on the source reconstruction of image 1.3. \textit{Red:} Same as previous, but from the continuum line image. \textit{Dashed line:} Exponential of scale length 0.98 kpc convolved with the MUSE PSF. Best match to the observed \Lya profile. \textit{Black line}: Same as before, but for the continuum profile. \textit{Grey line}: 2 $\sigma$ detection limit for \Lya alpha surface brightness}
\label{profile}
\end{figure}

\begin{figure}
	\begin{subfigure}[b]{0.49\textwidth}
		\includegraphics[width=\textwidth]{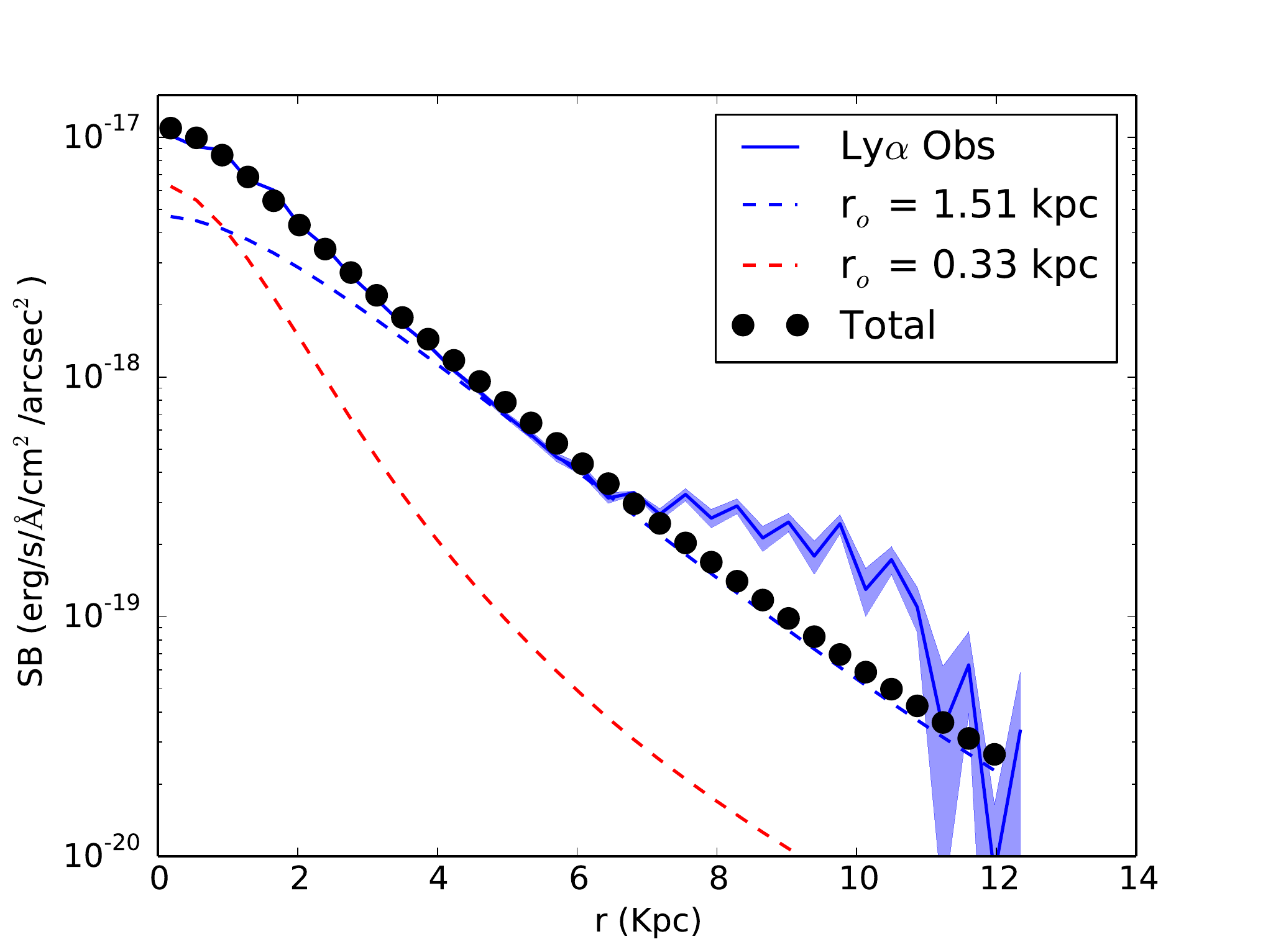}\\
	\end{subfigure}
\caption{\textit{Blue:} \Lya surface brightness profile measured in source plane, based on the source reconstruction of image 1.3. \textit{Dashed lines:}  Two components of the two exponential model. \textit{Dots:} Combined two-exponential model Compared with the previous (single exponential) tested model, we are able to better reproduce the 6-7 kpc region. Nevertheless, the outer region ($>$10 kpc) is still not correctly reproduced.}
\label{double_exp}
\end{figure}

\section{Lyman $\alpha$ modelling}
\label{Lyafit}

In this section we model the \Lya emission using a radiative transfer code predicting the spectra emerging from spherically expanding shells \citep{Verhamme+06}. In our first approach, we fit the \Lya line using the usual library produced by the radiative transfer code, deriving the physical parameters that better describe the total and companion spectra. In the second part we try, for the first time, to test if the same simple expanding shell model can simultaneously describe the three main body spectra (central, halo, total).

\subsection{Fitting of individual spectra with expanding shells}
\label{fit1}

 \begin{figure*}
 \begin{tabular}{cc}
  \includegraphics[width=0.5\textwidth]{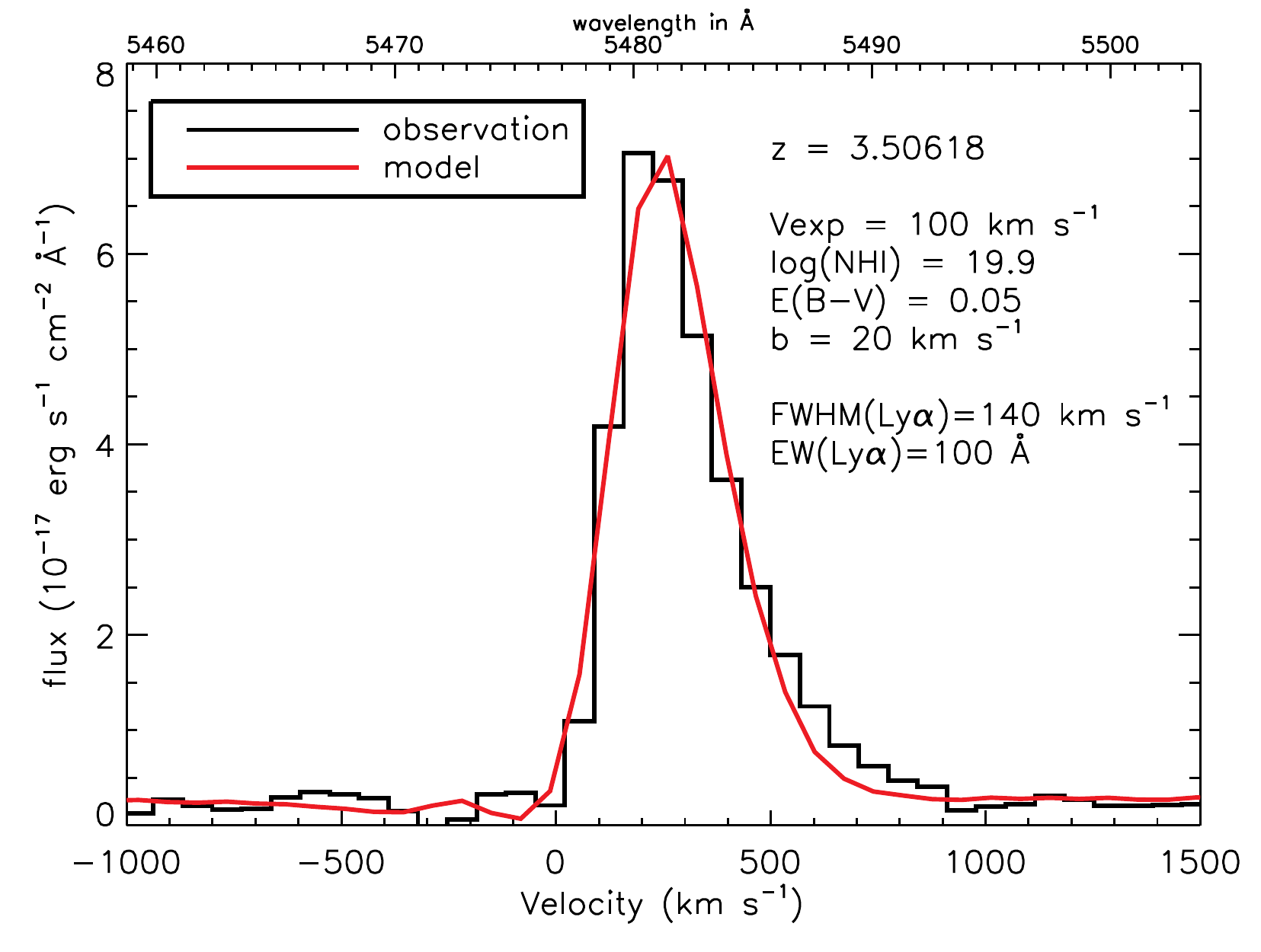} &
  \includegraphics[width=0.5\textwidth]{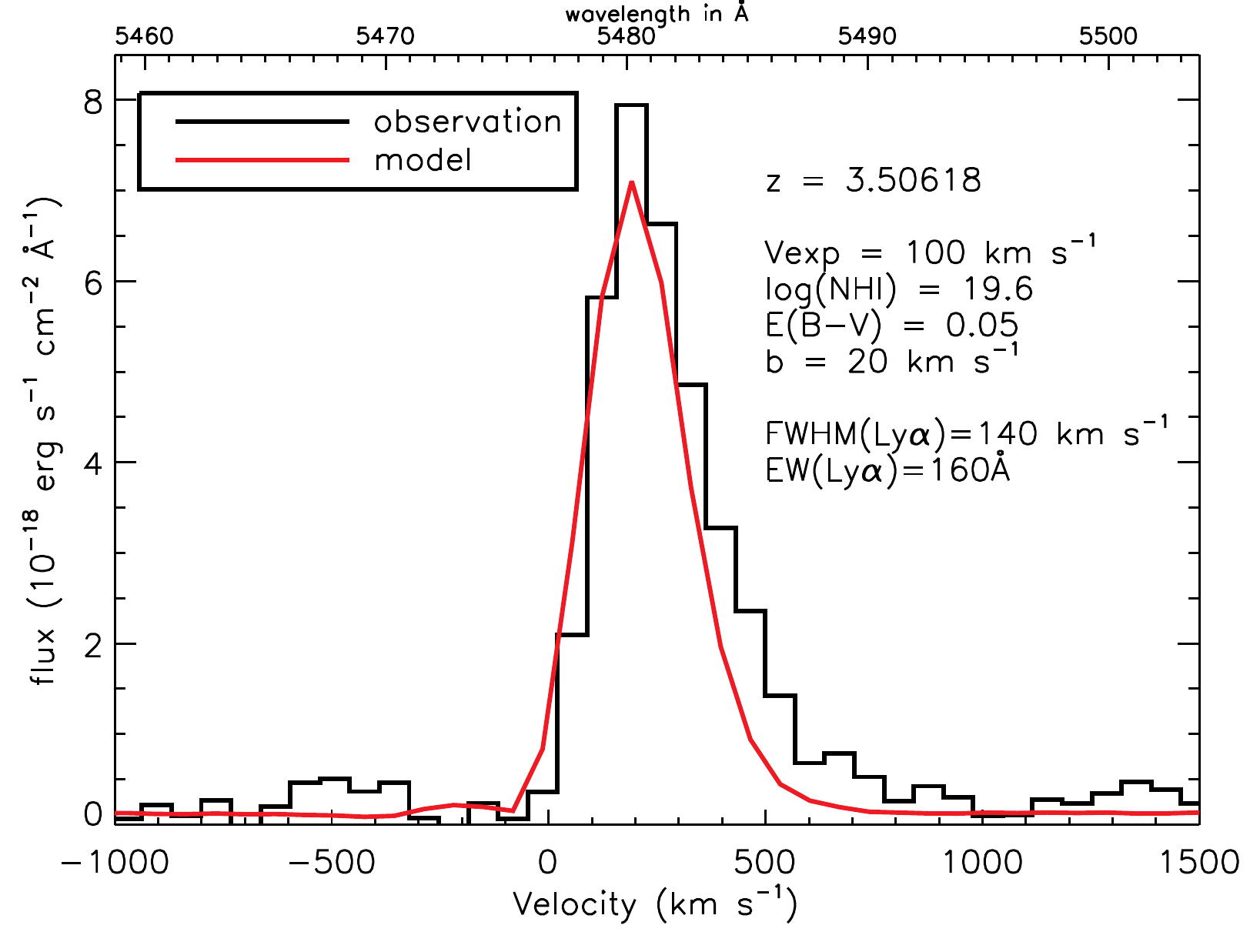} \\
  \end{tabular}
  \caption{Best fits of synthetic \Lya  spectra from homogeneous expanding shells to two different extracted MUSE spectra. {\bf Left:} Total \Lya  spectrum of System 1.3 in SMACS2031, {\bf Right:} \Lya  spectrum extracted from the companion. The models perform well on the bluer side of the peak emission, also correctly reproducing the peak, but fail to completely reproduce the red wing, predicting less flux than the one observed. Both models have similar physical parameters.}
  \label{lya_fit}
  \end{figure*}
  
We separately fit the spectra of the main body and the companion with our library of synthetic spectra \citep{Schaerer+11} which are characterised by four physical parameters: 

\begin{itemize}
\item the radial expansion velocity $v_{exp}$,
\item the radial column density $N_{H I}$, 
\item the Doppler parameter $b$ encoding thermal and turbulent motions,
\item the dust absorption optical depth $\tau_d$, linked to the extinction 
by $E(B-V)\,  \sim (0.06 ... 0.11) $ $\tau_d$, 
\end{itemize}
the lower numerical value corresponding to a Calzetti attenuation law for starbursts, the higher to the Galactic extinction law from \citet{Seaton+79}. The intrinsic \Lya  emission, before transfer, is modelled as a Gaussian plus continuum with two parameters: the intrinsic EW(\Lya ) and the intrinsic FWHM(\Lya ) of the profile. This library has already been successfully used to model \Lya  profiles from high redshift star-forming galaxies \citep{Verhamme+08, Dessauges-Zavadsky+10, Vanzella+10, Lidman+12, Hashimoto+15}, but also in the local Universe \cite[][Orlitova et al. in prep] {Atek+09, Leitherer+13}. 

In order to limit the number of free parameters, we redshift the synthetic spectra to the observed redshift $z=3.50618$; we fix the FWHM of the intrinsic \Lya   profile to 140 km\,s$^{-1}$, the mean FWHM of the nebular lines as measured in the observed total spectrum, and we allow for only low extinction values, $0<E(B-V)<0.05$, consistent with our stellar population results (Sect. \ref{stellarpop}). The results are presented in Fig~\ref{lya_fit}, and the best fit parameters are summarised in Table \ref{lya_fit_param}. 

Reasonable fits are obtained for both spectra, although the extension of the red wing is not well reproduced. The best fit values of the parameters are very similar for both spectra, which may be expected given their similar shapes. The obtained expansion velocity is 100 km\,s$^{-1}$ for both spectra and the column density is in the range $19<\log(N_{H\,{\sc i}})<20$ for the 10 best fits of each object, corresponding to typical values obtained in similar studies of high-redshift \Lya  emitters. Compared with the results obtained by \cite{Christensen+12b}, who used a combination of hydrogen clouds and an expanding shell, we derive similar parameters for the column density and shell properties (Table \ref{lya_fit_param}), but higher intrinsic equivalent width. This difference probably comes from the lower reddening in their model.

We should keep in mind that our model is an oversimplification of the circumgalactic gas around our object, that has certainly a more complex density and velocity distributions than the assumed homogeneous shell expanding at a single velocity. From the absorption profiles of the low ionisation lines (see Fig.~\ref{absorption}) we see a distribution of velocities and densities, with most of the gas being static or slowly expanding, but also outflowing material up to 500 km/s. The velocity and the column density values derived from the \Lya fitting can be interpreted as the effective velocity and column density seen by \Lya in this object and although they provide us with (at least) the right order of magnitude\footnote{Blind fitting of noisy synthetic spectra from expanding shells has been recently studied in detail by \citet{Gronke+15}, demonstrating that degeneracies exist among some parameters of the ``expanding shell'' model, but the column density and the expansion velocity are always accurately recovered.}, they are not direct measures and the real gas distribution is more complex than what can be learned from these models.

\begin{table*}
\begin{center}
    \begin{tabular}{l | c | c | c | c | c | c | c | c| }
                \hline  
        region    & $z$ & EW$_{int}$(\Lya ) & FWHM$_{int}$(\Lya ) & $v_{exp}$ (km\,s$^{-1}$) &  $\log$($N_{H I}$) (cm$^{-2}$) & $b$ (km\,s$^{-1}$) & $E(B-V)$ & $\chi^2_{ref}$ \\
                \hline \hline
	Christensen+12 	& 3.5073    	&  20 	& N/A 	& 110 	& 20.7$^{(*)}$  	 & 10 	& 0.01 &	N/A\\
        total 	       			& 3.50618 	&  100 	& 140 	& 100 	& 19.9 	& 20 		& 0.05 &	304\\
	companion      		& 3.50618 	& 144 	& 140 	& 100 	& 19.6 	& 20 		& 0.05 &	14\\
        \hline
        central	& 3.50618 	&  100 	& 140 	& 100 	& 19.9 	& 20 		& 0.05 &	245\\
        halo 		& 3.50618 	& 100	& 140 	& 100 	& 19.9 	& 20 		& 0.05 &	244\\
        \hline  
       \multicolumn{7}{l}{ $^{(*)}$ Equivalent column density from clouds model. }
     \end{tabular}
	 \caption[Lya]{Best fit values obtained by comparing the observed spectra to synthetic spectra emergent from a homogeneous expanding shell model containing 6 parameters: two that define the intrinsic \Lya emission (EW$_{int}(Ly\alpha)$ and FWHM$_{int}$(\Lya ) in km\,s$^{-1}$) and four others that describe the physical parameters of the shell (see text for a detailed description).}      
     \label{lya_fit_param}          
\end{center}
\end{table*}

\subsection{Comparision of synthetic spectra and surface brightness with observations}

We now propose to further test the expanding shell model, by computing the synthetic surface brightness profile and peak shift map as well as spectra from the central and halo regions from our best fit model, and directly comparing them with the available observables. The library of models used so far is not suitable to do this exercise since the spatial information (last scattering location of each photon before escape) is not computed and stored. For this reason we had to rerun the \Lya  radiation transfer simulation corresponding to the best fit model of the global spectrum
to retrieve this information. 

From this new simulation, we  build a synthetic data cube: 
for each photon, we compute the radius of the last scattering location 
onto the projected image of the shell perpendicular to the escape direction. 
Thanks to the spherical geometry of our problem, 
the contribution of all photons is stacked to obtain a distribution of projected radii.
Knowing the frequency and luminosity of each photon with a given $r$, we sample this 
distribution onto a 3D cube with two spatial dimensions and a spectral dimension. 
Our synthetic cube has $101\times101\times101$ pixels, with a spectral range from 
$\sim$1205 to 1232 $\AA$ in rest frame, but no physical spatial
scale, except for the fact that the ionising radiation is emitted from a 
central point source, infinitely small compare to the radius of the 
expanding shell. The last step before building meaningful observables is the convolution of the 
cube with the MUSE PSF and line spread function (LSF). For the PSF, it is necessary to 
to assume a model physical scale, since it has a fixed size
(that we convert to kpc in the source plane). The synthetic surface brightness profile will change depending on the assumed scale, 
and we use this fact to calculate a physical scale by fitting the observed \Lya surface brightness with the surface brightness
extracted from a synthetic \Lya image (5480 to 5490 \AA) following the exact same methodology used in the data (see Sect. \ref{sb_obs}).
After completing these steps, we are left with a synthetic data cube
spatially and spectrally comparable with the observations.

The next step is to define central and halo regions in the model equivalent to the the ones defined in the observations. We use the observed \Lya flux ratio obtained from the observed data --- the central region contains approximately 40\% of the total flux and the halo 60\% (see Sect. \ref{dataanalysis}) --- and integrate the escaping photons from the model in elliptical annuli of increasing size (to compensate for the asymmetry of the PSF) until the fluxes from the model 'central' and 'halo' regions match the observed fluxes. We fit the total observed spectrum with the total synthetic spectrum, obtaining a flux normalisation factor that we use to rescale the synthetic central and halo spectra to their expected flux and compare them with the observations (see Fig.~\ref{lya_fit2}). Both synthetic surface brightness profiles of continuum and \Lya are in good agreement with the observations, although the synthetic surface brightness profile is flatter than the observations at large radii. Interestingly, the synthetic central and halo spectra fit quality is very similar to the one of the total spectrum: the global shape is well reproduced, but they cannot fit the extension of the red wing. We assumed a radius for the gas shell of 54 kpc, which corresponds to a total hydrogen mass in the shell of $\sim$2 $\times 10{^9} M_{\odot}$, about a third of what we obtained for the stellar mass.  

 \begin{figure*}
 \begin{tabular}{cccc}
 \includegraphics[width=0.48\textwidth]{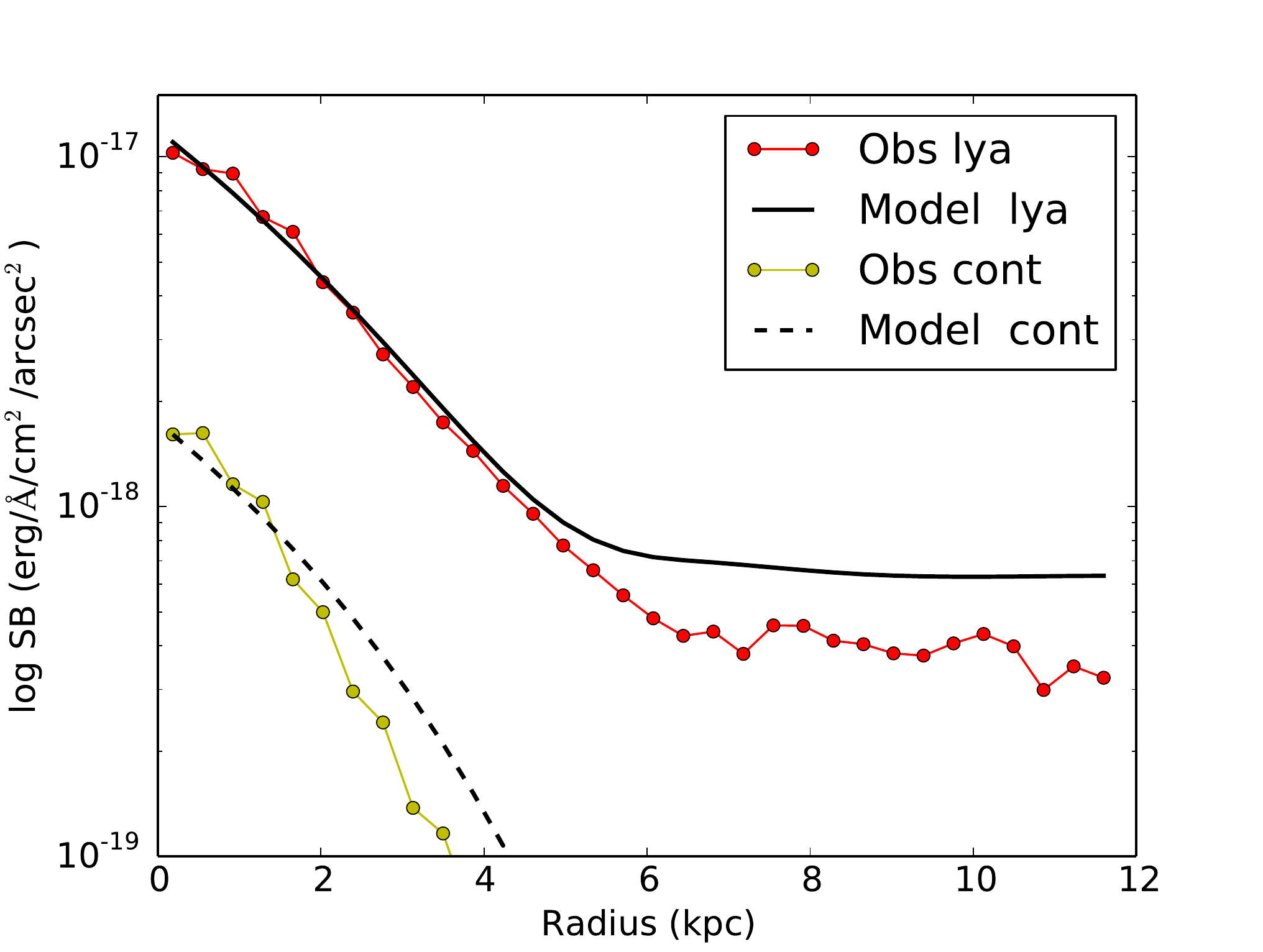} 
 \includegraphics[width=0.48\textwidth]{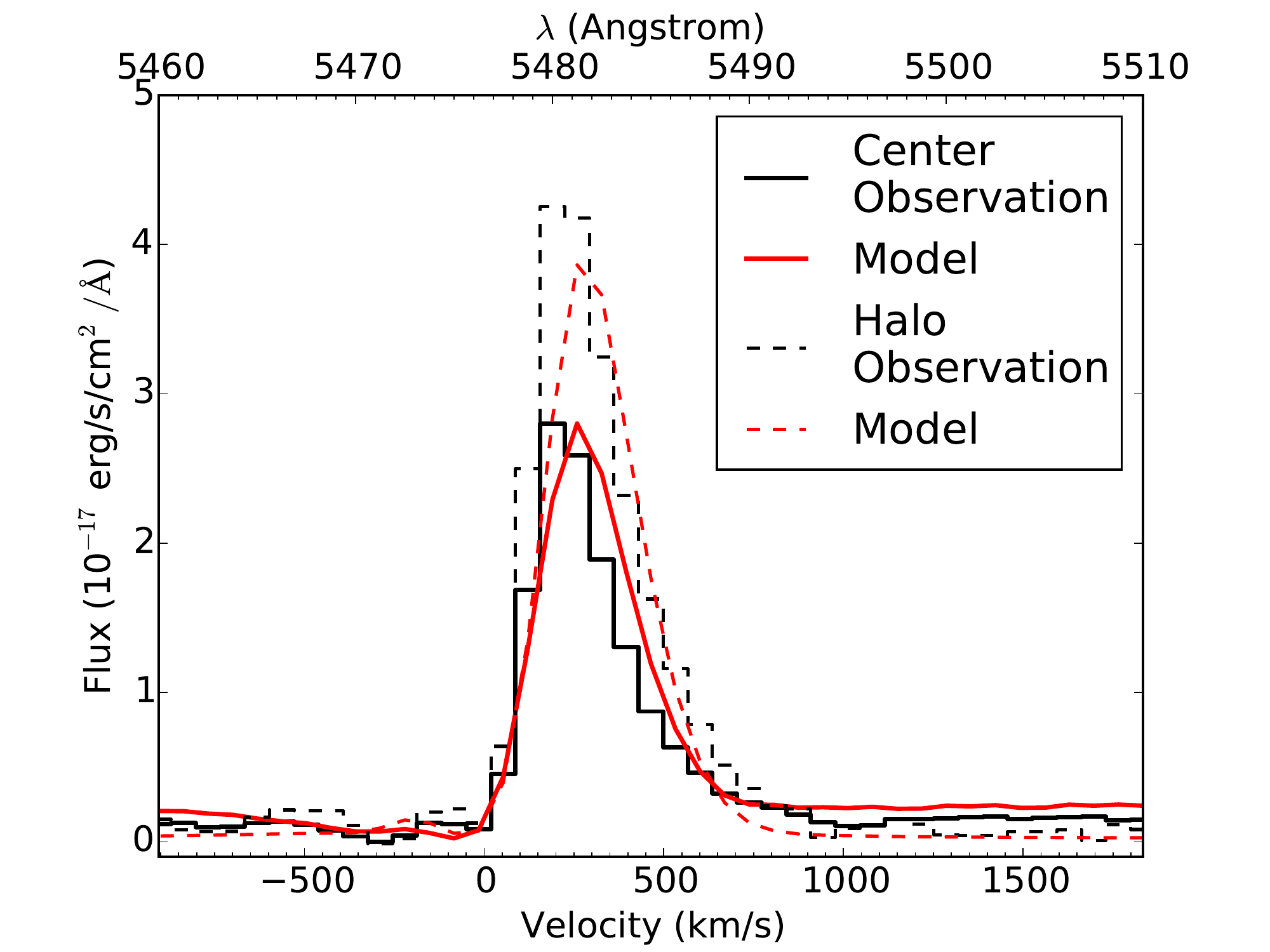} 
 \end{tabular}
 \caption{\textit{Left:} Surface brightness profiles from observations (dotted lines) and radiative transfer models in an expanding gas shell (black lines) after PSF convolution. A good agreement between model and observations is obtained in the inner region for \Lya and \CIII. However at larger radii the model predicts more \Lya flux than observed.  \textit{Right:} observed and best model central spectrum in full lines (black and red, respectively) and observed halo and best model halo spectra, in dashed. The small velocity difference between central and halo region is correctly predicted by the model and it is interesting to notice that these fits reproduce the observations with the same quality as the one obtained with the global spectrum.}
 \label{lya_fit2}
 \end{figure*}
  
Given the spherical symmetry of our geometry, the velocity maps show a radial gradient with the peak of the \Lya  profile slightly shifting further away from line centre with radius, with an increase on the order of $\sim50$ km\,s$^{-1}$ for our best fit. This variation between central and edge is compatible with what we observe in the upper-right edge of the velocity map. The main difference between the observed \Lya  maps and the model is the asymmetry seen in the observations, where the concentric distribution of velocities is not seen. This is not completely unexpected, due to the simplicity of the expanding shell model, that assumes, for example, a single shell of uniform density, a single ionising \Lya source, while HST data show at least two star forming regions in this galaxy, and does not account for possible interactions with nearby objects (does the companion interact with the gas?). 

Overall, we conclude that the strong and diverse constraints provided by this 
observation are generally reproduced by a simple expanding shell model, which can altogether fit  
the integrated spectrum, the spatially resolved spectra, the surface brightness profile 
and the shift of the line peak with radius with good accuracy, although not perfectly. Alternative, more complex models 
have been suggested, for example incorporating biconical outflows \citep{Mas-Hesse+03,Swinbank+07,Dijkstra+12, Laursen+13}, are beyond the scope of this study.

Finally, we combine information derived from the integrated spectrum and the expanding shell model to study the energetics of the gas outflow.
 We do so mainly following the considerations of \cite{Swinbank+07}, but  estimating the shell parameters (mass and size) from  the \Lya radiative transfer model. From the stellar population analysis, a recent ($\sim$10 Myr) period of star formation is seen. Assuming that this episode was responsible for the expansion of the shell, that the UV SFR (17.5 M$\odot$ yr$^{-1}$) does not change within this period and that SNe provide 10$^{49}$ ergs per M$_\odot$ (e.g. \citealt{Benson+03}), we estimate that there are  $\sim$2 $\times$ 10$^{57}$ ergs available from star formation. Assuming an expanding velocity of 100 km s$^{-1}$ and the shell mass derived in the previous section ($\sim$ 2 $\times$ 10$^9$ M$_\odot$) we obtain a kinetic energy of $\sim$2$\times$10$^{56}$ ergs. We also estimate the gravitational influence of a dark matter halo of 10$^{11.8}$ M$_\odot$ (typical for such a galaxy at these redshift, e.g. \citealt{Moster+13}) in the shell expansion. If we assume a simple point source scenario to calculate the loss of potential energy up to a 54 kpc radius, we obtain a loss of potential energy of $\sim$5$\times$10$^{56}$ ergs. Bearing in mind that these are rough estimations, we obtain a coupling factor between the radiative energy from star formation and the gas of $\sim$35\%.

\section{Summary}
\label{summary}

We present new observations of a high redshift ($z=3.5$) L$^*$ galaxy, obtaining exceptional spectra quality thanks to both to the magnification of gravitational lensing and the high-sensitivity spatial coverage and resolution of MUSE. We summarise our main results below:

\begin{itemize}

\item The high signal-to-noise rest-frame UV spectrum contains a wealth of emission and absorption lines, which we combine to derive physical properties (density, temperature, gas covering fraction), in the central kpc of this young $\sim$ 6\, $\times$\, 10$^9$ M$_\odot$ galaxy.

\item We spatially resolve the extended \Lya emission and its velocity, as well as, for the first time, the associated $\mbox{C\,{\sc iii}}]$ kinematics in the central region. The \Lya emission has a very uniform profile across $\sim10$ kpc, showing only a small velocity shift that is unrelated to the intrinsic kinematics of the nebular emission. 

\item We identify a companion object, previously unknown, located 11.2 kpc away, only detected in \Lya emission. The line profile is similar profile to the main object, although blueshifted by 20 km\,s$^{-1}$. 

\item We find that both stellar population fitting and direct estimations yield compatible stellar metallicities: 0.07 and 0.05 Z$_\odot$. We derive an ISM metallicity of 0.016--0.065 Z$_\odot$, which agrees with previously derived ISM metallicities and also confirms that this particular object is $\sim$2 $\sigma$ lower than the expected mass-metallicity relation. 

\item The \Lya  emission is more extended than the continuum emission, with length scales of 1.51 kpc and 0.34 kpc respectively. Comparing this with the results of stacked images, this galaxy seems more compact (\Lya  is less extended relatively to continuum) than what is obtained for bigger and more massive galaxies at equivalent redshifts. Nevertheless, this object is comparable with the local LARS sample and to some individual galaxies from the HUDF LAE sample.

\item The observed \Lya  line is well-fit by synthetic spectra generated from a spherically expanding $\sim2$ $\times$ 10$^9$ M$_\odot$ gas shell model. We further test this model by comparing the predicted surface brightness and predicted central and halo spectra and conclude that this simple model is compatible with the observations.

\end{itemize}

In our \Lya analysis, we  not only fit the \LyaÊspectral-profile, but also compare the predicted resolved properties, such as spatially resolved spectra, velocity maps and surface brightness profile with observations. The discrepancies seen between the simulated and observed \Lya properties, despite their general qualitative agreement, demonstrate that the quality of the presented observations represent a challenge to current models, and show the power of such a comparison and the need for further studies of resolved LAEs to improve our understanding of \Lya properties. Future observations of lensing clusters, as part of the MUSE guaranteed time observations, will allow us to find more examples of such sources at high redshift, which might constitute a rich testing sample not only for \Lya radiative transfer codes predictions, but also for other galaxy evolution features, such as structure formation and chemical enrichment.

\section*{Acknowledgments}
We thank the referee for providing useful comments on the submitted version that improve the clarity and content of this work as well as Claudia Scarlata, Dan Stark  and Thibault Garel for insightful and stimulating discussions. VP, JR, DL and BC acknowledge support from the ERC starting grant CALENDS. JR acknowledges support from the CIG grant 294074.  A.V. is supported by a Fellowship ÒBoursi\`ere d'ExcellenceÓ of Geneva University. RB acknowledges support from the ERC advanced grant 339659-MUSICOS. JR, TC and GS acknowledge support of the ANR FOGHAR (ANR-13-BS05-0010-02). LC is supported by YDUN/DFF Ð 4090-00079. 

\bibliographystyle{mn2e}
\bibliography{references}

\begin{thebibliography}{}

\bibitem[\protect\citeauthoryear{{Atek}, {Schaerer} \& {Kunth}}{{Atek}
  et~al.}{2009}]{Atek+09}
{Atek} H.,  {Schaerer} D.,    {Kunth} D.,  2009, \aap, 502, 791

\bibitem[\protect\citeauthoryear{{Bacon}, {Accardo}, {Adjali}, {Anwand},
  {Bauer}, {Biswas}, {Blaizot}, {Boudon}, {Brau-Nogue}, {Brinchmann} \&
  {Caillier}}{{Bacon} et~al.}{2010}]{Bacon+10}
{Bacon} R.,  {Accardo} M.,  {Adjali} L.,  {Anwand} H.,  {Bauer} S.,  {Biswas}
  I.,  {Blaizot} J.,  {Boudon} D.,  {Brau-Nogue} S.,  {Brinchmann} J.,
  {Caillier} P.,  2010, in Society of Photo-Optical Instrumentation Engineers
  (SPIE) Conference Series Vol.~7735 of Society of Photo-Optical
  Instrumentation Engineers (SPIE) Conference Series, {The MUSE
  second-generation VLT instrument}.
p.~8

\bibitem[\protect\citeauthoryear{{Bacon}, {Brinchmann}, {Richard}, {Contini},
  {Drake}, {Franx}, {Tacchella}, {Vernet}, {Wisotzki}, {Blaizot} \&
  {Bouch{\'e}}}{{Bacon} et~al.}{2015}]{Bacon+15}
{Bacon} R.,  {Brinchmann} J.,  {Richard} J.,  {Contini} T.,  {Drake} A.,
  {Franx} M.,  {Tacchella} S.,  {Vernet} J.,  {Wisotzki} L.,  {Blaizot} J.,
  {Bouch{\'e}} 2015, \aap, 575, A75

\bibitem[\protect\citeauthoryear{{Bayliss}, {Rigby}, {Sharon}, {Wuyts},
  {Florian}, {Gladders}, {Johnson} \& {Oguri}}{{Bayliss}
  et~al.}{2014}]{Bayliss+14}
{Bayliss} M.~B.,  {Rigby} J.~R.,  {Sharon} K.,  {Wuyts} E.,  {Florian} M.,
  {Gladders} M.~D.,  {Johnson} T.,    {Oguri} M.,  2014, \apj, 790, 144

\bibitem[\protect\citeauthoryear{{Behrens}, {Dijkstra} \& {Niemeyer}}{{Behrens}
  et~al.}{2014}]{Behrens+14}
{Behrens} C.,  {Dijkstra} M.,    {Niemeyer} J.~C.,  2014, \aap, 563, A77

\bibitem[\protect\citeauthoryear{{Benson}, {Bower}, {Frenk}, {Lacey}, {Baugh}
  \& {Cole}}{{Benson} et~al.}{2003}]{Benson+03}
{Benson} A.~J.,  {Bower} R.~G.,  {Frenk} C.~S.,  {Lacey} C.~G.,  {Baugh} C.~M.,
     {Cole} S.,  2003, \apj, 599, 38

\bibitem[\protect\citeauthoryear{{Bian}, {Fan}, {Bechtold}, {McGreer}, {Just},
  {Sand}, {Green}, {Thompson}, {Peng}, {Seifert}, {Ageorges}, {Juette},
  {Knierim} \& {Buschkamp}}{{Bian} et~al.}{2010}]{Bian+10}
{Bian} F.,  {Fan} X.,  {Bechtold} J.,  {McGreer} I.~D.,  {Just} D.~W.,  {Sand}
  D.~J.,  {Green} R.~F.,  {Thompson} D.,  {Peng} C.~Y.,  {Seifert} W.,
  {Ageorges} N.,  {Juette} M.,  {Knierim} V.,    {Buschkamp} P.,  2010, \apj,
  725, 1877

\bibitem[\protect\citeauthoryear{{Bouch{\'e}}, {Hohensee}, {Vargas},
  {Kacprzak}, {Martin}, {Cooke} \& {Churchill}}{{Bouch{\'e}}
  et~al.}{2012}]{Bouche+12}
{Bouch{\'e}} N.,  {Hohensee} W.,  {Vargas} R.,  {Kacprzak} G.~G.,  {Martin}
  C.~L.,  {Cooke} J.,    {Churchill} C.~W.,  2012, \mnras, 426, 801

\bibitem[\protect\citeauthoryear{{Brinchmann}, {Pettini} \&
  {Charlot}}{{Brinchmann} et~al.}{2008}]{Brinchmann+08}
{Brinchmann} J.,  {Pettini} M.,    {Charlot} S.,  2008, \mnras, 385, 769

\bibitem[\protect\citeauthoryear{{Cabanac}, {Valls-Gabaud} \&
  {Lidman}}{{Cabanac} et~al.}{2008}]{Cabanac+08}
{Cabanac} R.~A.,  {Valls-Gabaud} D.,    {Lidman} C.,  2008, \mnras, 386, 2065

\bibitem[\protect\citeauthoryear{{Calzetti}, {Armus}, {Bohlin}, {Kinney},
  {Koornneef} \& {Storchi-Bergmann}}{{Calzetti} et~al.}{2000}]{Calzetti+00}
{Calzetti} D.,  {Armus} L.,  {Bohlin} R.~C.,  {Kinney} A.~L.,  {Koornneef} J.,
    {Storchi-Bergmann} T.,  2000, \apj, 533, 682

\bibitem[\protect\citeauthoryear{{Christensen}, {Laursen}, {Richard}, {Hjorth},
  {Milvang-Jensen}, {Dessauges-Zavadsky}, {Limousin}, {Grillo} \&
  {Ebeling}}{{Christensen} et~al.}{2012b}]{Christensen+12b}
{Christensen} L.,  {Laursen} P.,  {Richard} J.,  {Hjorth} J.,  {Milvang-Jensen}
  B.,  {Dessauges-Zavadsky} M.,  {Limousin} M.,  {Grillo} C.,    {Ebeling} H.,
  2012, \mnras, 427, 1973

\bibitem[\protect\citeauthoryear{{Christensen}, {Richard}, {Hjorth},
  {Milvang-Jensen}, {Laursen}, {Limousin}, {Dessauges-Zavadsky}, {Grillo} \&
  {Ebeling}}{{Christensen} et~al.}{2012a}]{Christensen+12a}
{Christensen} L.,  {Richard} J.,  {Hjorth} J.,  {Milvang-Jensen} B.,  {Laursen}
  P.,  {Limousin} M.,  {Dessauges-Zavadsky} M.,  {Grillo} C.,    {Ebeling} H.,
  2012, \mnras, 427, 1953

\bibitem[\protect\citeauthoryear{{Cid Fernandes}, {Mateus}, {Sodr{\'e}},
  {Stasi{\'n}ska} \& {Gomes}}{{Cid Fernandes} et~al.}{2005}]{Cid+05}
{Cid Fernandes} R.,  {Mateus} A.,  {Sodr{\'e}} L.,  {Stasi{\'n}ska} G.,
  {Gomes} J.~M.,  2005, \mnras, 358, 363

\bibitem[\protect\citeauthoryear{{Dessauges-Zavadsky}, {D'Odorico}, {Schaerer},
  {Modigliani}, {Tapken} \& {Vernet}}{{Dessauges-Zavadsky}
  et~al.}{2010}]{Dessauges-Zavadsky+10}
{Dessauges-Zavadsky} M.,  {D'Odorico} S.,  {Schaerer} D.,  {Modigliani} A.,
  {Tapken} C.,    {Vernet} J.,  2010, \aap, 510, A26

\bibitem[\protect\citeauthoryear{{Dijkstra}}{{Dijkstra}}{2014}]{Dijkstra+14}
{Dijkstra} M.,  2014, 31, 40

\bibitem[\protect\citeauthoryear{{Dijkstra} \& {Kramer}}{{Dijkstra} \&
  {Kramer}}{2012}]{Dijkstra+12}
{Dijkstra} M.,  {Kramer} R.,  2012, \mnras, 424, 1672

\bibitem[\protect\citeauthoryear{{Ebeling}, {Edge} \& {Henry}}{{Ebeling}
  et~al.}{2001}]{Ebeling+01}
{Ebeling} H.,  {Edge} A.~C.,    {Henry} J.~P.,  2001, \apj, 553, 668

\bibitem[\protect\citeauthoryear{{Erb}, {Pettini}, {Shapley}, {Steidel}, {Law}
  \& {Reddy}}{{Erb} et~al.}{2010}]{Erb+10}
{Erb} D.~K.,  {Pettini} M.,  {Shapley} A.~E.,  {Steidel} C.~C.,  {Law} D.~R.,
   {Reddy} N.~A.,  2010, \apj, 719, 1168

\bibitem[\protect\citeauthoryear{{Erb}, {Steidel}, {Trainor},
  {Bogosavljevi{\'c}}, {Shapley}, {Nestor}, {Kulas}, {Law}, {Strom}, {Rudie},
  {Reddy}, {Pettini}, {Konidaris}, {Mace}, {Matthews} \& {McLean}}{{Erb}
  et~al.}{2014}]{Erb+14}
{Erb} D.~K.,  {Steidel} C.~C.,  {Trainor} R.~F.,  {Bogosavljevi{\'c}} M.,
  {Shapley} A.~E.,  {Nestor} D.~B.,  {Kulas} K.~R.,  {Law} D.~R.,  {Strom}
  A.~L.,  {Rudie} G.~C.,  {Reddy} N.~A.,  {Pettini} M.,  {Konidaris} N.~P.,
  {Mace} G.,  {Matthews} K.,    {McLean} I.~S.,  2014, \apj, 795, 33

\bibitem[\protect\citeauthoryear{{Fosbury}, {Villar-Mart{\'{\i}}n}, {Humphrey},
  {Lombardi}, {Rosati}, {Stern}, {Hook}, {Holden}, {Stanford}, {Squires},
  {Rauch} \& {Sargent}}{{Fosbury} et~al.}{2003}]{Fosbury+03}
{Fosbury} R.~A.~E.,  {Villar-Mart{\'{\i}}n} M.,  {Humphrey} A.,  {Lombardi} M.,
   {Rosati} P.,  {Stern} D.,  {Hook} R.~N.,  {Holden} B.~P.,  {Stanford} S.~A.,
   {Squires} G.~K.,  {Rauch} M.,    {Sargent} W.~L.~W.,  2003, \apj, 596, 797

\bibitem[\protect\citeauthoryear{{Garel}, {Guiderdoni} \& {Blaizot}}{{Garel}
  et~al.}{2015}]{Garel+15}
{Garel} T.,  {Guiderdoni} B.,    {Blaizot} J.,  2015, ArXiv e-prints

\bibitem[\protect\citeauthoryear{{Gronke}, {Bull} \& {Dijkstra}}{{Gronke}
  et~al.}{2015}]{Gronke+15}
{Gronke} M.,  {Bull} P.,    {Dijkstra} M.,  2015, ArXiv e-prints

\bibitem[\protect\citeauthoryear{{Hainline}, {Shapley}, {Kornei}, {Pettini},
  {Buckley-Geer}, {Allam} \& {Tucker}}{{Hainline} et~al.}{2009}]{Hainline+09}
{Hainline} K.~N.,  {Shapley} A.~E.,  {Kornei} K.~A.,  {Pettini} M.,
  {Buckley-Geer} E.,  {Allam} S.~S.,    {Tucker} D.~L.,  2009, \apj, 701, 52

\bibitem[\protect\citeauthoryear{{Hashimoto}, {Verhamme}, {Ouchi}, {Shimasaku},
  {Schaerer}, {Nakajima}, {Shibuya}, {Rauch}, {Ono} \& {Goto}}{{Hashimoto}
  et~al.}{2015}]{Hashimoto+15}
{Hashimoto} T.,  {Verhamme} A.,  {Ouchi} M.,  {Shimasaku} K.,  {Schaerer} D.,
  {Nakajima} K.,  {Shibuya} T.,  {Rauch} M.,  {Ono} Y.,    {Goto} R.,  2015,
  ArXiv e-prints

\bibitem[\protect\citeauthoryear{{Hayes}, {{\"O}stlin}, {Duval}, {Sandberg},
  {Guaita}, {Melinder}, {Adamo}, {Schaerer}, {Verhamme} \&
  {Orlitov{\'a}}}{{Hayes} et~al.}{2014}]{Hayes+14}
{Hayes} M.,  {{\"O}stlin} G.,  {Duval} F.,  {Sandberg} A.,  {Guaita} L.,
  {Melinder} J.,  {Adamo} A.,  {Schaerer} D.,  {Verhamme} A.,    {Orlitov{\'a}}
  I.,  2014, \apj, 782, 6

\bibitem[\protect\citeauthoryear{{Hayes}, {{\"O}stlin}, {Schaerer}, {Verhamme},
  {Mas-Hesse}, {Adamo}, {Atek}, {Cannon}, {Duval}, {Guaita}, {Herenz}, {Kunth},
  {Laursen}, {Melinder}, {Orlitov{\'a}}, {Ot{\'{\i}}-Floranes} \&
  {Sandberg}}{{Hayes} et~al.}{2013}]{Hayes+13}
{Hayes} M.,  {{\"O}stlin} G.,  {Schaerer} D.,  {Verhamme} A.,  {Mas-Hesse}
  J.~M.,  {Adamo} A.,  {Atek} H.,  {Cannon} J.~M.,  {Duval} F.,  {Guaita} L.,
  {Herenz} E.~C.,  {Kunth} D.,  {Laursen} P.,  {Melinder} J.,  {Orlitov{\'a}}
  I.,  {Ot{\'{\i}}-Floranes} H.,    {Sandberg} A.,  2013, \apjl, 765, L27

\bibitem[\protect\citeauthoryear{{Henry}, {Scarlata}, {Martin} \&
  {Erb}}{{Henry} et~al.}{2015}]{Henry+15}
{Henry} A.,  {Scarlata} C.,  {Martin} C.~L.,    {Erb} D.,  2015, \apj, 809, 19

\bibitem[\protect\citeauthoryear{{James}, {Pettini}, {Christensen}, {Auger},
  {Becker}, {King}, {Quider}, {Shapley} \& {Steidel}}{{James}
  et~al.}{2014}]{James+14}
{James} B.~L.,  {Pettini} M.,  {Christensen} L.,  {Auger} M.~W.,  {Becker}
  G.~D.,  {King} L.~J.,  {Quider} A.~M.,  {Shapley} A.~E.,    {Steidel} C.~C.,
  2014, \mnras, 440, 1794

\bibitem[\protect\citeauthoryear{{Jones}, {Ellis}, {Richard} \&
  {Jullo}}{{Jones} et~al.}{2013}]{Jones+13a}
{Jones} T.,  {Ellis} R.~S.,  {Richard} J.,    {Jullo} E.,  2013, \apj, 765, 48

\bibitem[\protect\citeauthoryear{{Jones}, {Ellis}, {Schenker} \&
  {Stark}}{{Jones} et~al.}{2013}]{Jones+13b}
{Jones} T.~A.,  {Ellis} R.~S.,  {Schenker} M.~A.,    {Stark} D.~P.,  2013,
  \apj, 779, 52

\bibitem[\protect\citeauthoryear{{Kennicutt}
  Jr.}{{Kennicutt}}{1998}]{Kennicutt98}
{Kennicutt} Jr. R.~C.,  1998, \apj, 498, 541

\bibitem[\protect\citeauthoryear{{Kroupa}}{{Kroupa}}{2001}]{Kroupa+01}
{Kroupa} P.,  2001, \mnras, 322, 231

\bibitem[\protect\citeauthoryear{{Laursen}, {Duval} \& {{\"O}stlin}}{{Laursen}
  et~al.}{2013}]{Laursen+13}
{Laursen} P.,  {Duval} F.,    {{\"O}stlin} G.,  2013, \apj, 766, 124

\bibitem[\protect\citeauthoryear{{Law}, {Steidel}, {Shapley}, {Nagy}, {Reddy}
  \& {Erb}}{{Law} et~al.}{2012}]{Law+12}
{Law} D.~R.,  {Steidel} C.~C.,  {Shapley} A.~E.,  {Nagy} S.~R.,  {Reddy} N.~A.,
     {Erb} D.~K.,  2012, \apj, 759, 29

\bibitem[\protect\citeauthoryear{{Leitherer}, {Chandar}, {Tremonti}, {Wofford}
  \& {Schaerer}}{{Leitherer} et~al.}{2013}]{Leitherer+13}
{Leitherer} C.,  {Chandar} R.,  {Tremonti} C.~A.,  {Wofford} A.,    {Schaerer}
  D.,  2013, \apj, 772, 120

\bibitem[\protect\citeauthoryear{{Leitherer}, {Schaerer}, {Goldader},
  {Delgado}, {Robert}, {Kune}, {de Mello}, {Devost} \& {Heckman}}{{Leitherer}
  et~al.}{1999}]{Leitherer+99}
{Leitherer} C.,  {Schaerer} D.,  {Goldader} J.~D.,  {Delgado} R.~M.~G.,
  {Robert} C.,  {Kune} D.~F.,  {de Mello} D.~F.,  {Devost} D.,    {Heckman}
  T.~M.,  1999, \apjs, 123, 3

\bibitem[\protect\citeauthoryear{{Leitherer}, {Tremonti}, {Heckman} \&
  {Calzetti}}{{Leitherer} et~al.}{2011}]{Leitherer+11}
{Leitherer} C.,  {Tremonti} C.~A.,  {Heckman} T.~M.,    {Calzetti} D.,  2011,
  \aj, 141, 37

\bibitem[\protect\citeauthoryear{{Lidman}, {Hayes}, {Jones}, {Schaerer},
  {Westra}, {Tapken}, {Meisenheimer} \& {Verhamme}}{{Lidman}
  et~al.}{2012}]{Lidman+12}
{Lidman} C.,  {Hayes} M.,  {Jones} D.~H.,  {Schaerer} D.,  {Westra} E.,
  {Tapken} C.,  {Meisenheimer} K.,    {Verhamme} A.,  2012, \mnras, 420, 1946

\bibitem[\protect\citeauthoryear{{Luridiana}, {Morisset} \& {Shaw}}{{Luridiana}
  et~al.}{2012}]{pyneb}
{Luridiana} V.,  {Morisset} C.,    {Shaw} R.~A.,  2012, in IAU Symposium
  Vol.~283 of IAU Symposium, {PyNeb: a new software for the analysis of
  emission lines}.
pp 422--423

\bibitem[\protect\citeauthoryear{{Mannucci}, {Salvaterra} \&
  {Campisi}}{{Mannucci} et~al.}{2011}]{Mannucci+11}
{Mannucci} F.,  {Salvaterra} R.,    {Campisi} M.~A.,  2011, \mnras, 414, 1263

\bibitem[\protect\citeauthoryear{{Mas-Hesse}, {Kunth}, {Tenorio-Tagle},
  {Leitherer}, {Terlevich} \& {Terlevich}}{{Mas-Hesse}
  et~al.}{2003}]{Mas-Hesse+03}
{Mas-Hesse} J.~M.,  {Kunth} D.,  {Tenorio-Tagle} G.,  {Leitherer} C.,
  {Terlevich} R.~J.,    {Terlevich} E.,  2003, \apj, 598, 858

\bibitem[\protect\citeauthoryear{{Matsuda}, {Yamada}, {Hayashino}, {Yamauchi},
  {Nakamura}, {Morimoto}, {Ouchi}, {Ono}, {Umemura} \& {Mori}}{{Matsuda}
  et~al.}{2012}]{Matsuda+12}
{Matsuda} Y.,  {Yamada} T.,  {Hayashino} T.,  {Yamauchi} R.,  {Nakamura} Y.,
  {Morimoto} N.,  {Ouchi} M.,  {Ono} Y.,  {Umemura} M.,    {Mori} M.,  2012,
  \mnras, 425, 878

\bibitem[\protect\citeauthoryear{{Momose}, {Ouchi}, {Nakajima}, {Ono},
  {Shibuya}, {Shimasaku}, {Yuma}, {Mori} \& {Umemura}}{{Momose}
  et~al.}{2014}]{Momose+14}
{Momose} R.,  {Ouchi} M.,  {Nakajima} K.,  {Ono} Y.,  {Shibuya} T.,
  {Shimasaku} K.,  {Yuma} S.,  {Mori} M.,    {Umemura} M.,  2014, \mnras, 442,
  110

\bibitem[\protect\citeauthoryear{{Morales-Luis}, {P{\'e}rez-Montero},
  {S{\'a}nchez Almeida} \& {Mu{\~n}oz-Tu{\~n}{\'o}n}}{{Morales-Luis}
  et~al.}{2014}]{Morales-Luis+14}
{Morales-Luis} A.~B.,  {P{\'e}rez-Montero} E.,  {S{\'a}nchez Almeida} J.,
  {Mu{\~n}oz-Tu{\~n}{\'o}n} C.,  2014, \apj, 797, 81

\bibitem[\protect\citeauthoryear{{Moster}, {Naab} \& {White}}{{Moster}
  et~al.}{2013}]{Moster+13}
{Moster} B.~P.,  {Naab} T.,    {White} S.~D.~M.,  2013, \mnras, 428, 3121

\bibitem[\protect\citeauthoryear{{Neufeld}}{{Neufeld}}{1990}]{Neufeld+90}
{Neufeld} D.~A.,  1990, \apj, 350, 216

\bibitem[\protect\citeauthoryear{{Peeples}, {Pogge} \& {Stanek}}{{Peeples}
  et~al.}{2009}]{Peeples+09}
{Peeples} M.~S.,  {Pogge} R.~W.,    {Stanek} K.~Z.,  2009, \apj, 695, 259

\bibitem[\protect\citeauthoryear{{Pettini}, {Rix}, {Steidel}, {Hunt}, {Shapley}
  \& {Adelberger}}{{Pettini} et~al.}{2002}]{Pettini+02}
{Pettini} M.,  {Rix} S.~A.,  {Steidel} C.~C.,  {Hunt} M.~P.,  {Shapley} A.~E.,
    {Adelberger} K.~L.,  2002, \apss, 281, 461

\bibitem[\protect\citeauthoryear{{Prescott}, {Momcheva}, {Brammer}, {Fynbo} \&
  {M{\o}ller}}{{Prescott} et~al.}{2015}]{Prescott+15}
{Prescott} M.~K.~M.,  {Momcheva} I.,  {Brammer} G.~B.,  {Fynbo} J.~P.~U.,
  {M{\o}ller} P.,  2015, \apj, 802, 32

\bibitem[\protect\citeauthoryear{{Quider}, {Pettini}, {Shapley} \&
  {Steidel}}{{Quider} et~al.}{2009}]{Quider+09}
{Quider} A.~M.,  {Pettini} M.,  {Shapley} A.~E.,    {Steidel} C.~C.,  2009,
  \mnras, 398, 1263

\bibitem[\protect\citeauthoryear{{Richard}, {Patricio}, {Martinez}, {Bacon},
  {Cl{\'e}ment}, {Weilbacher}, {Soto}, {Wisotzki}, {Vernet}, {Pello}, {Schaye},
  {Turner} \& {Martinsson}}{{Richard} et~al.}{2015}]{Richard+15}
{Richard} J.,  {Patricio} V.,  {Martinez} J.,  {Bacon} R.,  {Cl{\'e}ment} B.,
  {Weilbacher} P.,  {Soto} K.,  {Wisotzki} L.,  {Vernet} J.,  {Pello} R.,
  {Schaye} J.,  {Turner} M.,    {Martinsson} T.,  2015, \mnras, 446, L16

\bibitem[\protect\citeauthoryear{{Rivera-Thorsen}, {Hayes}, {{\"O}stlin},
  {Duval}, {Orlitov{\'a}}, {Verhamme}, {Mas-Hesse} \&
  {Schaerer}}{{Rivera-Thorsen} et~al.}{2015}]{Rivera-Thorsen+15}
{Rivera-Thorsen} T.~E.,  {Hayes} M.,  {{\"O}stlin} G.,  {Duval} F.,
  {Orlitov{\'a}} I.,  {Verhamme} A.,  {Mas-Hesse} J.~M.,    {Schaerer} D.,
  2015, \apj, 805, 14

\bibitem[\protect\citeauthoryear{{Rix}, {Pettini}, {Leitherer}, {Bresolin},
  {Kudritzki} \& {Steidel}}{{Rix} et~al.}{2004}]{Rix+04}
{Rix} S.~A.,  {Pettini} M.,  {Leitherer} C.,  {Bresolin} F.,  {Kudritzki}
  R.-P.,    {Steidel} C.~C.,  2004, \apj, 615, 98

\bibitem[\protect\citeauthoryear{{Schaerer}, {Hayes}, {Verhamme} \&
  {Teyssier}}{{Schaerer} et~al.}{2011}]{Schaerer+11}
{Schaerer} D.,  {Hayes} M.,  {Verhamme} A.,    {Teyssier} R.,  2011, \aap, 531,
  A12

\bibitem[\protect\citeauthoryear{{Seaton}}{{Seaton}}{1979}]{Seaton+79}
{Seaton} M.~J.,  1979, \mnras, 187, 73P

\bibitem[\protect\citeauthoryear{{Shapley}, {Steidel}, {Pettini} \&
  {Adelberger}}{{Shapley} et~al.}{2003}]{Shapley+03}
{Shapley} A.~E.,  {Steidel} C.~C.,  {Pettini} M.,    {Adelberger} K.~L.,  2003,
  \apj, 588, 65

\bibitem[\protect\citeauthoryear{{Sommariva}, {Mannucci}, {Cresci}, {Maiolino},
  {Marconi}, {Nagao}, {Baroni} \& {Grazian}}{{Sommariva}
  et~al.}{2012}]{Sommariva+12}
{Sommariva} V.,  {Mannucci} F.,  {Cresci} G.,  {Maiolino} R.,  {Marconi} A.,
  {Nagao} T.,  {Baroni} A.,    {Grazian} A.,  2012, \aap, 539, A136

\bibitem[\protect\citeauthoryear{{Stark}, {Schenker}, {Ellis}, {Robertson},
  {McLure} \& {Dunlop}}{{Stark} et~al.}{2013}]{Stark+13}
{Stark} D.~P.,  {Schenker} M.~A.,  {Ellis} R.,  {Robertson} B.,  {McLure} R.,
   {Dunlop} J.,  2013, \apj, 763, 129

\bibitem[\protect\citeauthoryear{{Stark}, {Swinbank}, {Ellis}, {Dye}, {Smail}
  \& {Richard}}{{Stark} et~al.}{2008}]{Stark+08}
{Stark} D.~P.,  {Swinbank} A.~M.,  {Ellis} R.~S.,  {Dye} S.,  {Smail} I.~R.,
  {Richard} J.,  2008, \nat, 455, 775

\bibitem[\protect\citeauthoryear{{Steidel}, {Bogosavljevi{\'c}}, {Shapley},
  {Kollmeier}, {Reddy}, {Erb} \& {Pettini}}{{Steidel}
  et~al.}{2011}]{Steidel+11}
{Steidel} C.~C.,  {Bogosavljevi{\'c}} M.,  {Shapley} A.~E.,  {Kollmeier} J.~A.,
   {Reddy} N.~A.,  {Erb} D.~K.,    {Pettini} M.,  2011, \apj, 736, 160

\bibitem[\protect\citeauthoryear{{Steidel}, {Erb}, {Shapley}, {Pettini},
  {Reddy}, {Bogosavljevi{\'c}}, {Rudie} \& {Rakic}}{{Steidel}
  et~al.}{2010}]{Steidel+10}
{Steidel} C.~C.,  {Erb} D.~K.,  {Shapley} A.~E.,  {Pettini} M.,  {Reddy} N.,
  {Bogosavljevi{\'c}} M.,  {Rudie} G.~C.,    {Rakic} O.,  2010, \apj, 717, 289

\bibitem[\protect\citeauthoryear{{Swinbank}, {Bower}, {Smith}, {Wilman},
  {Smail}, {Ellis}, {Morris} \& {Kneib}}{{Swinbank} et~al.}{2007}]{Swinbank+07}
{Swinbank} A.~M.,  {Bower} R.~G.,  {Smith} G.~P.,  {Wilman} R.~J.,  {Smail} I.,
   {Ellis} R.~S.,  {Morris} S.~L.,    {Kneib} J.-P.,  2007, \mnras, 376, 479

\bibitem[\protect\citeauthoryear{{Swinbank}, {Webb}, {Richard}, {Bower},
  {Ellis}, {Illingworth}, {Jones}, {Kriek}, {Smail}, {Stark} \& {van
  Dokkum}}{{Swinbank} et~al.}{2009}]{Swinbank+09}
{Swinbank} A.~M.,  {Webb} T.~M.,  {Richard} J.,  {Bower} R.~G.,  {Ellis} R.~S.,
   {Illingworth} G.,  {Jones} T.,  {Kriek} M.,  {Smail} I.,  {Stark} D.~P.,
  {van Dokkum} P.,  2009, \mnras, 400, 1121

\bibitem[\protect\citeauthoryear{{Swinbank}, {Vernet}, {Smail}, {De Breuck},
  {Bacon}, {Contini}, {Richard}, {Rottgering}, {Urritia} \&
  {Venemans}}{{Swinbank} et~al.}{2015}]{Swinbank+15}
{Swinbank} M.,  {Vernet} J.,  {Smail} I.,  {De Breuck} C.,  {Bacon} R.,
  {Contini} T.,  {Richard} J.,  {Rottgering} H.,  {Urritia} T.,    {Venemans}
  B.,  2015, ArXiv e-prints

\bibitem[\protect\citeauthoryear{{Vanzella}, {Giavalisco}, {Dickinson},
  {Cristiani}, {Nonino}, {Kuntschner}, {Popesso}, {Rosati}, {Renzini}, {Stern},
  {Cesarsky}, {Ferguson} \& {Fosbury}}{{Vanzella} et~al.}{2009}]{Vanzella+09}
{Vanzella} E.,  {Giavalisco} M.,  {Dickinson} M.,  {Cristiani} S.,  {Nonino}
  M.,  {Kuntschner} H.,  {Popesso} P.,  {Rosati} P.,  {Renzini} A.,  {Stern}
  D.,  {Cesarsky} C.,  {Ferguson} H.~C.,    {Fosbury} R.~A.~E.,  2009, \apj,
  695, 1163

\bibitem[\protect\citeauthoryear{{Vanzella}, {Giavalisco}, {Inoue}, {Nonino},
  {Fontanot}, {Cristiani}, {Grazian}, {Dickinson}, {Stern}, {Tozzi},
  {Giallongo}, {Ferguson}, {Spinrad}, {Boutsia}, {Fontana}, {Rosati} \&
  {Pentericci}}{{Vanzella} et~al.}{2010}]{Vanzella+10}
{Vanzella} E.,  {Giavalisco} M.,  {Inoue} A.~K.,  {Nonino} M.,  {Fontanot} F.,
  {Cristiani} S.,  {Grazian} A.,  {Dickinson} M.,  {Stern} D.,  {Tozzi} P.,
  {Giallongo} E.,  {Ferguson} H.,  {Spinrad} H.,  {Boutsia} K.,  {Fontana} A.,
  {Rosati} P.,    {Pentericci} L.,  2010, \apj, 725, 1011

\bibitem[\protect\citeauthoryear{{Verhamme}, {Orlitov{\'a}}, {Schaerer} \&
  {Hayes}}{{Verhamme} et~al.}{2015}]{Verhamme+15}
{Verhamme} A.,  {Orlitov{\'a}} I.,  {Schaerer} D.,    {Hayes} M.,  2015, \aap,
  578, A7

\bibitem[\protect\citeauthoryear{{Verhamme}, {Schaerer}, {Atek} \&
  {Tapken}}{{Verhamme} et~al.}{2008}]{Verhamme+08}
{Verhamme} A.,  {Schaerer} D.,  {Atek} H.,    {Tapken} C.,  2008, \aap, 491, 89

\bibitem[\protect\citeauthoryear{{Verhamme}, {Schaerer} \&
  {Maselli}}{{Verhamme} et~al.}{2006}]{Verhamme+06}
{Verhamme} A.,  {Schaerer} D.,    {Maselli} A.,  2006, \aap, 460, 397

\bibitem[\protect\citeauthoryear{{Villar-Mart{\'{\i}}n}, {Cervi{\~n}o} \&
  {Gonz{\'a}lez Delgado}}{{Villar-Mart{\'{\i}}n}
  et~al.}{2004}]{Villar-Martin+04}
{Villar-Mart{\'{\i}}n} M.,  {Cervi{\~n}o} M.,    {Gonz{\'a}lez Delgado} R.~M.,
  2004, \mnras, 355, 1132

\bibitem[\protect\citeauthoryear{{Weijmans}, {Bower}, {Geach}, {Swinbank},
  {Wilman}, {de Zeeuw} \& {Morris}}{{Weijmans} et~al.}{2010}]{Weijmans+10}
{Weijmans} A.-M.,  {Bower} R.~G.,  {Geach} J.~E.,  {Swinbank} A.~M.,  {Wilman}
  R.~J.,  {de Zeeuw} P.~T.,    {Morris} S.~L.,  2010, \mnras, 402, 2245

\bibitem[\protect\citeauthoryear{{Wisotzki}, {Bacon}, {Blaizot}, {Brinchmann},
  {Herenz} \& {Schaye}}{{Wisotzki} et~al.}{2015}]{Wisotzki+15}
{Wisotzki} L.,  {Bacon} R.,  {Blaizot} J.,  {Brinchmann} J.,  {Herenz} E.~C.,
   {Schaye} 2015, astro-ph/1509.05143

\bibitem[\protect\citeauthoryear{{Wuyts}, {Rigby}, {Sharon} \&
  {Gladders}}{{Wuyts} et~al.}{2012}]{Wuyts+12b}
{Wuyts} E.,  {Rigby} J.~R.,  {Sharon} K.,    {Gladders} M.~D.,  2012, \apj,
  755, 73

\bibitem[\protect\citeauthoryear{{Yuan}, {Kewley}, {Swinbank}, {Richard} \&
  {Livermore}}{{Yuan} et~al.}{2011}]{Yuan+11}
{Yuan} T.-T.,  {Kewley} L.~J.,  {Swinbank} A.~M.,  {Richard} J.,    {Livermore}
  R.~C.,  2011, \apjl, 732, L14

\end{thebibliography}

\section*{Appendix A}

We describe here the information that can be retrieved from the intervening absorbers detected in the spectrum of System 1. By reconstructing the source plane image at each redshift we were able to derive the impact parameters for the three systems, 33 kpc for s2, 41 kpc for system 12 and 26 kpc for system 4 (proper coordinates). From the same MUSE data set, the spectrum of the intervening systems galaxies was also extracted and examined (see figure 3 from \cite{Richard+15} for systems 4 and 12). System 4 has a very faint continuum, and only the \Lya line can be identified, so not much information could be derived. As for system 12 and s2, both had enough signal in the continuum to allow us to measure systemic redshift from other lines than \Lya (\mbox{C\,{\sc iv}}  doublet in absorption for both systems) as well as to derive stellar mass and SFR (Table \ref{sys12_s2}) using the same procedure as described in Sect. \ref{stellarpop}.

One of the questions that can be addressed by the study of intervening absorbers is whether we are probing in-falling gas or an outflow (or galactic wind) and what are their characteristics. It has been proposed that the azimuthal angle (the angle between the disk of the absorbing galaxy and the direction of the illuminating system) can be used to make the distinction between an outflow or an inflow through geometrical arguments: moving gas close to the disk vertical direction (azimuthal angles close to 90 degrees) has its most probable origin in galactic winds, since this is the only mechanism that can produce cool material systematically along the minor axis \citep{Bouche+12}. Comparing our azimuthal angles with the bimodal distribution obtained in \cite{Bouche+12}, we see that s2 is consistent with being in the inflow group but that system 12 is still compatible with the galactic wind hypothesis. This last system gas has a $\Delta$v of 144 km\,s$^{-1}$ and a specific SFR of 9.94 Gyr$^{-1}$. We briefly compare our results to those from \citet{Steidel+10}, who used stacks of low-resolution spectra from star-forming galaxies at z$\sim$2.5 to obtain the typical equivalent widths of intervening absorption lines as a function of impact parameter. While we find that our rest-frame  \mbox{Si\,{\sc iv}} equivalent width (0.45$\pm$0.02$\AA$) agrees well with theirs (0.39$\pm$0.08 $\AA$ at a mean impact parameter of 31 proper kpc), system 12 shows a significantly lower equivalent width of \mbox{C\,{\sc iv}}. Specifically, since \citet{Steidel+10} do not resolve the \mbox{C\,{\sc iv}} doublet, to facilitate comparison we add our two components together to obtain a rest-frame CIV EW of 0.498$\pm$0.001$\AA$ (while \citealt{Steidel+10} obtain 2.13$\pm$0.15 $\AA$). This discrepancy could be attributed to, i.e., differences in ionisation state or relative abundances, however we leave a more detailed analysis to a future work.

\begin{table}
    \begin{tabular}{l r | r | r | r | r | }
                \hline  
$ \lambda_{obs}  $ & EW   &  line & $z$ &  system \\  
  $[\AA$] & [$\AA$] & [$\AA$]   &    & \\
          \hline \hline 
5550.68	&$	1.15	\pm	0.074	$&  \mbox{Si\,{\sc iv}} & 	2.984 	& s2	\\
5737.58	&$	0.40	\pm	0.198	$&	(1) 			&			& 	\\
5766.45	&$	0.63	\pm	0.034	$&	(2) 			&			& 	\\ 	
5775.84	&$	2.15	\pm	0.127	$&  H$\gamma$ 	& 	0.331	& cluster	 \\
6081.24	&$	0.46	\pm	0.021	$&  \mbox{Si\,{\sc ii}} & 	2.984	&	s2	\\
6153.12	&$	1.99	\pm	0.074	$&  \mbox{Si\,{\sc iv}}  & 	3.414	&	system 12	\\
6166.40	&$	2.71	\pm	0.027	$&  \mbox{C\,{\sc iv}}  &	2.984	&	s2	\\
6176.66	&$	2.04	\pm	0.012	$&   \mbox{C\,{\sc ii}}  &	2.984	&	s2 \\
6192.28	&$	1.44	\pm	0.076	$&  \mbox{Si\,{\sc iv}}  &	3.414	&	system 12	\\
6374.71	&$	0.92	\pm	0.056	$&	(3) 			&			& 	\\
6622.59	&$	0.22	\pm	0.090	$&	\mbox{Si\,{\sc ii}}  &	1.158	&	s4	\\
6654.27	&$	0.51	\pm	0.052	$&   \mbox{Al\,{\sc ii}}  & 2.984		&	s2	\\
6718.53	&$	0.30	\pm	0.090	$& \mbox{C\,{\sc iv}}  & 3.340		&	system 4	\\
6729.72	&$	0.03	\pm	0.090	$& \mbox{C\,{\sc iv}}  & 3.340		&	system 4	\\
6835.88	&$	1.10	\pm	0.002	$& \mbox{C\,{\sc iv}}  &	3.414	&	system 12	\\
6847.26	&$	1.10	\pm	0.004	$& \mbox{C\,{\sc iv}}  &	3.414	&	system 12	\\
9127.71	&$	2.84	\pm	0.383	$& \mbox{Mg\,{\sc ii}}  & 2.2653	& $^{(*)}$	\\
9152.61	&$	2.75	\pm	0.383	$&  \mbox{Mg\,{\sc ii}}  & 2.2653	& $^{(*)}$	\\        
               \hline  
     \end{tabular}
     
     $^{(*)}$ Absorber probably related to system 11 ($z=2.256$).
	 \caption[inter]{Absorption lines not identified as system 1 lines. Given wavelengths are in observed frame and in air and equivalent widths in observed frame. Last columns  matches the observed line with one of the intervening systems taken from \citep{Richard+15}, when the identification was possible, otherwise a label is given.}      
     \label{table_inter}          
\end{table}

\begin{table}
    \begin{tabular}{| r l r | r | r |}
                \hline  
      Parameter & Unit & s2 & sys12 \\
             \hline \hline       
      $z_{sys} $ & & 2.9840	 &  3.4162 \\
      $z_{int}$ & & 2.9850    & 3.4141	 \\
      $\Delta$v  & [km\,s$^{-1}$] & -79$\pm$40 & 144	$\pm$ 32\\
      Mass  & [$\log$(M$_\star$/M$_\odot$)] & 8.40 & 9.29 \\
      SFR & [M$_\odot$/yr]  & 2.1 & 5.1 \\
      $\theta $ & [deg] & 46 & 27 \\
      $\alpha$ & [deg]  & 25.5 & 64 \\
      b   &  [kpc] & 33 & 41 \\ 
               \hline  
     \end{tabular}
	 \caption[]{Physical parameters derived for the s2 and system 12 absorbers, respectively.  $z_{sys}$ was measured in absorption lines (and emission lines whenever possible) on the absorber spectrum and $z_{int}$ from the absorption lines seen in the spectrum of system 1. Mass and SFR were estimated with the same techniques as for system 1 (see Sect. \ref{stellarpop}). Inclination ($\theta$), azimuthal angle ($\alpha$), and impact parameter (b) were measured on the reconstructed source plane images of each system.}      
     \label{sys12_s2}          
\end{table}


\label{lastpage}

\end{document}